\documentclass[aps,nofootinbib,preprintnumbers]{revtex4}
\usepackage{lipsum}
\usepackage{amsfonts}
\usepackage{amsmath}
\usepackage{amssymb}
\usepackage[english]{babel}
\usepackage{stackrel}
\usepackage{mathtools}
\usepackage{graphicx}
\usepackage{epsfig}
\usepackage{bm}
\usepackage{verbatim}
\usepackage[utf8]{inputenc}
\usepackage{booktabs}
\usepackage{multirow}
\usepackage{subfig}
\usepackage{slashed}
\usepackage{xcolor}
\usepackage[colorlinks=true,urlcolor=red,citecolor=red]{hyperref}
\usepackage[font=small]{caption}
\usepackage{float}
\usepackage{soul}
\usepackage{blindtext}
\usepackage{placeins}
\usepackage[utf8]{inputenc}
\usepackage{bbm}
\usepackage[colorinlistoftodos]{todonotes}
\usepackage{tikz}
\usepackage[compat=1.1.0]{tikz-feynman}
\usetikzlibrary{arrows.meta, decorations.pathmorphing}
\usepackage{csquotes}
\usepackage[compat=1.1.0]{tikz-feynman}
\newenvironment{Eqnarray}{\arraycolsep 0.14em\begin{eqnarray}}{\end{eqnarray}}
\newcommand{\ba}{\begin{Eqnarray}}
\newcommand{\ea}{\end{Eqnarray}}
\newcommand{\be}{\begin{equation}}
\newcommand{\ee}{\end{equation}}
\newcommand{\bal}{\begin{aligned}}
\newcommand{\eal}{\end{aligned}}
\newcommand{\bea}{\begin{eqnarray}}
\newcommand{\eea}{\end{eqnarray}}
\newcommand{\ben}{\begin{enumerate}}
\newcommand{\een}{\end{enumerate}}
\newcommand{\bit}{\begin{itemize}}
\newcommand{\eit}{\end{itemize}}
\newcommand{\bde}{\begin{widetext}}
\newcommand{\ede}{\end{widetext}}

\def\lsim{\mathrel{\rlap{\lower4pt\hbox{\hskip1pt$\sim$}}
    \raise1pt\hbox{$<$}}}
\def\gsim{\mathrel{\rlap{\lower4pt\hbox{\hskip1pt$\sim$}}
    \raise1pt\hbox{$>$}}}

\def\3211{$\mathrm{SU(3) \otimes SU(2)_L \otimes U(1)_R \otimes U(1)_{B-L}}$ }
\def\321{$\mathrm{SU(3) \otimes SU(2) \otimes U(1)}$ }
\def\422{$\mathrm{SU(4) \otimes SU(2) \otimes SU(2)_R}$ }

\newcommand{\abs}[1]{\left| #1 \right| }

\newcommand{\mathsym}[1]{{}}

\definecolor{bostonuniversityred}{rgb}{0.8, 0.0, 0.0}

\definecolor{Mycolor1}{RGB}{0,128,0}
\begin{document}

\title{A novel two loop inverse seesaw model}
%\vspace{-0.2cm}
\author{Gonzalo Ben\'itez-Irarr\'azabal}
\email{gibi1e25@soton.ac.uk}
\affiliation{School of Physics and Astronomy, University of Southampton, Highfield, Southampton SO17 1BJ, UK}

\author{Roc\'io Branada Balbont\'in}
\email{rocio.branada@sansano.usm.cl}
\affiliation{Instituto de F\'{\i}sica, Pontificia Universidad Cat\'olica de Valpara\'iso, Casilla 4950, Valpara\'{\i}so 2373223, Chile}
\affiliation{Departamento de F\'{\i}sica, Universidad T\'ecnica Federico Santa Mar\'{\i}a, Casilla 110-V, Valpara\'{\i}so, Chile}

\author{Cesar Bonilla}
\email{cesar.bonilla@ucn.cl}
\affiliation{Departamento de F\'{\i}sica, Universidad Cat\'olica del Norte, Avenida
Angamos 0610, Casilla 1280, Antofagasta, Chile}
\author{A. E. C\'arcamo Hern\'andez}
\email{antonio.carcamo@usm.cl}
\affiliation{{Universidad T\'ecnica Federico Santa Mar\'{\i}%
a, Casilla 110-V, Valpara\'{\i}so, Chile}}
\affiliation{{Centro Cient\'{\i}fico-Tecnol\'ogico de Valpara\'{\i}so, Casilla 110-V,
Valpara\'{\i}so, Chile}}
\affiliation{{Millennium Institute for Subatomic Physics at High-Energy Frontier
(SAPHIR), Fern\'andez Concha 700, Santiago, Chile}}
\author{Sergey Kovalenko}
\email{sergey.kovalenko@unab.cl}
\affiliation{Departamento de F\'isica y Astronomía, Universidad Andr\'es Bello, Sazi\'e
2212, Piso 7, Santiago, Chile}

\affiliation{Millennium Institute for Subatomic Physics at High-Energy Frontier (SAPHIR),
Fern\'andez Concha 700, Santiago, Chile}

\author{Juan Marchant Gonz\'{a}lez}
\email{juan.marchant@upla.cl}
\affiliation{Laboratorio de C\'omputo de F\'isica (LCF-UPLA), Facultad de Ciencias
Naturales y Exactas, Universidad de Playa Ancha, Subida Leopoldo Carvallo
270, Valpara\'iso, Chile.}

\begin{abstract}
We propose a Standard Model (SM) extension where neutrinos get masses through a two-loop inverse seesaw mechanism. This naturally explains the smallness of the neutrino masses and allows seesaw mediators to be at the TeV scale with testable phenomenology. The model adds two real singlet scalars and four electrically neutral leptons to the SM. The extension considers the existence of two global Abelian symmetries, a continuous $U(1)$ and a discrete $Z_3$. The latter, remains unbroken after spontaneous symmetry breaking and forbids tree-level and one-loop neutrino masses, and stabilizes the dark matter (DM) candidates. This setup accommodates neutrino-oscillation data, yields two pseudo-Dirac heavy pairs with small active–sterile mixing, and predicts an effective Majorana mass $m_{ee}$ in the $2.1$–$4.4~\mathrm{meV}$ range for normal ordering. Charged-lepton flavor violation is naturally suppressed yet testable: for a representative benchmark we obtain $\mathrm{BR}(\mu\!\to\!e\gamma)\simeq 1.6\times 10^{-14}$, with correlated signals in $\mu\!\to\! eee$ and $\mu$–$e$ conversion within next-generation experimental reach. Altogether, the radiative origin of neutrino masses links low-energy flavor observables to collider signatures, delineating discovery targets for MEG~II, Mu2e/COMET, and the HL-LHC and distinguishing this framework from conventional inverse- and radiative-seesaw models.
 Moreover, the $Z_3$ guarantees a stable DM candidate, either scalar ($\rho$) or fermionic ($\Omega$). Then, here we analyze and identify the viable parameter space that is consistent with the observed DM relic abundance for both situations.
%}
\end{abstract}
% Optional, if your journal supports it:
\vspace{-0.2cm}

%%%%%%%%%%%%%%%%%%%%

\maketitle

\section{Introduction}
The presence of dark matter (DM) and the tiny masses of active neutrinos stand out as strong motivations for physics beyond the Standard Model (BSM). A minimal BSM extension that introduces right-handed ({\color{purple} RH}) Majorana neutrinos can generate neutrino masses via the type-I seesaw mechanism \cite{Minkowski:1977sc,Yanagida:1979as,Glashow:1979nm,Mohapatra:1979ia,Gell-Mann:1979vob, Schechter:1980gr, Schechter:1981cv}. %{\bf \color{purple} Refs}. Done. 
However, achieving the correct neutrino mass scale in this framework typically requires either extremely large (RH) Majorana masses or highly suppressed Dirac Yukawa couplings. Consequently, the mixing between light and heavy neutrinos becomes negligible, leading to extremely suppressed rates for charged lepton flavor violating (CLFV) processes—well below the reach of current experimental detection. In addition, such a minimal seesaw model does not provide a stable dark matter (DM) candidate to explain the observed DM relic density.

A natural connection between the neutrino mass generation and a stable dark sector comes out in  theories of radiative neutrino masses,  %radiative neutrino mass generation models
 see for instance Ref.~\cite{Cai:2017jrq} for a review and Refs.~\cite{Jana:2019mgj,Arbelaez:2022ejo} for comprehensive studies of one- and two-loop radiative neutrino mass models. The minimal scenario introduces $Z_2$-odd particles to the SM, RH-neutrinos and a $SU(2)$ scalar doublet which does not develop a vacuum expectation value (vev) as described in Ref. \cite{Tao:1996vb} for the first time. The $Z_2$ symmetry remains intact after spontaneous symmetry breaking, stabilizes the DM candidates, and prevents neutrino masses at tree-level.

An attractive alternative is offered by models that realize the inverse seesaw mechanism ~\cite{Wyler:1982dd, Mohapatra:1986bd, GonzalezGarcia:1988rw, Akhmedov:1995ip, Akhmedov:1995vm, Malinsky:2005bi, Ma:2009gu, Malinsky:2009df, Bazzocchi:2010dt, Law:2012mj, Das:2012ze, Abada:2014vea, Fraser:2014yha, Ahriche:2016acx, CarcamoHernandez:2013krw, Das:2017ski, CarcamoHernandez:2017owh, CarcamoHernandez:2018hst, CarcamoHernandez:2018iel, Bertuzzo:2018ftf, Mandal:2019oth, Das:2019pua, CarcamoHernandez:2019eme, CarcamoHernandez:2019pmy, CarcamoHernandez:2019vih, CarcamoHernandez:2019lhv, Hernandez:2021uxx,  Hernandez:2021xet, Hernandez:2021kju, Nomura:2021adf, Hernandez:2021mxo, Abada:2021yot, Abada:2023zbb, Bonilla:2023egs, Bonilla:2023wok, Binh:2024lez, CarcamoHernandez:2024hll, Huong:2025uwx} . These scenarios attribute the smallness of active neutrino masses to a tiny Majorana mass term that softly breaks lepton number symmetry. This term can itself be generated radiatively at one or more loop levels. Notably, the inverse seesaw allows for sizable mixings between active and heavy neutrinos, enabling CLFV rates that may be accessible in forthcoming experiments. When embedded in radiative or “scotogenic” setups, inverse seesaw models can simultaneously address dark matter and neutrino mass generation. In such models, some of the neutral seesaw mediators can serve as viable dark matter candidates, stabilized by the same discrete symmetries that forbid tree-level neutrino masses. The correct relic abundance can be achieved through dark matter annihilations into Standard Model or additional BSM particles within suitable regions of parameter space.

The inverse seesaw mechanism offers a viable framework for resonant leptogenesis, facilitated by the small mass splitting of heavy pseudo-Dirac neutral leptons ~\cite{Pilaftsis:1997jf,Gu:2010xc,Dolan:2018qpy,Dib:2019jod,Blanchet:2009kk,Blanchet:2010kw,Hernandez:2021xet,Hernandez:2021kju,Hernandez:2021uxx,Abada:2023zbb,Binh:2024lez}
In this paper, we present an SM extension where active neutrino masses arise radiatively via a two-loop inverse seesaw. The model extends the SM particle content with just two gauge-singlet scalars and a minimal set of neutral leptons, while incorporating an additional global $U(1)_X$ symmetry and a discrete $\mathbb{Z}_{3}$ symmetry. In our proposed model, the spontaneous breaking of the $U(1)_X$ symmetry occurs, whereas the $\mathbb{Z}_{3}$ symmetry remains intact, enabling the radiative generation of a lepton-number-violating Majorana mass terms at two loops with the same topology of the Zee-Babu mechanism. The preserved $\mathbb{Z}_{3}$ symmetry in our model also ensures stable dark matter candidates, both scalar and fermionic. The proposed framework aligns with neutrino oscillation data and satisfies constraints arising from charged lepton flavor violation, dark matter relic abundance, and direct/indirect detection dark matter experiments. Our proposed model, has the following advantages with respect with the Zee-Babu model: 1) It incorporates Dark Matter candidates in the neutrino mass generation mechanism, which corresponds to a scotogenic feature, absent in the Zee Babu model. 2) Our model has two pairs of nearly degenerate Pseudo-Dirac leptons, whose small mass splitting, corresponding to the two loop level induced lepton number violating mass parameter is crucial for achieving successfull TeV scale leptogenesis. 
The paper is organized as follows. We discuss the proposed model in Sec. \ref{model}. The scalar potential is discussed in Sec. \ref{scalar potential}. In Sec. \ref{neutrino masses}. we discuss the neutrino sector and how the masses arise. In Sec. \ref{dark matter}. we presented a framework in which a singlet scalar and a fermionic DM candidate are studied separately and are contrasted with experimental measurements and projections. In Sec. \ref{clfv}. we discuss some sizeble CLFV rates. Our conclusions are stated in section \ref{sec:conclusions}. In Appendix \ref{App:2loop-1} we provide a discussion of the two–loop integral for the inverse seesaw $\mu$-entry in our model.

\section{The model}

\label{model} 
We propose the Standard Model (SM)  extension with  a  spontaneously broken global
$U(1)_X$ and an exact $\mathbb{Z}_3$ discrete symmetries. The field content is enlarged with two electrically neutral, gauge singlet scalars  $\rho$ and $\sigma$, and
with two generations ($n=1,2$) of neutral leptons $\nu_{nR}$, $N_{nR}$, $\Omega_{nL}$, and $\Omega_{nR}$.  In this setup, the tiny masses of the active neutrinos arise from an 
\emph{inverse seesaw mechanism} realized at the two-loop level. The preserved
$\mathbb{Z}_3$ symmetry plays a dual role: it (i) forbids any tree-level
mixing between $N$ and $\Omega$, thereby eliminating a tree-level (and
one-loop) source for the lepton-number–violating (LNV) submatrix $\mu$ in the
inverse-seesaw mass matrix, fixing the dominant two-loop topology for $\mu$ and (ii) stabilizes the Dark Matter candidates of the model. The model matter content and its charge assignments under the new symmetries is summarized in Table~\ref{model} .The Standard Model quarks are assumed to transform trivially under the new symmetries. Notice that $\mathbb{Z}_3$ is the smallest Abelian group that promotes the two-loop origin of $\mu$.
 
The scalar fields are written as follows,
\begin{equation}
\phi =\left( 
\begin{array}{c}
\phi ^{+} \\ 
\frac{v_{\phi }+\phi_R+i\xi }{\sqrt{2}}
\end{array}
\right), \ \ \ \sigma =\frac{v_{\sigma }+\sigma _{R}+i\sigma _{I}}{\sqrt{2}}, \ \ \ \rho =%
\frac{\rho _{R}+i\rho _{I}}{\sqrt{2}},
\label{eq: scalar_fields}
\end{equation}
where $\phi$ is the SM Higgs doublet,  and $\sigma,\rho$ are singlets. Here, $v_\phi\approx 246$ GeV. Then, only the SM gauge doublet $\phi$ and the gauge singlet scalar $\sigma$ participate in the spontaneous symmetry breaking after getting vacuum expectation values (VEVs) given by, 
\begin{equation}
\left\langle \phi \right\rangle =\left( 
\begin{array}{c}
0 \\ 
\frac{v_{\phi }}{\sqrt{2}}
\end{array}
\right), \ \ \ \left\langle \sigma \right\rangle =\dfrac{v_{\sigma }}{\sqrt{2}}
\end{equation}
we get the following sequential symmetry breaking pattern: 
\begin{eqnarray}
&&\mathcal{G}\equiv \mathrm{SU}(3)_{\mathrm{C}}\times \mathrm{SU}(2)_{%
\mathrm{L}}\times \mathrm{U}(1)_{\mathrm{Y}}\times \mathrm{U}(1)_{\mathrm{X}%
}\times \mathbb{Z}_{3}{\xrightarrow{v_\sigma}}  \notag \\
&&\hspace{7mm}\mathrm{SU}(3)_{\mathrm{C}}\times \mathrm{SU}(2)_{\mathrm{L}%
}\times \mathrm{U}(1)_{\mathrm{Y}}\times \mathbb{Z}_{3}{\xrightarrow{v_\phi}}
\notag \\
&&\hspace{7mm}\mathrm{SU}(3)_{\mathrm{C}}\times \mathrm{U}(1)_{\mathrm{Q}%
}\times \mathbb{Z}_{3}  \label{SB}
\end{eqnarray}%
%It is assumed a sequential symmetry breaking pattern as
%\begin{equation}
%\mathrm{SU}(2)_{%
%\mathrm{L}}\times \mathrm{U}(1)_{\mathrm{Y}}\times \mathrm{U}(1)_{\mathrm{X}%
%}\times \mathbb{Z}_{3}\,\xrightarrow{\,\,v_\sigma\,\,}\,
%\mathrm{SU}(2)_{\mathrm{L}%
%}\times \mathrm{U}(1)_{\mathrm{Y}}\times \mathbb{Z}_{3}\,\xrightarrow{\,\,v_\phi\,\,}\,\mathrm{U}(1)_{\mathrm{em}%
%}\times \mathbb{Z}_{3}  \label{SB}
%\end{equation}%
where the global $\mathrm{U}(1)_{\mathrm{X}}$ symmetry is spontaneously broken by the scalar singlet VEV $v_\sigma $, then the SM model gauge symmetry gets broken by the scalar doublet VEV $v_\phi $. We assume $v_\sigma \gg v_\phi $ The Abelian $\mathbb{Z}_{3}$ symmetry is preserved after spontaneous symmetry breaking just like color and electric charge.

\begin{table}[tbp]
\centering%
\begin{tabular}{c|ccccccccc}
\hline\hline
& $l_{iL}$ & $l_{iR}$ & $\nu _{kR}$ & $N_{kR}$ & $\Omega _{kL}$ & $\Omega
_{kR}$ & $\phi $ & $\rho $ & $\sigma $ \\ \hline\hline
$SU(3)_{C}$ & \textbf{$1$} & \textbf{$1$} & \textbf{$1$} & \textbf{$1$} & 
\textbf{$1$} & \textbf{$1$} & \textbf{$1$} & \textbf{$1$} & \textbf{$1$} \\ 
$SU(2)_{L}$ & \textbf{$2$} & \textbf{$1$} & \textbf{$1$} & \textbf{$1$} & 
\textbf{$1$} & \textbf{$1$} & \textbf{$2$} & \textbf{$1$} & \textbf{$1$} \\ 
$U(1)_{Y}$ & $-\frac{1}{2}$ & $-1$ & $0$ & $0$ & $0$ & $0$ & $\frac{1}{2}$ & $0$ & $0$
\\ 
$U(1)_{X}$ & $1$ & $1$ & $1$ & $-1$ & $-1$ & $0$ & $0$ & $0$ & $-1$ \\ 
$\mathbb{Z}_{3}$ & $0$ & $0$ & $0$ & $0$ & $-1$ & $-1$ & $0$ & $-1$ & $0$ \\ 
%$U(1)_{ac}$ & $0$ & $0$ & $0$ & $0$ & $-1$ & $-1$ & $0$ & $2$ & $0$ \\
\hline\hline
\end{tabular}%
\caption{Charge assignments of leptonic and scalar fields of the model under
the group $\mathrm{SU}(3)_{\mathrm{C}}\times \mathrm{SU}(2)_{\mathrm{L}%
}\times \mathrm{U}(1)_{\mathrm{Y}}\times \mathrm{U}(1)_{\mathrm{X}}\times 
\mathbb{Z}_{3}$. Here $i=1,2,3$ and $k=1,2$.}
\label{tab:charge}
\end{table}

\section{Scalar potential}
\label{scalar potential}
The most general scalar potential invariant under the symmetry described in the Table \ref{tab:charge}, reads as:
\begin{align}
    V=&-\mu_\phi^2(\phi^\dagger\phi)-\mu_\sigma^2(\sigma^\dagger\sigma)+\mu_\rho^2(\rho^\dagger\rho)+A\left[\rho^3+h.c.\right]+\lambda_{\phi}(\phi^\dagger\phi)^{2} 
    + \lambda_{\sigma}(\sigma^\dagger\sigma)^{2} \nonumber \\
    &+\lambda_{\rho}(\rho^\dagger\rho)^{2}
    + \lambda_{\phi \sigma}(\phi^\dagger\phi)(\sigma^\dagger\sigma) + \lambda_{\phi \rho}(\phi^\dagger\phi)(\rho^\dagger\rho) + \lambda_{\sigma \rho}(\sigma^\dagger\sigma)(\rho^\dagger\rho).
    \label{eq:scalar_potential}
\end{align}

Due to the global $U(1)_X$ and the discrete $Z_3$ symmetries, there is only one trilinear interaction in the scalar potential, i.e., $A\,(\rho^3+\text{h.c.})$. As a result  and due to the fact that  $\langle\rho\rangle=0$, there is no splitting between $m_{\rho_R}$ and $m_{\rho_I}$ neither Higgs--$\rho$ mixing in our model. The former means that $\rho$ is a  complex SM-singlet scalar $\rho$ with the mass 
\begin{eqnarray}
\label{eq:mRho-1}
&&m^2_{\rho}=\mu_\rho^2+\dfrac{1}{2}\lambda_{\rho\sigma}v_\sigma^2+\dfrac{1}{2}\lambda_{\phi\rho}v_\phi^2.
\end{eqnarray}

The absence of the Higgs--$\rho$ mixing, as will be explained in Section~\ref{sec:FermionicDM-DDM}, results in loop suppression of the scattering of our fermionic DM candidate $\Omega$ on nuclei. This allows  relaxation of the limits from direct DM detection on the relic density.

From the minimization conditions for the scalar potential \eqref{eq:scalar_potential} we find
\begin{align*}
    \mu_\phi^2=\lambda_\phi v_\phi^2+\dfrac{1}{2}\lambda_{\phi\sigma} v_\sigma^2, &&\mu_\sigma^2&=\lambda_\sigma v_\sigma^2+\dfrac{1}{2}\lambda_{\phi \sigma}v_\phi^2.
\end{align*}

Taking into account these relationships 
we obtain the squared mass matrix for the CP-even states 
% Taking into account the minimization conditions of the scalar potential , using as a matrix basis: $(h , \sigma_R )$, one part of the squared even mass is given by:
\begin{align}
    M_{\phi_R\sigma_R}^2=\left(\begin{array}{cc}
         2\lambda_\phi v_\phi^2&\lambda_{\phi\sigma}v_\sigma v_\phi  \\
         \lambda_{\phi\sigma}v_\sigma v_\phi&2\lambda_\sigma v_\sigma^2 
    \end{array}\right),
\end{align}
in the basis $(\phi_R, \sigma_R )$.
Diagonalizing this mass matrix with  orthogonal rotation of these fields  
\begin{equation}
\label{eq:scalar-mixing}
    \left(\begin{array}{c}
          \Tilde{\sigma} \\
          h
    \end{array}\right)=\left(\begin{array}{cc}
         \cos{\theta_{\phi\sigma}}&-\sin{\theta_{\phi\sigma}}  \\
         \sin{\theta_{\phi\sigma}}&\cos{\theta_{\phi\sigma}}   \end{array}\right)\left(\begin{array}{c}
          \sigma_R \\
          \phi_R\end{array}\right),
\end{equation}
we find the physical fields $\tilde\sigma, h$ with the masses 
\begin{equation}
    m^2_{\Tilde{\sigma},h}=\lambda_\phi v_\phi^2+\lambda_\sigma v_\sigma^2\pm\sqrt{(\lambda_\sigma v_\sigma^2-\lambda_\phi v_\phi^2)^2+\lambda^2_{\phi\sigma}v_\phi^2v_\sigma^2}
    \label{mass_new_fields}
\end{equation}
and 
%where $\theta_{\phi\sigma}$ is 
the mixing angle $\theta_{\phi\sigma}$, which depends on the scalar potential parameters
\begin{align}
\tan 2\theta_{\phi\sigma}=\dfrac{\lambda_{\phi\sigma}v_\sigma v_\phi}{\lambda_\sigma v_\sigma^2-\lambda_\phi v_\phi^2}.
\label{eq:mixing-angle}
\end{align}
The imaginary (CP-odd) part $\sigma_I$ of the SM-singlet scalar $\sigma$, as defined in \eqref{eq: scalar_fields}, is the massless Nambu--Goldstone boson resulting from the spontaneous breaking of the global $U(1)_X$ symmetry with the vev $\langle\sigma\rangle =v_\sigma$. 
% {\color{blue}On the other hand, the masses of the new fields $\Tilde{\sigma}$ and $h$ are

% \begin{equation}
%     m^2_{\Tilde{\sigma},h}=\lambda_\phi v_\phi^2+\lambda_\sigma v_\sigma^2\pm\sqrt{(\lambda_\sigma v_\sigma^2-\lambda_\phi v_\phi^2)^2+\lambda^2_{\phi\sigma}v_\phi^2v_\sigma^2}
%     \label{mass_new_fields}
% \end{equation}
% \
% }

%of the spontaneously broken global $U(1)_X$.
\subsection{ $U(1)_X$ Nambu--Goldstone boson as Dark radiation }

We denote this Majoron-like Nambu--Goldstone boson by $\sigma_I\equiv J$. 
%(majoron-like state). 
This particle may have  phenomenological and cosmological implications. 
Here we address two of the most direct ones: its contribution to the invisible Higgs decay $h\to JJ$ \cite{Hirsch:2004higgsMajoron, Bonilla:2015uwa} and  to the radiation density of the universe. In Sec.~\ref{scalar_dark_matter} we will also study its possible contribution to the DM relic abundance.  
%
%For this analysis it is convenient to 
Let us use a nonlinear parametrization for the field $\sigma$ of the form
\begin{equation}
\sigma=\frac{1}{\sqrt2}\left(v_\sigma+\sigma_R\right)\exp\!\left(\frac{iJ}{v_\sigma}\right).
\label{eq:sigma_nonlinear}
\end{equation}
From the kinetic term $|\partial_\mu\sigma|^2$ one finds
\begin{equation}
|\partial_\mu\sigma|^2 \supset \frac12(\partial_\mu J)(\partial^\mu J)\left(1+\frac{\sigma_R}{v_\sigma}\right)^2
\;\supset\; \frac{\sigma_R}{v_\sigma}(\partial_\mu J)(\partial^\mu J).
\label{eq:sRJ2}
\end{equation}
%The $\phi_R$--$\sigma_R$ mixing, specified in Eq.~(\ref{eq:scalar-mixing}), and taken into account in Eq.~\eqref{eq:sRJ2} induces an $hJJ$ interaction proportional to $\sin\theta_{\phi\sigma}/v_\sigma$, and hence an invisible Higgs decay channel.
The $\phi_R$–$\sigma_R$ mixing, specified in Eq.~(\ref{eq:scalar-mixing}) and taken into account in Eq.~\eqref{eq:sRJ2}, induces an $hJJ$ interaction proportional to $\sin\theta_{\phi\sigma}/v_\sigma$. This interaction opens an invisible Higgs decay channel. Using $\sigma_R = \cos\theta_{\phi\sigma}\,\tilde\sigma + \sin\theta_{\phi\sigma}\,h$, one obtains the corresponding rate 
\begin{equation}
\Gamma(h\to JJ)=\frac{m_h^3}{128\pi}\,\frac{\sin^2\theta_{\phi\sigma}}{v_\sigma^2}.
\label{eq:hJJ}
\end{equation}
Therefore, the present collider bound on ${\rm BR}(h\to{\rm inv})$ translates into an upper limit on
$\sin\theta_{\phi\sigma}/v_\sigma$, which can always be satisfied by taking a sufficiently small $\phi$--$\sigma$
mixing (without affecting the neutrino sector and the dark-matter analysis presented below).
Using the LHC bound on invisible/exotic Higgs  decays \cite{PDG2024,ATLAS:2023invH} 
\begin{eqnarray}
    \mathrm{ATLAS}\, (95\%~{\rm CL}): \mathrm{BR}(h\to \mathrm{inv}) < 0.107\ \ \  
\end{eqnarray}
we obtain from \eqref{eq:hJJ} the limit 
$\sin\theta_{\phi\sigma}/v_\sigma < 4.5\times10^{-4}\,\mathrm{GeV}^{-1}$, which is satisfied for $v_\sigma\geq 2.2$ TeV and arbitrary $\theta_{\phi\sigma}$.

Cosmologically, the massless boson $J$ contributes appreciably to the radiation density only if it thermalizes with the SM
plasma. In our setup, the dominant SM-$J$ interaction is again controlled by the Higgs--singlet mixing, and hence by the same
combination $\sin\theta_{\phi\sigma}/v_\sigma$ entering     Eq.~\eqref{eq:hJJ}. For small mixing, $J$ decouples early,
leading to a suppressed contribution to $\Delta N_{\rm eff}$. 
%
% Parametrically, for decoupling at temperature $T_D$,
% \begin{equation}
% \Delta N_{\rm eff}=\frac{4}{7}\left(\frac{g_{*s}(T_{\nu\,{\rm dec}})}{g_{*s}(T_D)}\right)^{4/3},
% \label{eq:DeltaNeff}
% \end{equation}
% with $g_{*s}(T_{\nu\,{\rm dec}})=10.75$.
% For instance, if $J$ decouples well above the electroweak scale (so that $g_{*s}(T_D)\simeq 106.75$),
% Eq.~\eqref{eq:DeltaNeff} yields $\Delta N_{\rm eff}\simeq 2.7\times 10^{-2}$, which is safely below current CMB bounds.
%
Thermalization of $J$ through Higgs decays/inverse decays occurs for
\begin{eqnarray}
    \sin\theta_{\phi\sigma}/v_\sigma \gtrsim 2.9\times10^{-9}\,\mathrm{GeV}^{-1},
\end{eqnarray}
so that $J$ is generically thermalized in the parameter space consistent with collider data.
After decoupling at temperature $T_D$, entropy conservation gives \cite{PDGBBN2025}
\begin{equation}
\Delta N_{\rm eff}=\frac{4}{7}\left(\frac{g_{*s}(T_{\nu\,{\rm dec}})}{g_{*s}(T_D)}\right)^{4/3},
\label{eq:DeltaNeff}
\end{equation}
with $g_{*s}(T_{\nu\,{\rm dec}})=10.75$ \cite{deSalas:2016Neff}. By definition 
    $\Delta N_{\text{eff}} \equiv N_{\text{eff}} - N_{\text{eff}}^{\text{SM}}
    $
where the SM prediction is $N_{\text{eff}}^{\text{SM}} \simeq 3.045$ \cite{PDGNeutrinosCosmology2025}.
In a step-function approximation for $g_{*s}(T)$ \cite{Husdal:2016gstar,Drees:2015qcdEoS}, decoupling above the QCD epoch yields
$\Delta N_{\rm eff}\simeq 0.03$--$0.06$, comfortably below current CMB limits \cite{Planck:2018params}
\begin{eqnarray}
    \mathrm{Planck}\,  2018: N_{\rm eff} = 2.99 \pm 0.17\ \ 
\end{eqnarray}
so $\Delta N_{\rm eff}$ is constrained at the $\mathcal{O}(0.1)$--$\mathcal{O}(0.3)$ level, depending on the dataset.
For completeness, if $\sin\theta_{\phi\sigma}/v_\sigma \lesssim 3\times10^{-9}\,\mathrm{GeV}^{-1}$,
the Goldstone never thermalizes and its freeze-in abundance \cite{Hall:2009freezein} yields a negligible $\Delta N_{\rm eff}$.

Finally, since $U(1)_X$ is broken by a SM-singlet VEV, the Majoron-like couplings to light neutrinos are suppressed
by $1/v_\sigma$ (and, in the mass basis, scale with neutrino-mass insertions), implying that stellar-cooling and
supernova limits are automatically satisfied for the TeV-scale singlet sector considered in this work.

{\it UV option (escape hatch).}
If one nevertheless wishes to avoid an exactly massless Goldstone, two standard ultraviolet (UV) possibilities exist:
(i) promote $U(1)_X$ to a gauge symmetry so that the Goldstone is eaten by the longitudinal component of the associated $Z'$ gauge boson,
with $m_{Z'}\sim g_X v_\sigma$; in this case anomaly cancellation must be ensured for the charge assignment of Table~I, e.g.\ by adding suitable chiral fermions with appropriate $U(1)_X$ charges;
(ii) introduce a tiny explicit (soft or higher-dimensional) $U(1)_X$ breaking, which makes the Goldstone a pseudo-NGB with a small mass while leaving the low-energy phenomenology essentially unchanged. The implementations of such UV extensions is beyond the scope of the present work.

% The mixing angle $\theta_{\phi\sigma}$ is small for  $
% |\lambda_{\phi \sigma}|\ll\sqrt{\lambda_{\phi}\lambda_{\sigma}}$  in  .
% Choosing $
% |\lambda_{\phi \sigma}|\ll\sqrt{\lambda_{\phi}\lambda_{\sigma}}$ ensures  $|\theta_{\phi\sigma}|\ll 1$
% We thus take Higgs couplings to SM states to be SM-like.
% }
\subsection{Stability conditions}
In order to determine the stability conditions of the scalar potential, one has to analyze its quartic terms because they will dominate the behavior of the scalar potential in the region of very large values of the fields components. Then, to study these conditions we introduce the following Hermitian bilinear combination of the scalar fields
$$
\begin{aligned}
& a=\phi^{\dagger} \phi, \quad b=\sigma^* \sigma, \quad c=\rho^* \rho.
\end{aligned}
$$
Substituting these bilinears in the scalar potential (\ref{eq:scalar_potential}), the quartic terms get the form,
\begin{equation}
V_{4}=\lambda_{\phi}a^{2} + \lambda_{\sigma}b^{2} + \lambda_{\rho}c^{2} +  \lambda_{\phi \sigma}ab + \lambda_{\phi \rho}ac + \lambda_{\sigma \rho}bc,
\end{equation}
which can be recast as
\begin{equation}
\begin{aligned} V_{4}= &  
\left(\sqrt{\lambda_{\sigma}} b - \sqrt{\lambda_\rho} c\right)^2 +\left(\sqrt{\lambda_{\sigma}} b - \sqrt{\lambda_\phi} a\right)^2 \\
&  + \left(\sqrt{\lambda_\rho} c - \sqrt{\lambda_\phi} a\right)^2 + \left(\lambda_{\phi \rho} + 2 \sqrt{\lambda_{\rho} \lambda_{\phi}}\right)ac \\
&  + \left(\lambda_{\sigma \rho} + 2 \sqrt{\lambda_{\sigma} \lambda_{\rho}}\right)bc + \left(\lambda_{\phi \sigma} + 2\sqrt{\lambda_{\sigma}\lambda_{\phi}}\right)ab \\
& -2\left(\lambda_{\phi}a^{2} + \lambda_\sigma b^2 +\lambda_{\rho}c^{2}  \right) .
\end{aligned}
\end{equation}
Following the criteria described in Refs. \cite{BHATTACHARYYA_2016, Maniatis_2006}, we find that the stability conditions for this model are
\begin{equation}
\label{eq:V-stab-1}
\begin{aligned}
& \lambda_{\phi} \geq 0, \quad \lambda_\sigma \geq 0, \quad \lambda_\rho \geq 0,\\
& \lambda_{\phi \sigma} + 2 \sqrt{\lambda_{\phi}\lambda_{\sigma}} \geq 0, \quad \lambda_{\phi \rho} + 2 \sqrt{\lambda_{\phi}\lambda_{\rho}} \geq 0, \quad \lambda_{\sigma \rho} + 2 \sqrt{\lambda_{\sigma} \lambda_{\rho}} \geq 0.\\
\end{aligned}
\end{equation}

\section{Neutrino masses}
\label{neutrino masses}
With the particle content and symmetries specified in Table \ref{tab:charge}, the Yukawa Lagrangian for the neutrino sector is given by
\begin{align}
-\mathcal{L}^{(\nu)}_Y &=
\sum_{i=1}^3\sum_{k=1}^2 (y_\nu)_{ik}\, \overline{\ell_{iL}}\tilde\phi\, \nu_{kR}
+\sum_{n,k=1}^2 M_{nk}\, \overline{\nu_{nR}}\, N^{C}_{kR}
+\sum_{n,k=1}^2 (y_N)_{nk}\, \overline{\Omega_{kL}}\, \rho\, N_{nR} \nonumber\\
&\quad
+\sum_{n,k=1}^2 (z_\Omega)_{nk}\, \overline{\Omega^C_{nR}}\, \rho\, \Omega_{kR}
+\sum_{n,k=1}^2 (y_\Omega)_{nk}\, \overline{\Omega_{nL}}\, \sigma\, \Omega_{kR}
+\text{h.c.}
\label{eq:lag-neutrino}
\end{align}

After spontaneous breaking (\ref{SB}) the neutrino mass matrix turns out to be 
%
% After the spontaneous breaking of the Lagrangian symmetry of Eq.~%
% \eqref{eq:lag-neutrino}, we obtain the following mass matrix for the
% neutrino sector 
\begin{equation}
M_{\nu }=\left( 
\begin{array}{ccc}
0_{3\times 3} & m_D & 0_{3\times 2} \\ 
m_D^{T} & 0_{2\times 2} & M \\ 
0_{2\times 3} & M^{T} & \mu%
\end{array}
\right) ,  \label{Mnu}
\end{equation}
where $m_D$ is a Dirac mass matrix, while $M$ and $\mu$ are Majorana matrices parametrized as
\begin{align}
m_D&=\frac{v_\phi}{\sqrt{2}}%
\begin{pmatrix}
y_{\nu_{11}} &  & y_{\nu_{12}} \\ 
y_{\nu_{21}} &  & y_{\nu_{22}} \\ 
y_{\nu_{31}} &  & y_{\nu_{32}}%
\end{pmatrix},
& M&= 
\begin{pmatrix}
M_{11} & M_{12} \\ 
M_{12} & M_{22}%
\end{pmatrix},
& \mu&= 
\begin{pmatrix}
\mu_{11} & \mu_{12} \\ 
\mu_{12} & \mu_{22}%
\end{pmatrix},%
\label{eq:mD_M}
\end{align}
where we have assumed the matrix $M$ as diagonal.

\begin{figure}[th]
\includegraphics[width=0.55\textwidth]{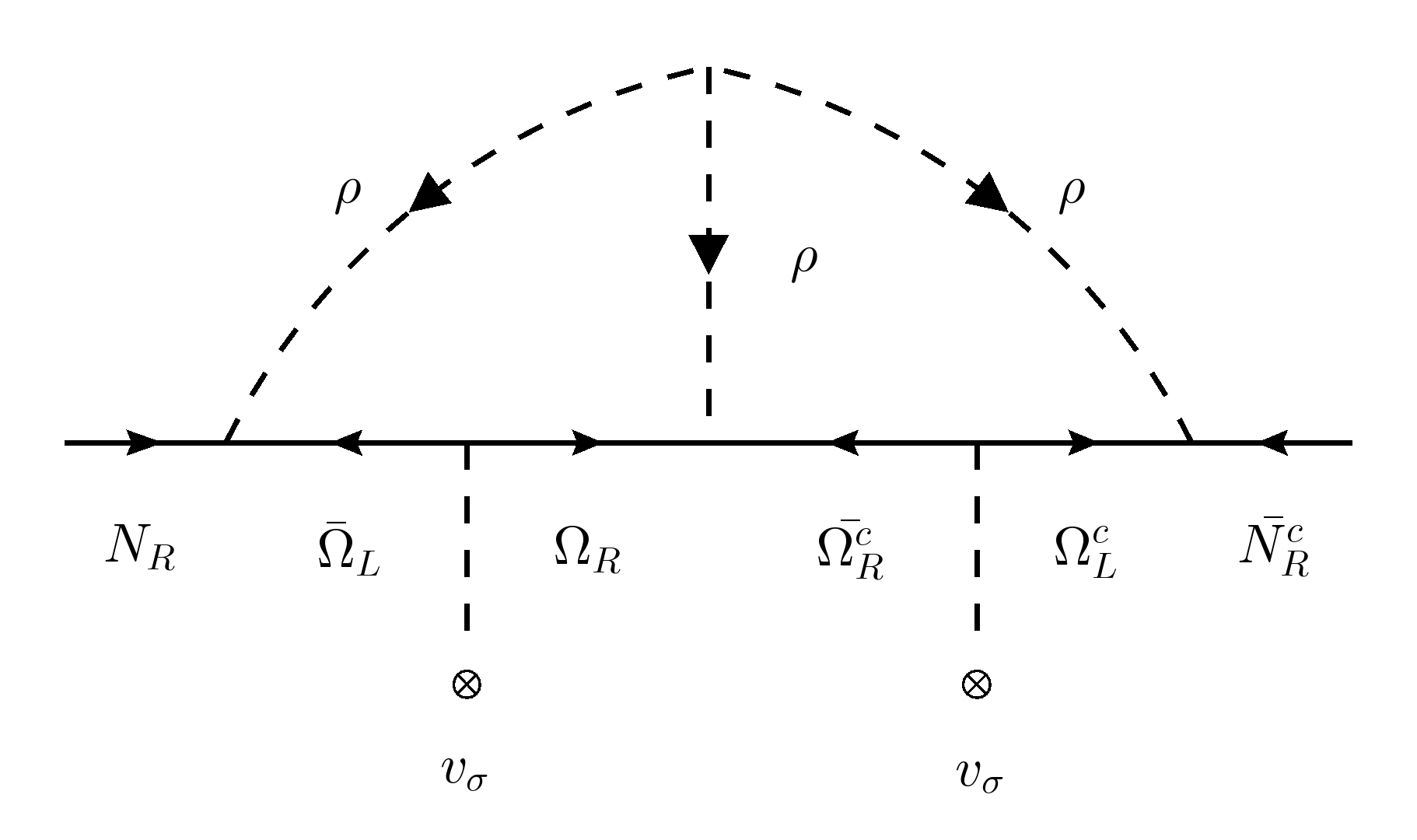}
\caption{Two-loop Feynman diagram contributing to the Majorana neutrino mass
submatrix $\protect\mu $.}
\label{Neutrinodiagram}
\end{figure}

The two-loop contribution to the Majorana mass submatrix $\mu$, shown in Figure~\ref{Neutrinodiagram}, can be expressed in terms of the model parameters as
\begin{equation}
    \mu = 3 A\, y_{N}M_{\Omega }z_{\Omega
 }^{\ast }M_{\Omega}^{T}y_{N}^{T}  I(m_\Omega,m_\rho),
    \label{eq:mu-2loop-23}
\end{equation}
%{\color{blue}
where the two-loop integral $I(m_\Omega,m_\rho)$ is defined   and evaluated in Appendix \ref{App:2loop-1}. 
Using \eqref{eq:I-reduction-1}, we get
\begin{equation}
    \mu \approx  
    \frac{6
    }{(16\pi^{2})^{2}} A\, y_{N}z_{\Omega
 }^{\ast }y_{N}^{T} 
 \frac{m^4_\rho}{m^4_\Omega}\, z,
    \label{eq:mu-2loop-234}
\end{equation}
where 
%{$\kappa =3$ is a factor, and 
the constant $z\approx 1.56$ is derived in Appendix \ref{App:2loop-1}. For simplicity, we assumed that a $2\times 2$ mass matrix $M_\Omega$ of $\Omega$ field is approximately diagonal with equal entries $m_\Omega = y_\Omega v_\sigma$. 
From this formula, the parameter $\mu$ is small due to the two loop suppression of order 
$(16\pi^2)^{-2}\sim 10^{-5}$ without requiring small Yukawa couplings $y_N, z_\Omega$. Sizable Yukawa couplings have important consequences in the DM phenomenology. This is discussed below in Section~\ref{fermionic_dark_matter}. 
% As follows from Eqs. \eqref{eq:mRho-1} and \eqref{eq:V-stab-1} this region is 

Note that the trilinear term $A\rho^3$ in the scalar potential \eqref{eq:scalar_potential} softly breaks an accidental global $U(1)'$ symmetry with charge assignments $Q'(\rho)=2, Q'(\Omega_R)=-1, Q'(\Omega_L)=2, Q'(\sigma)=3$. Consequently, a small value for parameter A is technically natural and can therefore be consistently chosen.
% }

 %
 %\hspace{%1cm}
%in which \cite{Herrero-Garcia:2014hfa,McDonald:2003zj}, 
% \begin{eqnarray}
% m_{S} &=&\max \left( m_{\chi },m_{\rho }\right) ,  \notag \\
% J\left( \varkappa \right)  &=&\left\{ 
% \begin{array}{l}
% 1+\frac{3}{\pi ^{2}}\left( \ln ^{2}\varkappa -1\right) \hspace{0.1cm}\,%
% \mbox{ for},\hspace{0.2cm}\varkappa \gg 1 \\ 
% \\ 
% 1\hspace{0.1cm}\,\mbox{for},\hspace{0.2cm}\varkappa \rightarrow 0.%
% \end{array}%
% \right. ,
% \end{eqnarray}
The neutrino mass matrix from Eq.~\eqref{Mnu} can be diagonalized with the rotation matrix \cite{Catano:2012kw,Abada:2023zbb}:
\begin{equation}
\mathbb{U}\simeq \begin{pmatrix}
        \left(1_{3\times 3}-\frac{R_1R_1^\dagger }{2}\right) \left(1_{3\times 3}-\frac{R_2R_2^\dagger }{2}\right) R_\nu  & R_1R_{M}^{(1) } & R_2R_{M}^{(2)} \\ 
        -\frac{R_1^\dagger +R_2^\dagger }{\sqrt{2}}R_\nu  & \left(1_{3\times 3}-\frac{R_1R_1^\dagger }{2}\right) \frac{1-S}{\sqrt{2}} R_{M}^{(1) } & \frac{1+S}{\sqrt{2}}R_{M}^{(2)} \\ 
        -\frac{R_1^\dagger -R_2^\dagger }{\sqrt{2}}R_\nu  & -\frac{1+S}{\sqrt{2}}R_{M}^{(1) } & \left(1_{3\times 3} - \frac{R_2R_2^\dagger }{2}\right) \frac{1-S}{\sqrt{2}}R_{M}^{(2)}%
    \end{pmatrix} 
\label{eq:rotation}
\end{equation}

where
\begin{eqnarray}
S= -\frac{1}{4}M^{-1}\mu, \quad\quad R_1\simeq R_2\simeq \frac{1}{\sqrt{2}}m_D^{*}M^{-1} \equiv R.
\end{eqnarray}
The masses of the light active neutrinos are generated via a two-loop inverse seesaw mechanism, with the corresponding mass matrices expressed as follows~\cite{Mohapatra:1986bd,Malinsky:2005bi}: 
\begin{eqnarray}
\label{eq:InvSeesaw-1}
\widetilde{M}_{\nu} &=& m_D\left(M^T\right)^{-1}\mu M^{-1}m_D^T, \\
\widetilde{M}_{\nu}^{(-)} &=& \frac{1}{2}\mu-\frac{1}{2}\left(M+M^T\right),
\\
\widetilde{M}_{\nu}^{(+)} &=& \frac{1}{2}\mu+\frac{1}{2}\left(M+M^T\right),
\end{eqnarray}
The matrix $\widetilde{M}_{\nu}$ represents the (predominantly) active neutrino mass matrix, while $\widetilde{M}_{\nu}^{(-)}$ and $\widetilde{M}_{\nu}^{(+)}$ describe the matrices associated with the exotic neutrino sectors.  The resulting physical spectrum consists of three light active neutrinos and four exotic neutrinos, which form two pseudo-Dirac pairs. These pairs have masses approximately given by $\sim \pm \frac{1}{2} \left(\widetilde{M}_{\nu}^{(+)} + \left(\widetilde{M}_{\nu}^{(-)}\right)^T\right)$, with a small mass splitting $\mu$ within each pair.  The rotation matrix $R_\nu$, which diagonalizes the light neutrino mass matrix $\widetilde{M}_{\nu}$, is closely related to the Pontecorvo-Maki-Nakagawa-Sakata (PMNS) matrix $U_{\text{PMNS}}$. However, due to mixing between active and heavy exotic neutrinos (encoded in $R_1$ and $R_2$, the PMNS matrix may exhibit slight deviations from unitarity.

We analyzed a specific benchmark scenario with a diagonal  
matrix $M = \mbox{diag}(M_{1}, M_2)$ and  
fitted the effective parameters of the neutrino sector specified in 
Eqs.~\eqref{eq:mD_M}-\eqref{eq:mu-2loop-234} to reproduce the experimental data on
$\Delta m_{i1}^2$ (with $i= 2, 3)$  neutrino mass squared differences, $\sin\theta_{jk}$ ($j,k=1,2,3$, $j\neq k$) mixing angles, and the leptonic Dirac CP violating phase $\delta_{\text{CP}}$.
%
% To simplify our analysis and parameter adjustment, we consider a benchmark scenario with a diagonal  
% matrix $M = \mbox{Diag}(M_{1}, M_2)$. 
%
% Then, we fitted the effective parameters of the neutrino sector according to 
% Eqs.~\eqref{eq:mD_M}-\eqref{eq:mu-2loop-234} to reproduce the experimental data on
% $\Delta m_{i1}^2$ (with $i= 2, 3)$  neutrino mass squared differences, $\sin\theta_{jk}$ mixing angles, and the CP violating phase $\delta_{\text{CP}}$.
%
% Therefore, to fit the parameters of the effective neutrino sector according to Eqs.~\eqref{eq:mD_M}-\eqref{eq:mu-2loop-234} and successfully reproduce the experimental values of the neutrino mass-squared splittings, the leptonic mixing angles, and the leptonic Dirac CP phase., 
%
We performed a generic scan where we randomly varied the parameters of the matrix $m_D$ between $[10^{-3}, 1]$; for the matrix $M$ 
%they were 
we varied between $[100, 1000]$ GeV whereas for $\mu$ the range $[10^{-6}, 10^{-3}]$ eV was considered. 
The phases of $m_D$ and $\mu$ were varied in the interval $[0, 2\pi]$.
%
%
% By performing the numerical analysis of our model, randomly varying the magnitude of each free, where the parameters of the matrix $m_D$ were varied between $[10^{-3}, 1]$, for the matrix $M$ they were varied between $[100, 1000]$ GeV and for $\mu$ the range $[10^{-6}, 10^{-3}]$ eV was considered. 
% The phases of $m_D$ and $\mu$ were varied in the interval $[0, 2\pi]$.
% %
% Also, the parameters of the matrices $m_D$ and $\mu$ were considered complex, where the phases of each were varied between $[0, 2\pi]$ rad.
The values of the model parameters representing the studied benchmark point are shown in Table~\ref{tab:best-fit}.
\begin{table}
\centering
\begin{tabular}{l|l} 
\hline\hline
\multicolumn{2}{c}{\textbf{Benchmark parameters}}                                               \\ 
\hline\hline
$y_{\nu_{11}}= 3.84\times 10^{-3}e^{0.958i}$  & $\mu_{11}= 6.75\times 10^{-3}e^{-1.35i}\ \text{eV}$  \\
$y_{\nu_{12}}= 4.32\times 10^{-3}e^{-2.02i}$  & $\mu_{12}= \mu_{21}= 3.27\times 10^{-5}e^{-1.33i}\ \text{eV}$   \\
$y_{\nu_{21}}= 1.20\times 10^{-3}e^{-0.623i}$ & $\mu_{22}= 2.91\times 10^{-6}e^{-1.51i}\ \text{eV}$  \\
$y_{\nu_{22}}= 9.78\times 10^{-3}e^{0.846i}$  &  $M_1= 6536.9\ \text{GeV}$  \\
$y_{\nu_{31}}= 5.95\times 10^{-3}e^{-2.42i}$  &   $M_2= 531.1\ \text{GeV}$                         \\
$
y_{\nu_{32}}
= 1.23\times 10^{-2}e^{0.854i}$  &                              \\
\hline\hline
\end{tabular}
\caption{Effective parameter values for the benchmark point of the model.}
\label{tab:best-fit}
\end{table}
After performing the fit of the effective parameters and deriving the benchmark point, we obtained the values shown in Table \ref{table:neutrinos_value}, alongside the experimental values of neutrino oscillation parameters within the $1\sigma$ and $3\sigma$ ranges, as reported in Refs.~\cite{deSalas:2020pgw, Esteban:2024eli}. 
% {\color{blue}Despite the obtained values are not exactly equal to their corresponding central experimental values, they are within the $2\sigma$ and $3\sigma$ experimentally allowed ranges. This is a consequence of the specific benchmark scenario chosen in our model. The exact central experimental values can be perfectly reproduced by perturbing around the above given benchmark point.}
Although the obtained values do not exactly match their corresponding central experimental values, they lie within the $2\sigma$ and $3\sigma$ experimentally allowed ranges. This is a consequence of the specific benchmark scenario chosen in our model. The exact central experimental values can be reproduced by perturbing around the above-given benchmark point.

In Table \ref{table:neutrinos_value}, we see that the neutrino mass-squared differences ($\Delta m_{21}^2$, $\Delta m_{31}^2$) and the solar and reactor mixing angles ($\sin^2\theta_{12}^{(l)}$, $\sin^2\theta_{13}^{(l)}$) lie within the $1\sigma$ range. The atmospheric mixing angle ($\sin^2\theta_{23}^{(l)}$) and the leptonic Dirac CP-violating phase ($\delta_{\text{CP}}$) are within the $2\sigma$ range.

Furthermore, our model can predict the effective Majorana neutrino mass parameter of the neutrinoless double beta decay ($0\nu\beta\beta$), which gives us information about the Majorana nature of neutrino. This observable depends on the mixing parameters and the masses of the light-active neutrinos as follows, 
\begin{equation}
m_{ee}=\left| \sum_i \mathbf{U}_{ei}^2m_{\nu i}\right|\;,  \label{ec:mee}
\end{equation}

where $\mathbf{U}_{ei}$ and $m_{\nu i}$ are the matrix elements of the PMNS leptonic mixing matrix and the light active neutrino masses, respectively. From the equation \eqref{ec:mee}, we can see that the neutrinoless double beta ($0\nu\beta\beta$) decay amplitude is proportional to $m_{ee}$. \\

\begin{figure}[tbp]
\centering
\subfloat[]{\includegraphics[scale=0.35]{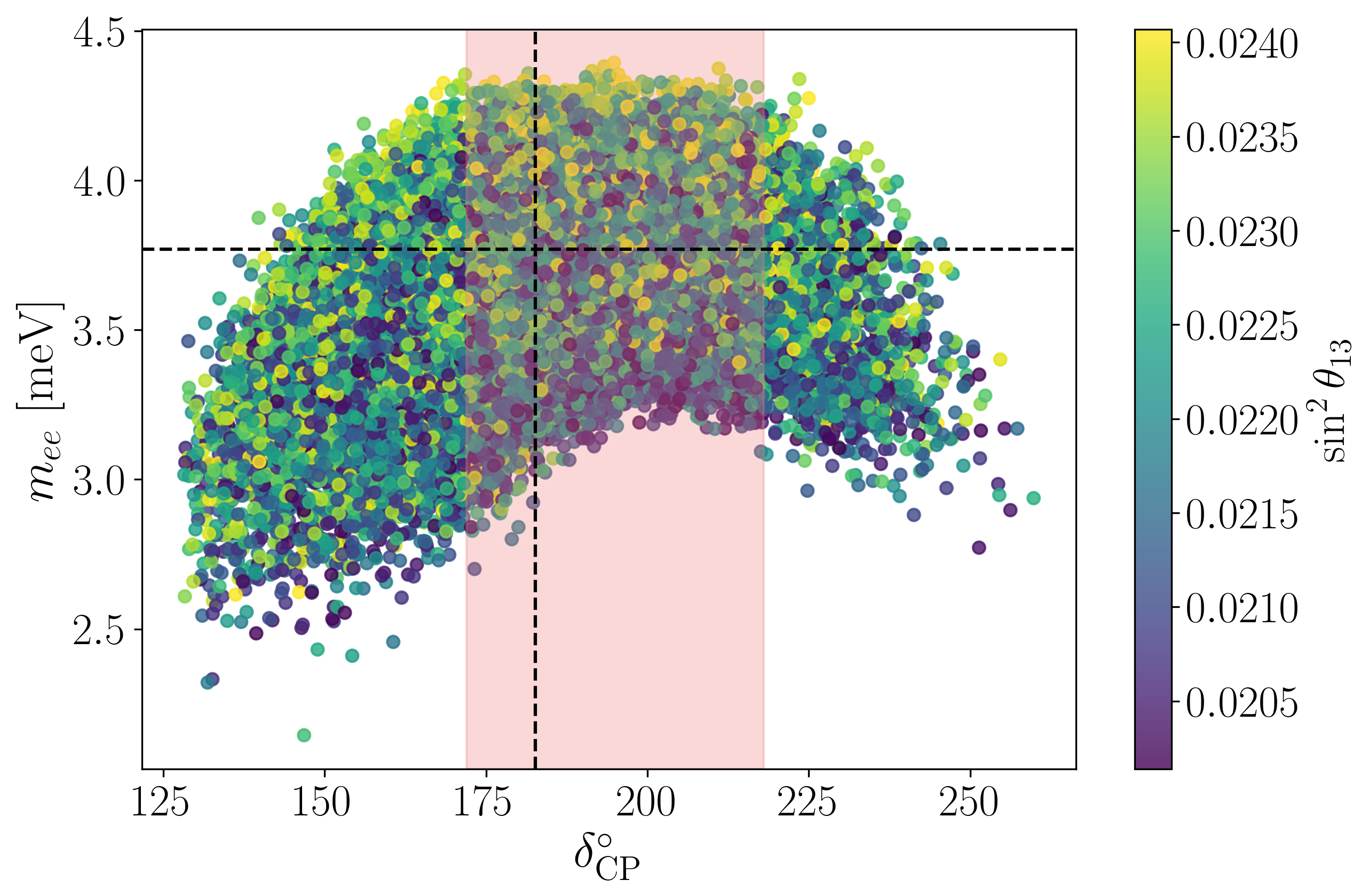}\label{fig:delmee}}%
\quad \subfloat[]{\includegraphics[scale=0.35]{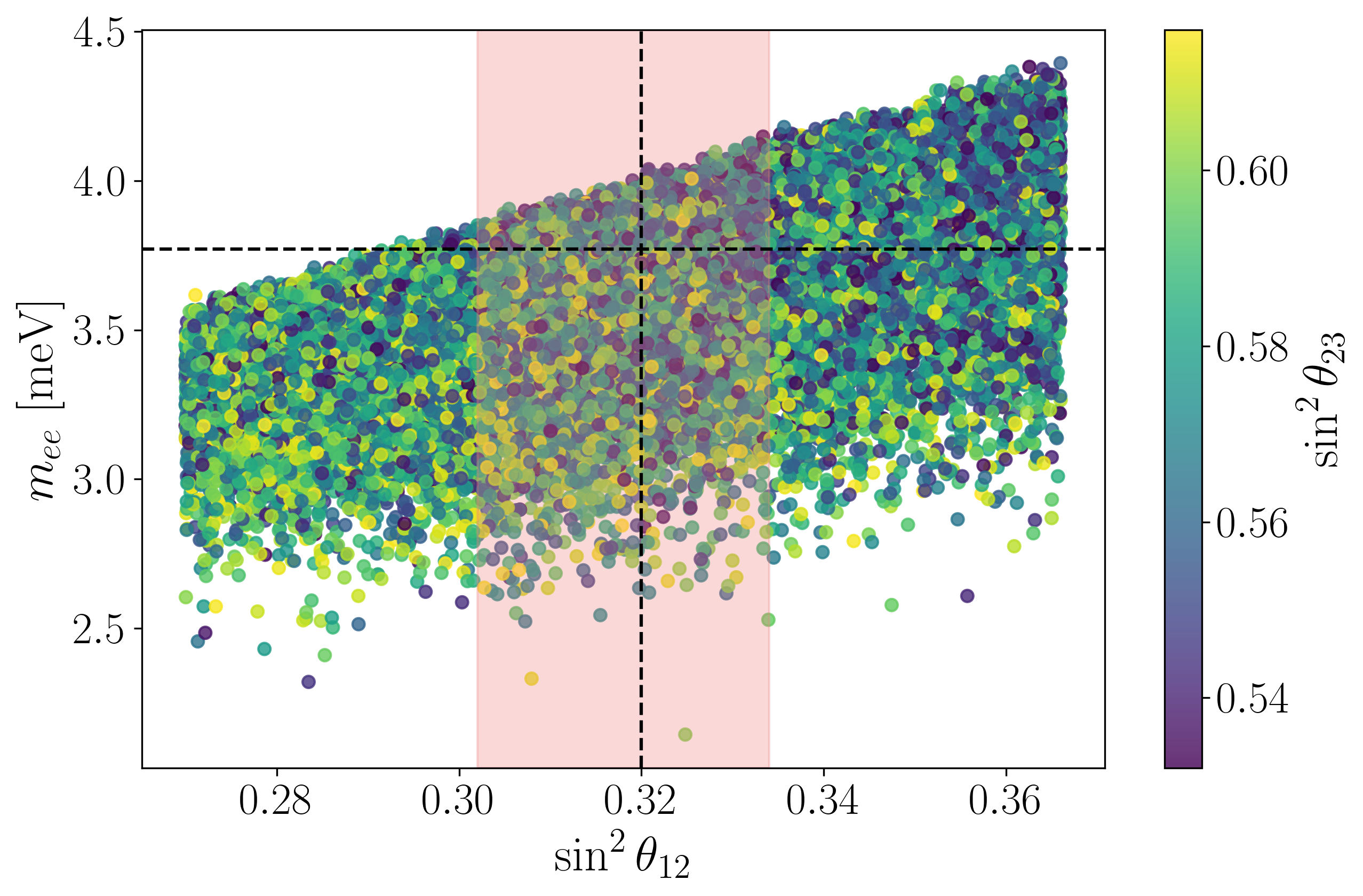}\label{fig:sin12mees}}%
\caption{Correlation plot between the effective mass parameter of the neutrinoless double beta decay $(0\nu \beta \beta)$ $m_{ee}$ and a) CP violation phase $\delta^{(l)}_{\rm CP}$, for different values of the mixing angle $\sin^2\protect\theta_{13}$ and b) for the mixing angle $\sin^2\protect\theta_{12}$ and different values of the mixing angle $\sin^2\protect\theta_{13}$. A pink shaded region corresponds to the $1\sigma$ experimentally allowed range for the leptonic Dirac CP violating phase (left-panel) and the solar mixing parameter $\sin^2\protect\theta_{12}$ (right-panel). The black dotted lines represent our benchmark point for each observable.}
\label{fig:corr-meee}
\end{figure}

Figures \ref{fig:delmee} and \ref{fig:sin12mees} show the correlation between the effective mass parameter $m_{ee}$ of $0\nu \beta \beta$, the CP violation phase and the mixing angle $\sin^2\protect\theta_{12}$ for different values of the mixing angle $\sin^2\protect\theta_{13}$ and $\sin^2\protect\theta_{13}$, respectively. The red background fringe represent the $1\sigma$ range of the experimental values and the black dotted lines represent our benchmark point for each observable. 
%%%%%%%%%%%%

%These correlations should be understood as correlations within the restricted scan adopted in this work, rather than as sharp predictions of the most general parameter space of the model.
These correlations are specific to the restricted parameter scan employed in this work and should not be interpreted as sharp predictions for the most general parameter space of the model.
A dedicated global fit can, in principle, reproduce the experimental central values
to essentially arbitrary precision; our goal here is to illustrate representative viable regions and the corresponding range for $m_{ee}$ in this framework.

%%%%%%%%%%%
In Figure \ref{fig:corr-meee}, we see that for the mixing angles and the CP violating phase, we obtain values up to $3\sigma$, where each lepton sector observable is obtained in the following range of values: $0.532\lesssim \sin^2\theta_{23}\lesssim 0.616$, $0.0201\lesssim \sin^2\theta_{13}\lesssim 0.0241$, $0.27\lesssim \sin^2\theta_{12}\lesssim 0.366$ and $128\lesssim \delta_{\text{CP}}\lesssim 260$.

Therefore, assuming normal neutrino mass hierarchy, one finds that the Majorana neutrino mass parameter is bounded to be in the range $2.14\  \text{meV}\lesssim m_{ee}\lesssim 4.39\ \text{meV}$. This region is clearly shown in Figure \ref{fig:corr-meee}, where all values are below the current experimental limit. The current most stringent experimental upper bound on the effective Majorana neutrino mass parameter, i.e., %m_{ee}\leq 50\ \text{meV}$ 
%\st{$\langle m_{2\beta}\rangle<(28-122)$meV $(90\%C.L.)$,} 
$m_{ee}<(28-122)$ meV (90\% C.L.), arises from the KamLAND-Zen limit on the $^{136}X_e\; 0\nu\beta\beta$ decay half-life $T_{1/2}^{0\nu\beta\beta}(^{136}X_e) >3.8\times 10^{26}$ yr (90\% C.L.)\citep{KamLAND-Zen:2024eml}.
\begin{table}[tp]
%\resizebox{13cm}{!}{
\begin{tabular}{c|c|cccccc}
%\toprule[0.13em]
\hline\hline
Observable & range & $\Delta m_{21}^{2}$ [$10^{-5}$eV$^{2}$] 
& $\Delta m_{31}^{2}$ [$10^{-3}$eV$^{2}$] & $\sin^2\theta^{(l)}_{12}/10^{-1}$
& $\sin^2\theta^{(l)}_{13}/10^{-2}$ & $\sin^2\theta^{(l)}_{23}/10^{-1}$ & $%
\delta^{(l)}_{\text{CP}} (^{\circ })$ \\
\hline\hline
Experimental & $1\sigma$ & $7.50_{-0.20}^{+0.22}$ & $2.55_{-0.03}^{+0.02}$ & 
$3.18\pm 0.16$ & $2.200_{-0.062}^{+0.069}$ & $5.74\pm 0.14$ & $%
194_{-22}^{+24}$ \\ 
Value~\cite{deSalas:2020pgw} & $3\sigma$ & $6.94-8.14$ & $2.47-2.63 $ & $2.71-3.69$ & $2.000-2.405$
& $4.34-6.10$ & $128-359$ \\ \hline
Experimental & $1\sigma$ & $7.49\pm 0.19$ & $2.513_{-0.019}^{+0.021}$ & $3.08_{-0.11}^{+0.12}$ & $2.215_{-0.056}^{+0.058}$ & $4.7_{-0.13}^{+0.17}$ & $212_{-41}^{+26}$ \\ 
Value~\cite{Esteban:2024eli} & $3\sigma$ & $6.92-8.05$ & $2.451-2.578 $ & $2.75-3.45$ & $2.03-2.388$
& $4.35-5.85$ & $124-364$ \\ \hline
Fit & $1\sigma-3\sigma$ & $7.69$ & $2.54$ & $3.41$ & $2.24$ & $5.73$ & $219.7$\\ 
\hline\hline
%\bottomrule[0.13em] &  &  &  &  &  &  & 
\end{tabular}
\caption{Numerical Results for the Normal-Ordering (NO) neutrino-mass scenario.}
\label{table:neutrinos_value}
\end{table}
\section{Dark matter}
\label{dark matter}

Due to the preserved $Z_3$ symmetry, our model contains stable scalar and fermionic dark matter candidates, whose stability is ensured by this discrete symmetry. The dark matter candidate corresponds to the lightest electrically neutral particle transforming non-trivially under the preserved $Z_3$ symmetry.

\begin{figure}[h!]
    \centering
    \includegraphics[width=0.25\linewidth]{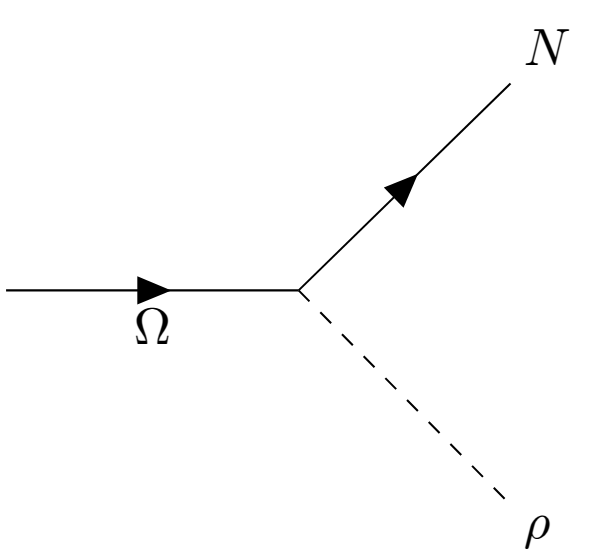}
    \caption{Feynman diagram for the $\Omega$ decay into the $\rho$-singlet and the fermion $N$.}
    \label{rho_decay}
\end{figure}

Depending on the mass hierarchy among the new particles, either a scalar or a fermionic dark matter scenario can be realized. {\it Scalar dark matter scenario}: This occurs when  $m_{\rho} < m_{\Omega}$, with the $\rho$-singlet as the dark matter candidate. In this regime, the decay channel $ \Omega \to N\rho$ is kinematically accessible (see Figure~\ref{rho_decay}). {\it Fermionic dark matter scenario:} This is realized when $m_{\rho} > m_{\Omega}$. Here, $ \Omega$ acts as a fermionic dark matter candidate. In this case, we consider a benchmark scenario where $m_{\Omega} > 2m_N$ in order to kinematically open the DM annihilation channel $\Omega\Omega\rightarrow NN$, which is crucial to successfully reproduce the experimentally measured amount of dark matter relic abundance.

In this section, we analyze the implications of our model for both scalar and fermionic dark matter scenarios, taking into account the experimentally allowed range for the dark matter relic abundance measured by the Planck collaboration~\cite{Planck:2018vyg}:
\begin{equation}
    \Omega h^2 = 0.1200 \pm 0.0012 .
\end{equation}

We will also address the compatibility of our model with the direct DM detection experiments
XENON1T \cite{XENON1T:2018voc}, XENONnT  \cite{XENONnT:2020kmp}, LUX-ZEPLIN  \cite{LZ:2024zvo} and DARWIN \cite{DARWIN:2016hyl} 
deriving the corresponding constraints on the model parameter space.

\subsection{Scalar dark matter}
\label{scalar_dark_matter}
\begin{figure}
    \centering
    \includegraphics[width=0.5\linewidth]{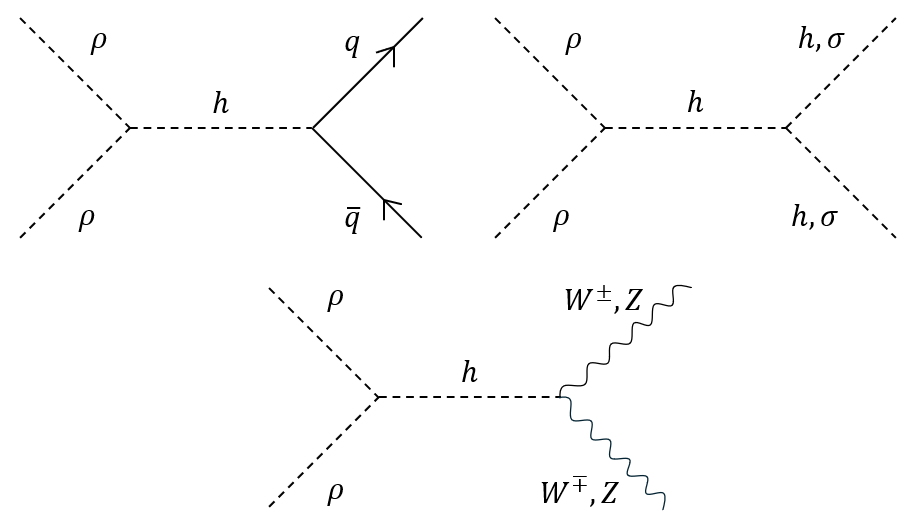}
    \caption{Some Feymann diagrams contributing to $\rho\rho$ annihilation to SM particles.}
    \label{scalar_DM_images}
\end{figure}

The physical masses for the CP even ($\rho_R$) and CP odd ($\rho_I$) components of the $\rho$-singlet are degenerate, as seen in section \ref{scalar potential}. Therefore, we consider the SM singlet $\rho$ a viable complex scalar 
DM candidate with mass $m_\rho$ as given in \eqref{eq:mRho-1} and satisfying $m_\Omega > m_\rho$. In this scenario, the decay channels $\rho\to\Omega N $ and $\rho\to\Omega\Omega$ are forbidden, making the $\rho$-scalar stable as required for a DM candidate. 
The mixing of $\rho$ with the Higgs boson, arising after electroweak symmetry breaking, 
 contributes to dark matter annihilation by enabling $s$-channel processes mediated by the CP-even component of the Higgs field (see Figure \ref{scalar_DM_images}).
 The list of annihilation cross sections relevant for the $\rho$-DM candidate scenario is 
\begin{eqnarray}
v_{\text{rel}}\sigma (\rho\rho \to WW ) &= & \frac{\lambda_{\phi \rho}^2}{8\pi} 
\frac{s \left(1 + \frac{12m_W^4}{s^2} - \frac{4m_W^2}{s}\right)}{\left(s - m_h^2\right)^2 + m_h^2\Gamma_h^2} 
\sqrt{1 - \frac{4m_W^2}{s}}, \\
v_{\text{rel}}\sigma (\rho\rho \to ZZ) &=& \frac{\lambda_{\phi \rho}^2}{16\pi} 
\frac{s \left(1 + \frac{12m_Z^4}{s^2} - \frac{4m_Z^2}{s}\right)}{\left(s - m_h^2\right)^2 + m_h^2\Gamma_h^2} 
\sqrt{1 - \frac{4m_Z^2}{s}}, \\
v_{\text{rel}}\sigma (\rho\rho \to q\bar{q}) &=& \frac{N_c\lambda_{\phi \rho}^2 m_q^2}{4\pi} 
\frac{\left(1 - \frac{4m_q^2}{s}\right)^{3/2}}{\left(s - m_h^2\right)^2 + m_h^2\Gamma_h^2}, \\
v_{\text{rel}}\sigma (\rho\rho \to hh) &=& \frac{\lambda_{\phi \rho}^2}{16\pi s} 
\left[1 + \frac{3m_h^2}{s - m_h^2} - \frac{4\lambda_{\phi \rho}v^2}{s - 2m_h^2}\right]^2 
\sqrt{1 - \frac{4m_h^2}{s}},\\
v_{\text{rel}}\sigma (\rho\rho \to \tilde{\sigma}\tilde{\sigma}) &=& \dfrac{\sqrt{s-4m_{\tilde{\sigma}}^2}}{8\pi s^\frac{3}{2}}\left[ \dfrac{v_\phi^4\lambda_{\phi\rho}^2\lambda_{\phi\sigma}^2}{\left(s - m_h^2\right)^2 + m_h^2\Gamma_h^2}+\dfrac{2(s-m_h^2)v_\phi^2\lambda_{\phi\rho}\lambda_{\phi\sigma}\lambda_{\sigma\rho}}{\left(s - m_h^2\right)^2 + m_h^2\Gamma_h^2}+\lambda_{\sigma\rho}^2\right],\\
v_{\text{rel}}\sigma (\rho\rho \to JJ) &=& \dfrac{\sqrt{s-4m_{G_\sigma}^2}}{\sqrt{s-4m_{\tilde{\sigma}}^2}} v_{\text{rel}}\sigma (\rho\rho \to \tilde{\sigma}\tilde{\sigma}), 
\end{eqnarray}
% v_{\text{rel}}\sigma (\rho\rho \to JJ) &=\dfrac{\sqrt{s-4m_{G_\sigma}^2}}{8\pi s^\frac{3}{2}}\left[ \dfrac{v_\phi^4\lambda_{\phi\rho}^2\lambda_{\phi\sigma}^2}{\left(s - m_h^2\right)^2 +   m_h^2\Gamma_h^2}+\dfrac{2(s-m_h^2)v_\phi^2\lambda_{\phi\rho}\lambda_{\phi\sigma}\lambda_{\sigma\rho}}{\left(s - m_h^2\right)^2 + m_h^2\Gamma_h^2}+\lambda_{\sigma\rho}^2\right]
%
where $\sqrt{s}$ is the center-of-mass energy, $N_c = 3$ is the color factor, $m_h = 125.7\,\text{GeV}$ and $\Gamma_h = 4.1\,\text{MeV}$ are the SM Higgs boson $h$ mass and its total decay width, respectively. The final-state $q \bar{q}$ stands for $t$ and $b$ quarks.

In the present analysis we neglect the semi-annihilation process $\rho\rho\to\rho h$. The corresponding thermally
averaged cross section is controlled by the trilinear scalar coupling $A$ as 
$\langle\sigma v\rangle_{\rm semi}\propto |A|^2$. Since $A$ is taken to be small in our setup (cf.\
Sec.~\ref{neutrino masses}), this channel is numerically irrelevant in the parameter region considered here.
We emphasize that the channel $\rho\rho\to JJ$ is present in our model. 
For sufficiently large portal couplings,  its thermally averaged cross section can reach values as large as 
$\langle\sigma v\rangle_{\rho\rho\to JJ}\sim 10^{-24}\ {\rm cm^3/s}$, i.e., well above the canonical thermal value
$\langle\sigma v\rangle_{\rm th}\sim 3\times 10^{-26}\ {\rm cm^3/s}$ required to reproduce the measured dark matter abundance $\Omega_{\rm DM}h^2\simeq 0.12$.
In this regime $\rho$ would be underabundant, and therefore our relic-density-saturating benchmark points lie in the region
where $\rho\rho\to JJ$ is subdominant compared to the SM final states considered in Fig.~8.
Independently, since $J$ couples to the SM through the $hJJ$ vertex induced by the Higgs--singlet mixing,
it can thermalize with the SM plasma in the early Universe and subsequently contribute as dark radiation, with its impact on
$\Delta N_{\rm eff}$ determined by the $J$ decoupling temperature.

% Below, in our analysis we neglect the $v^{2}$ dependent terms.

\subsubsection{Direct Detection}
\label{sec:Direct-DM}

\begin{figure}[h]
    \centering
    \includegraphics[width=0.25\linewidth]{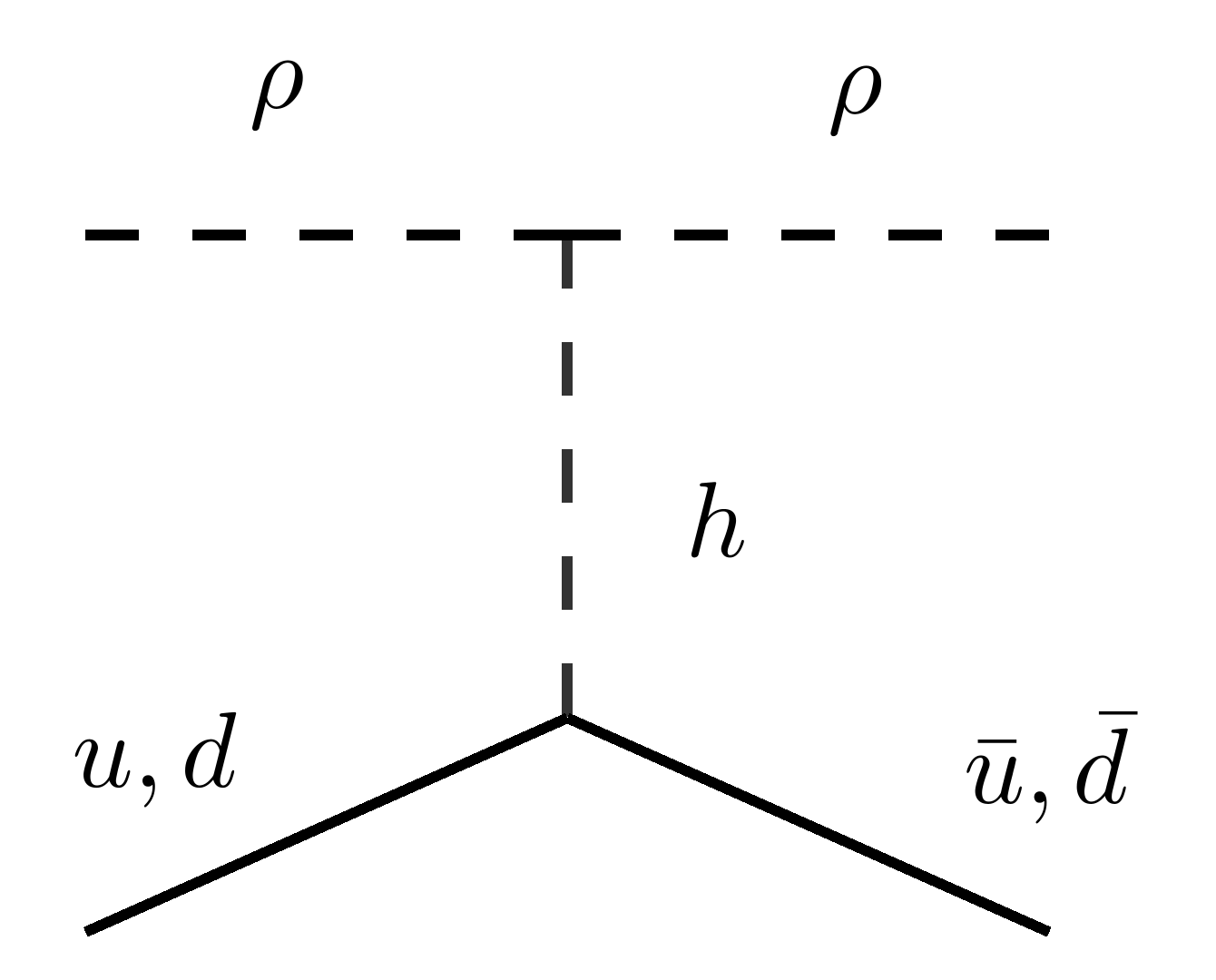}
    \caption{Feynman diagram for the dark matter contribution to spin-independent direct detection at tree-level.}
    \label{figure_SI_scalar_dark_matter}
\end{figure}

The spin-independent cross section for the $\rho$-nucleon elastic scattering is dominated by the contribution shown in Figure~\ref{figure_SI_scalar_dark_matter}. It is induced by the same Higgs portal interaction as in the $\rho-\rho$ annihilation, resulting in the well-known correlation between these observables. For this contribution, the total cross section is given by \cite{Bernal:2017xat} 
\begin{equation}
    \sigma_{SI}=\dfrac{ f_n^2m_n^4\lambda_{\phi\rho}^2}{8 \pi   m_h^4 m_\rho^2};
\end{equation}
Here, $m_{n}$ is the nucleon mass and $f_{n}\simeq 1/3$ corresponds to the form factor \cite{Farina_2010, Giedt_2009}.
 We derive constraints on the $\rho$-DM scenario by comparing the above expression with the latest limits on $\sigma_{SI}$ from XENON1T \cite{XENON:2018voc}, XENONnT \cite{XENON:2023cxc}, and the projected sensitivity of LUX-ZEPLIN \cite{LZ:2020sensitivity} and DARWIN \cite{DARWIN:2016whitepaper}.
 
\begin{figure}[h!]
\centering
\subfloat[]{\includegraphics[scale=0.4]{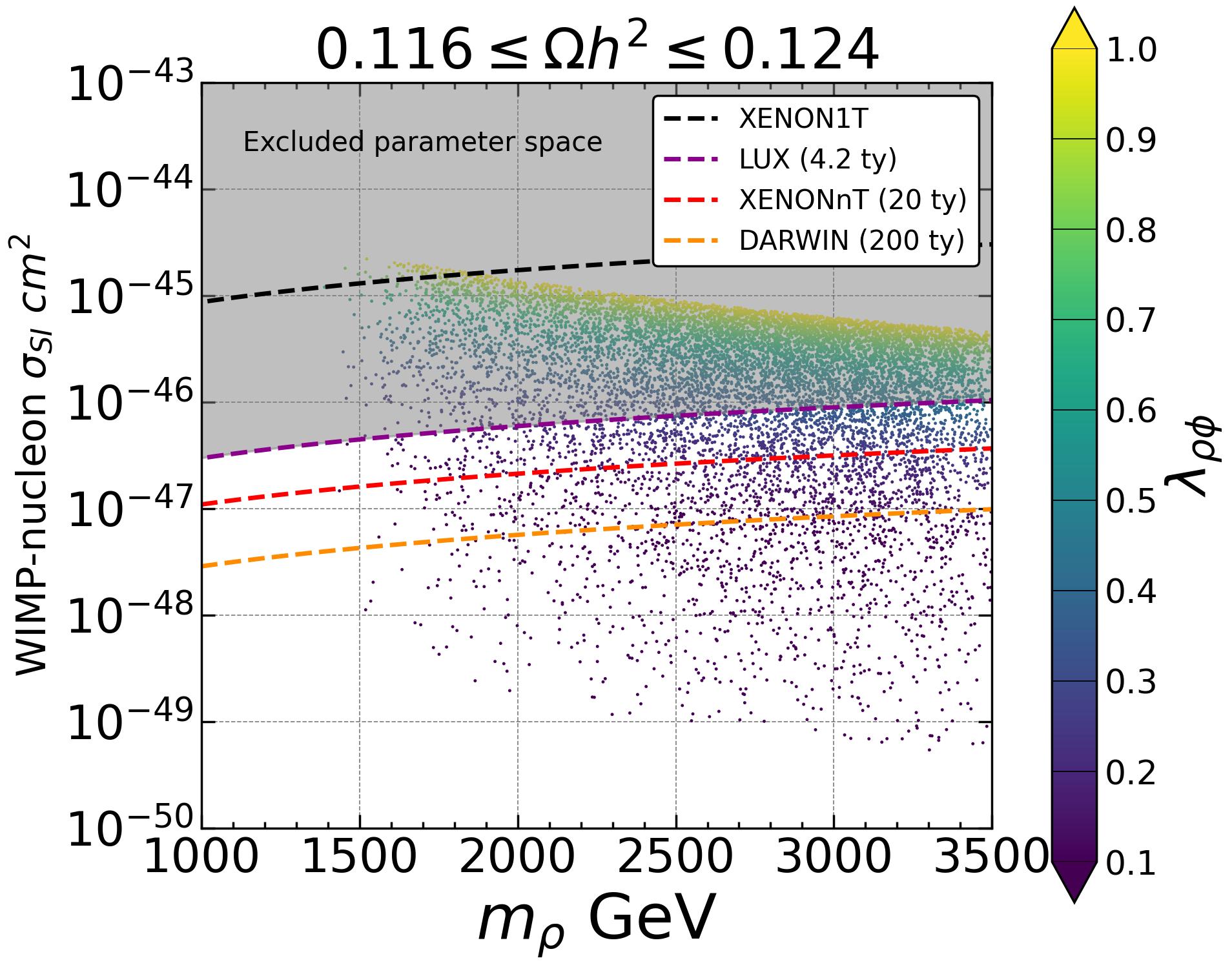}}\quad
\subfloat[]{\includegraphics[scale=0.4]{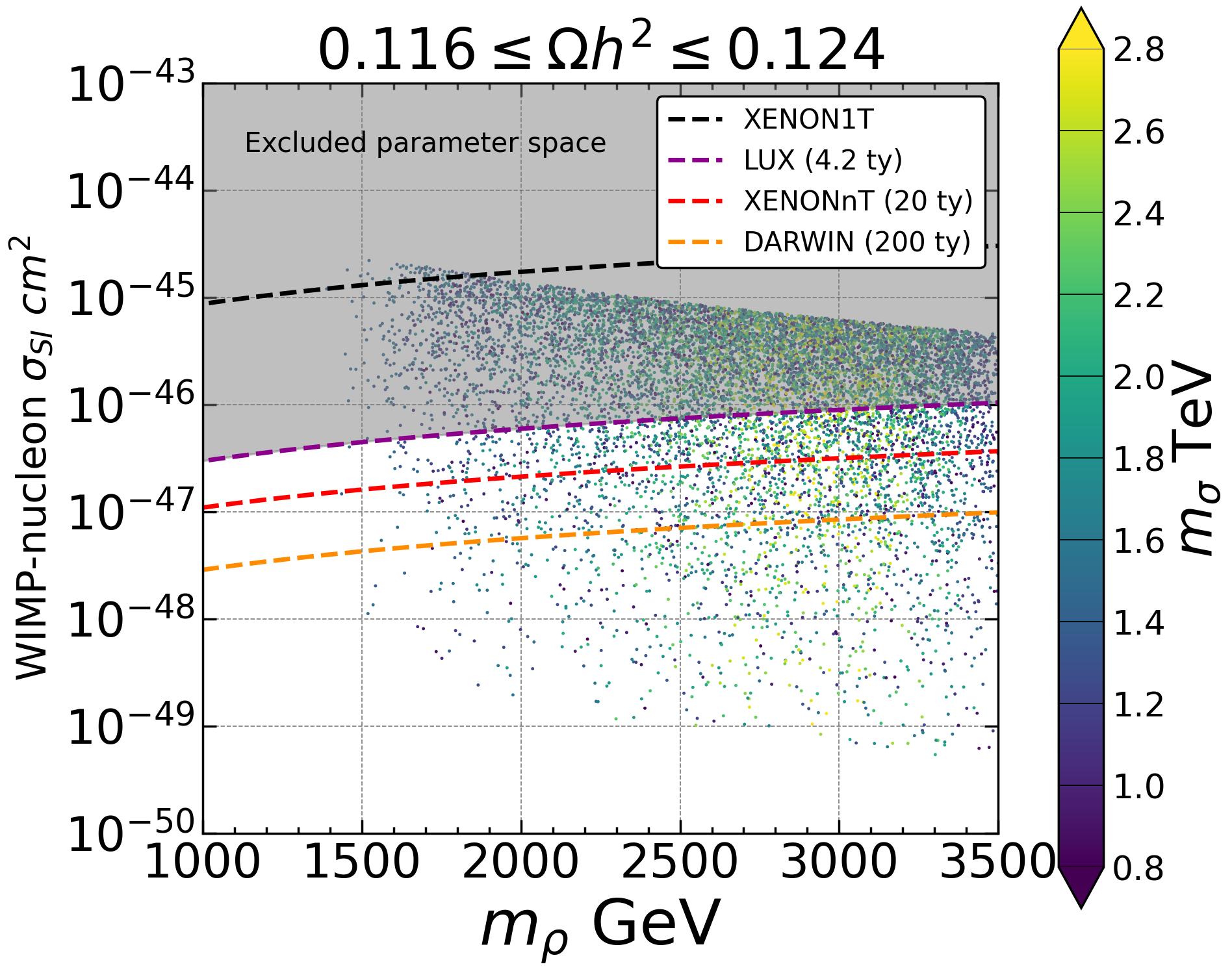}}\quad
\subfloat[]{\includegraphics[scale=0.4]{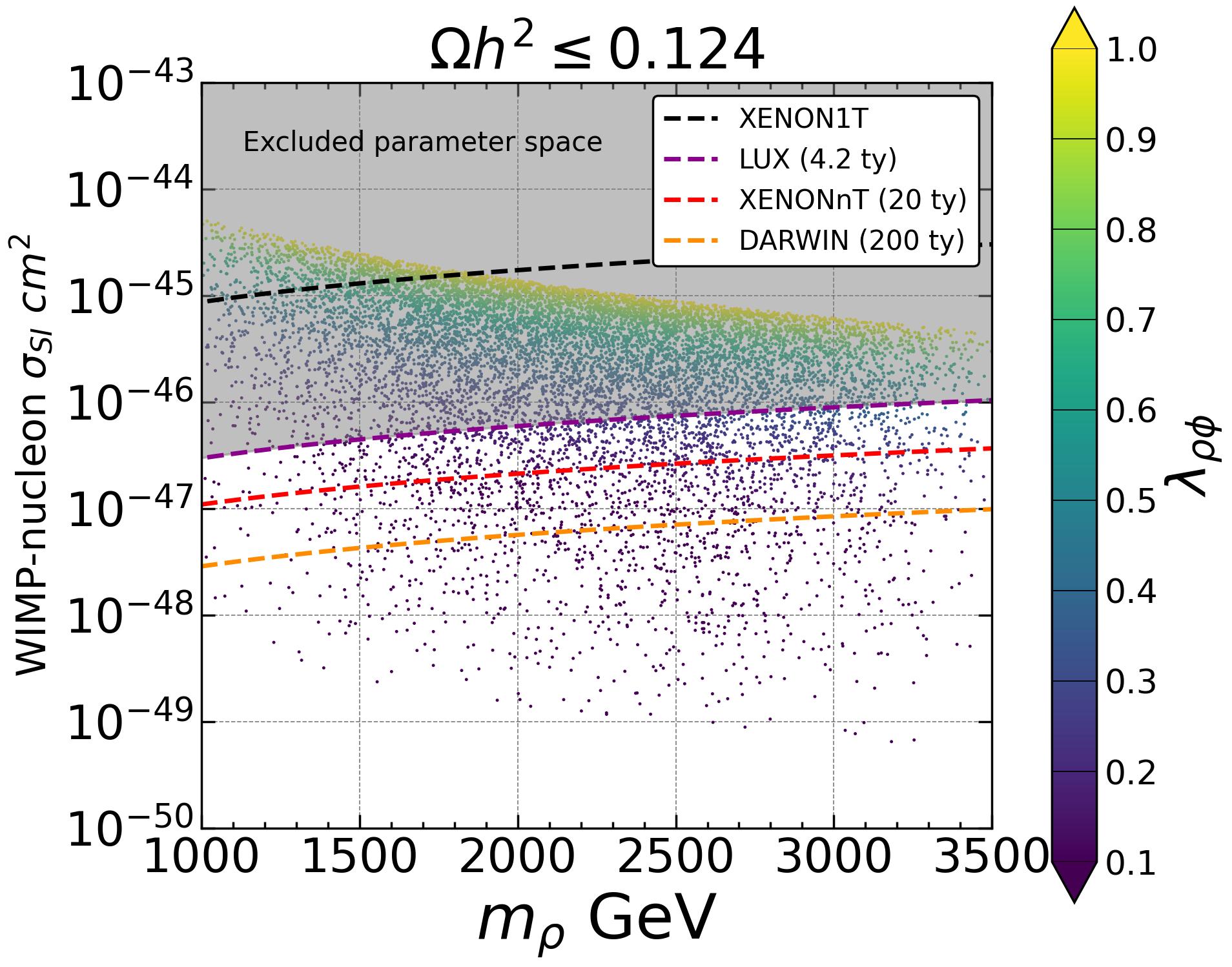}}\quad
\subfloat[]{\includegraphics[scale=0.4]{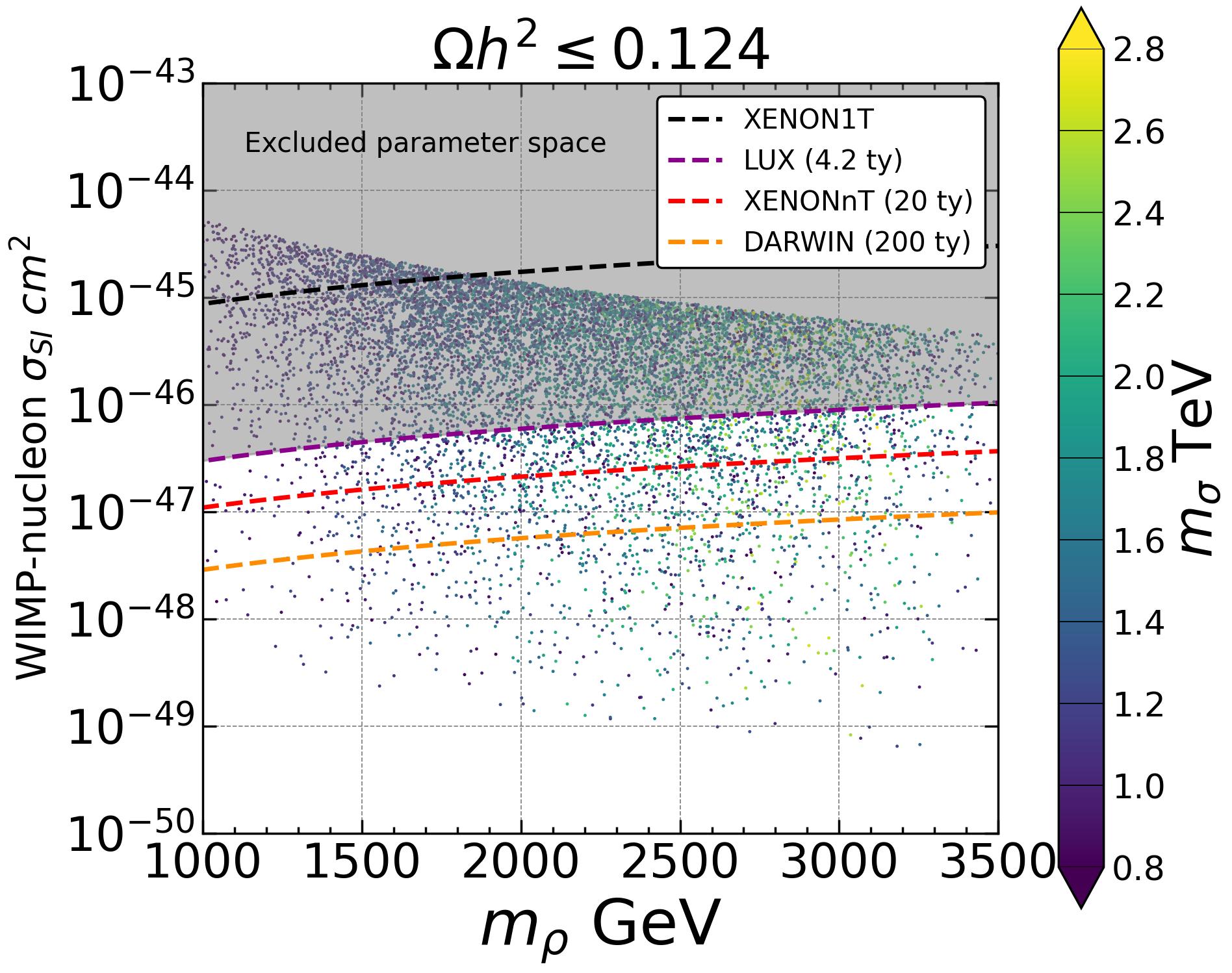}}
\caption{Spin-independent elastic DM-nucleon scattering cross section as a function of the DM mass. The gray area represents the excluded parameter space given by the XENON1T and LUX-ZEPLIN at 4.2 ton-year represented by the black and purple lines respectively. While the red line shows the future projections from XENONnT at 20 ton-year and the orange line the projections for DARWIN at 200 ton-year. Points in plots (a) and (b) satisfy the relic density saturation constraint within $3\sigma$, whereas plots (c) and (d) correspond to sub-abundant dark matter scenarios. The color map in (a) and (c) indicates the $\rho$–Higgs coupling, while in (b) and (d) it represents the mass of the $\sigma$-singlet.}
\label{dd_mass_DMscalar}
\end{figure}

\begin{figure}[h!]
\centering
\subfloat[]{\includegraphics[scale=0.35]{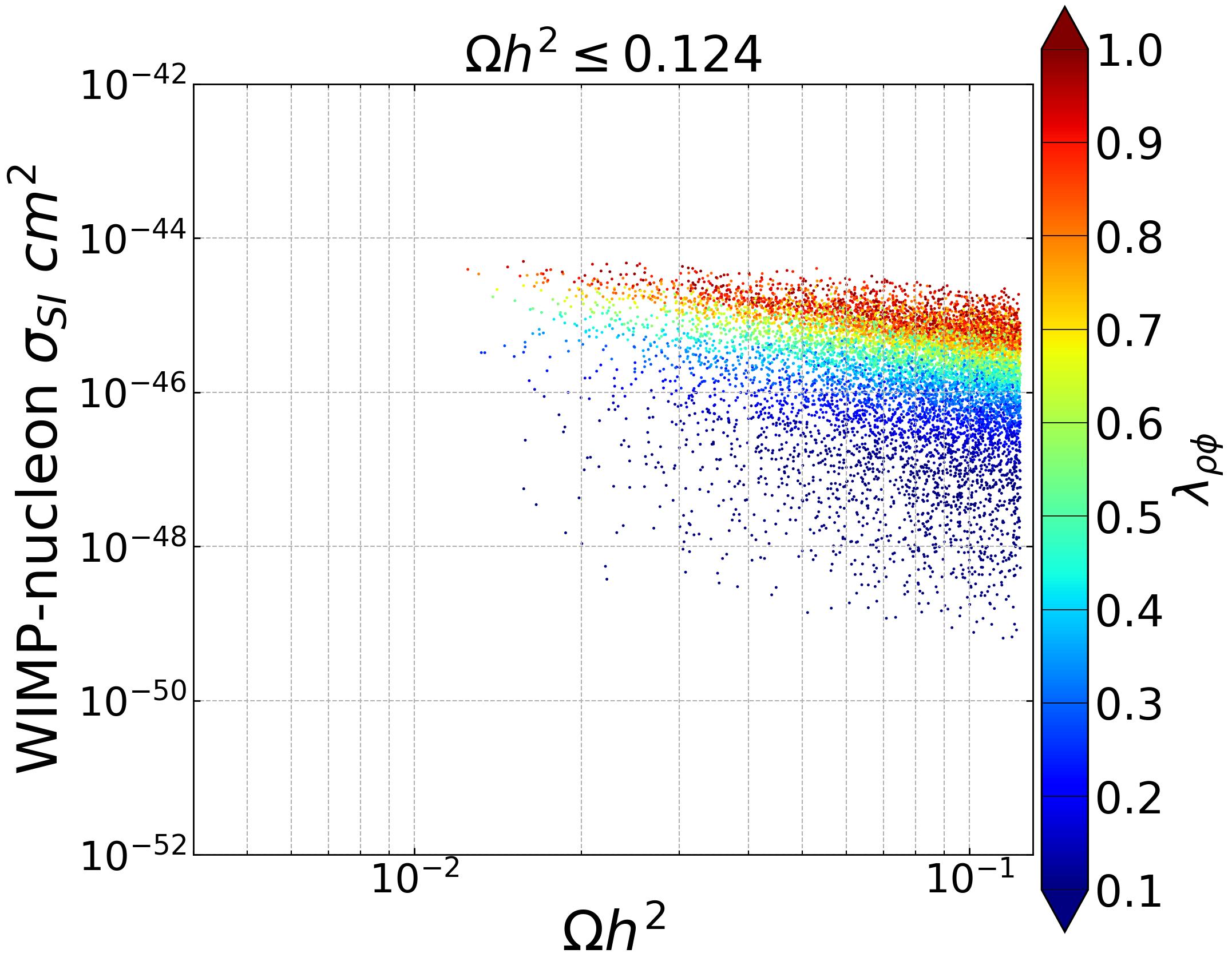}}\quad
\subfloat[]{\includegraphics[scale=0.35]{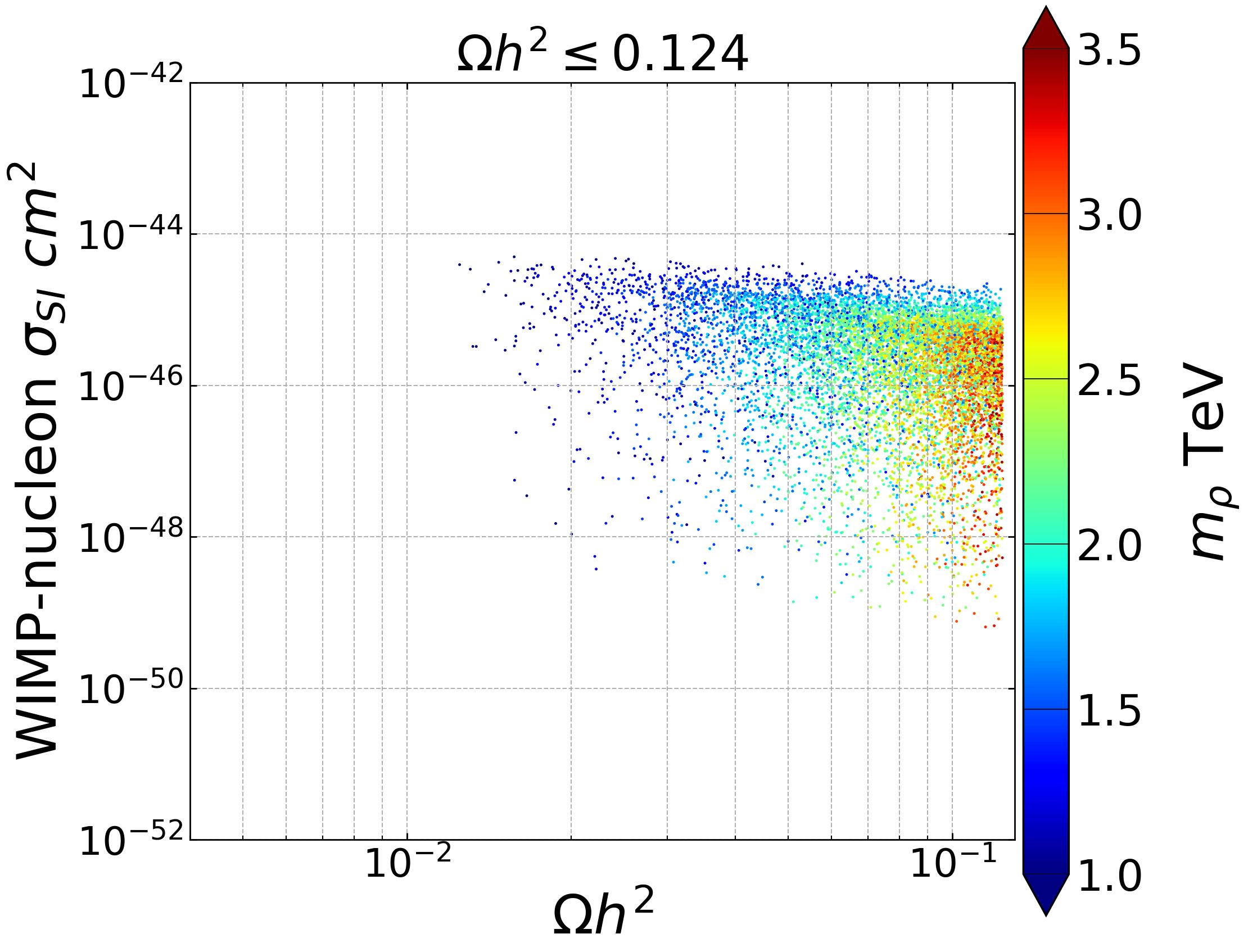}}
\caption{Spin-independent elastic DM-nucleon scattering cross section as a function of the relic density. The color map represents the $\rho$-Higgs coupling and the mass of the DM in (a) and (b) respectively.}
\label{dd_relic_DMscalar}
\end{figure}

We focus on the region of parameter space where $\lambda_{\phi\rho} \lesssim 1$, $\lambda_{\sigma\rho} \neq 0$, $\lambda_{\phi\sigma} \neq 0$, $0.8$ TeV $\lesssim m_\sigma \lesssim 2.8$ TeV, and $1$~TeV~ $\lesssim m_\rho \lesssim 3.5$ TeV. Within this region, relic density saturation is achieved, except for $m_\rho \lesssim$ 1.4 TeV as indicated by the points in Figure \ref{dd_mass_DMscalar} (a) and (b). Interestingly, values of $\lambda_{\phi\rho} \gtrsim 0.5$ are entirely excluded across the full dark matter mass range. Moreover, only values $\lambda_{\phi\rho} \lesssim 0.1$ can avoid the projected limits from XENONnT (20 ton-year) and DARWIN (200 ton-year), even in the case of sub-abundant dark matter. Furthermore, Figure \ref{dd_relic_DMscalar} (a) shows more clearly that, in order to elude a large spin-independent scattering cross section, it is necessary to increase the dark matter mass, which in turn tends to bring the model closer to relic density saturation.

On the other hand, the mass of the dark matter candidate does not appear to be excluded by current experiments, allowing for values in the range $1$ TeV $\lesssim m_\rho \lesssim 3.5$ TeV. However, a heavier $\rho$-singlet can more easily escape the direct detection bounds and saturate the measured relic density, as shown in Figure \ref{dd_relic_DMscalar} (b). In contrast, lighter $\rho$ masses exhibit larger spin-independent cross sections, even when $\Omega h^2 \simeq 0.12$. Meanwhile, the mass of the $\sigma$-singlet appears to play a minor role in the allowed parameter space, although the plots indicate that it typically falls below 2.7 TeV for relic density saturation and below 2.4 TeV for sub-abundant dark matter.

Additionally, another possible scenario arises when the $\rho$-singlet is lighter than the $\sigma$-singlet, effectively this closes the annihilation channel $\rho\rho\to\sigma\sigma$. While this configuration is viable in principle, it presents significant phenomenological challenges. This will suppress the total average annihilation cross sections pushing the relic density to generate it's overabundance. Moreover reducing the $\rho$-singlet mass due to this scenario, typically enhances the spin-independent scattering cross section, making the scenario increasingly constrained by the XENON1T and LUX-ZEPLIN experiments. Taken together, these effects render this benchmark less favored in light of current experimental bounds.

\subsubsection{Indirect Detection}

In the previous sub-section, we discussed the challenges faced by the model in producing observable signals in direct detection experiments. We emphasized that the coupling $\lambda_{\rho\phi}$ plays a crucial role in connecting the dark matter particles to the Standard Model sector via the Higgs boson, effectively acting as a Higgs portal. This interaction not only affects the spin-independent DM-nucleon cross section, but also governs the dominant annihilation channels of the dark matter candidate. In particular, we focus on the annihilation cross sections $\langle\sigma v\rangle_{\rho\rho\to b\bar{b}}$ and $\langle\sigma v\rangle_{\rho\rho\to W^+W^-}$, 
where the most stringent current bounds on weak-scale DM annihilation are given in \cite{Cirelli:2024ssz}. %and on the annihilation cross section given by the Goldstone $\langle\sigma v\rangle_{\rho\rho\to JJ}$.

\begin{figure}[h!]
\centering
\subfloat[]{\includegraphics[scale=0.3]{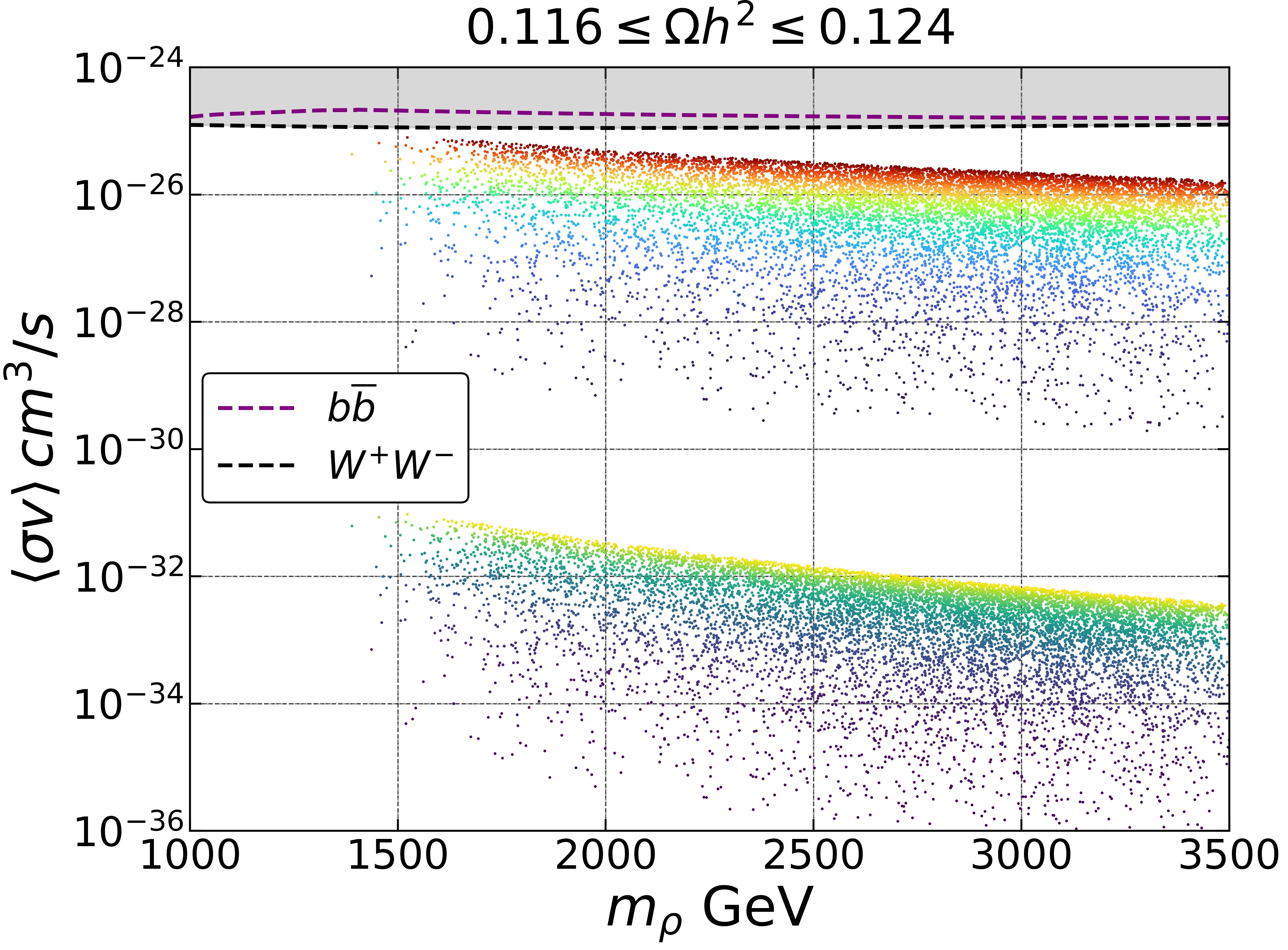}}\quad
\subfloat[]{\includegraphics[scale=0.3]{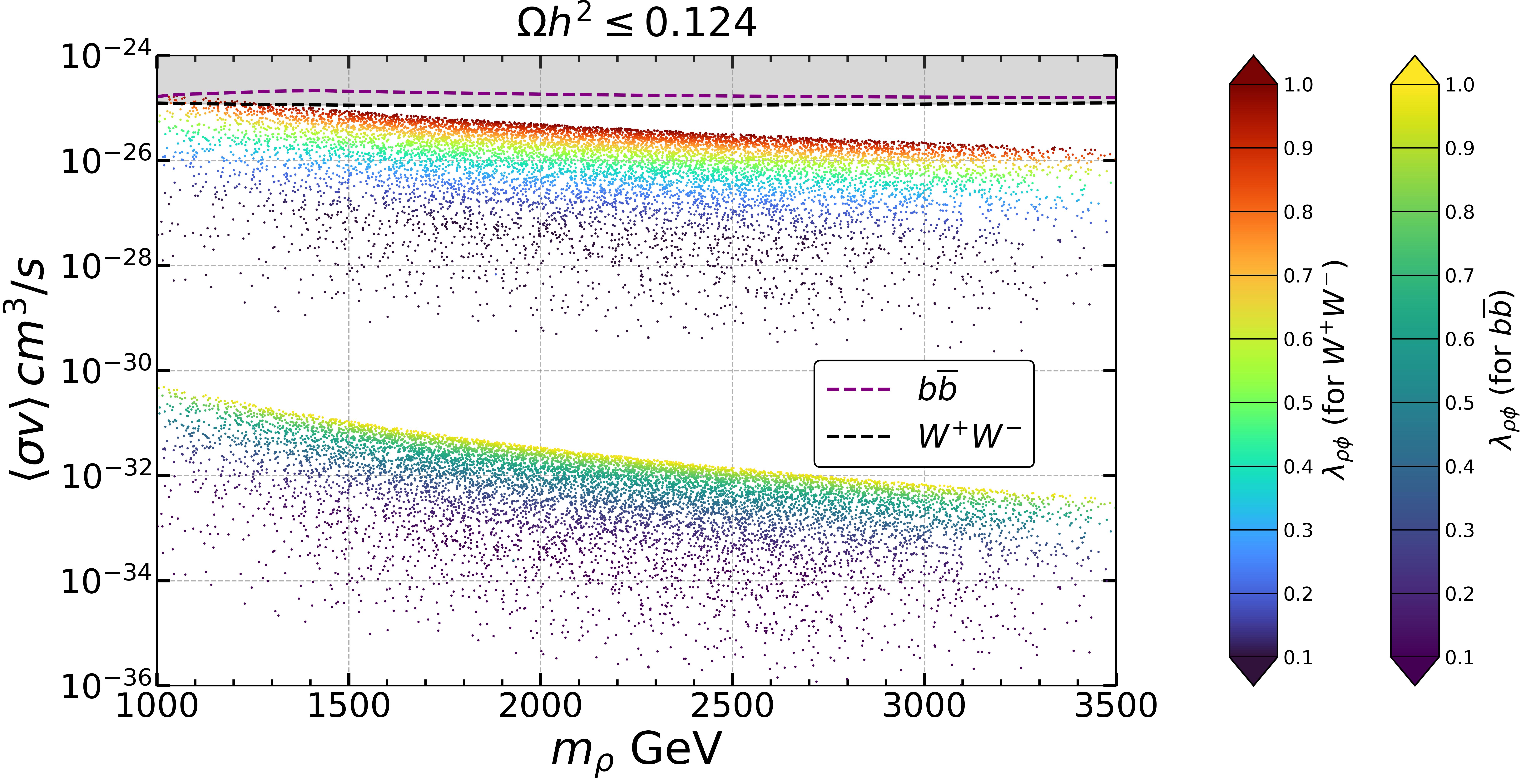}}
\caption{Thermally averaged total annihilation cross-section as a function of the DM mass. Both color maps represent the $\rho$-Higgs coupling $\lambda_{\rho \phi}$, the left one is for the annihilation cross-section with $W^+W^-$ as final states, while the right one is with $b\overline{b}$. The dashed purple and black lines represent the experimental limits for the annihilation cross-sections taken from Ref.~\cite{Cirelli:2024ssz}. All the points in the graph (a) satisfy the relic density saturation constraint within 3$\sigma$, whereas in the graph (b) correspond to sub-abundant dark matter. }
\label{id_scalarDM}
\end{figure}

We observe in both plots of Figure \ref{id_scalarDM} a small region excluded by current experimental bounds, specifically due to the annihilation channel with $b\bar{b}$ in the final state. Even when considering both relic density saturation and sub-abundant dark matter, the viable parameter space is limited to values around $m_\rho \lesssim 1.1$ TeV and $\lambda_{\rho\phi}\sim 1$, the latter is already ruled out by direct detection constraints. Conversely, the annihilation cross section $\langle\sigma v\rangle_{\rho\rho\to b\bar{b}}$ remains sufficiently suppressed to evade current experimental sensitivity.
%masses of 1 keV, 1 MeV and 1 GeV

\subsection{Fermionic dark matter}
\label{fermionic_dark_matter}
\begin{figure}[h!]
    \centering
    \includegraphics[width=0.25\linewidth]{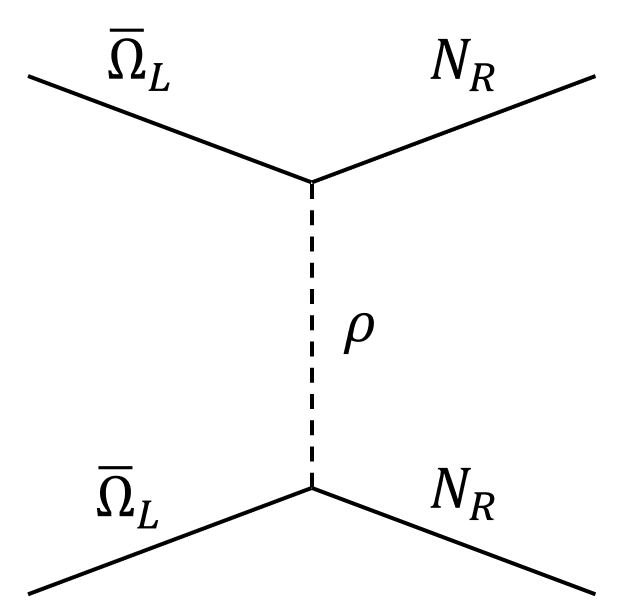}
    \caption{Feymann diagram for the DM $\Omega$ annihilation to the sterile neutrino.}
    \label{fermionic_DM_image}
\end{figure}

As mentioned, the model also admits a \emph{fermionic} dark–matter candidate. This happens when $\Omega$ is the lightest $Z_3$–nontrivial state. In practice we consider
$m_\Omega < m_\rho,$
which ensures $\Omega$’s stability under the exact $Z_3$ symmetry. The process that maintains $\Omega$ in thermal equilibrium with the primordial plasma is
$\Omega\,\overline{\Omega}\ \to\ N\,\overline{N}$, mediated by a $t$–channel exchange of the SM–singlet scalar $\rho$, as shown in Fig.~\ref{fermionic_DM_image}. The final state contains the lightest sterile neutrinos $N$ introduced in Sec.~\ref{neutrino masses}.

The corresponding thermally averaged cross section is given by \cite{Goyal:2016zeh}
\begin{equation}
 \left<\sigma v\right>\simeq\dfrac{y_N^4m_\Omega^2  \sqrt{1-\dfrac{4 m_N^2}{m_\Omega^2}}}{32 \pi  \left(m_\rho ^2+m_\Omega ^2-m_N ^2\right)^2},
\end{equation}
where the $y_{N}$ is the $\Omega N\rho$ Yukawa coupling introduced in \eqref{eq:lag-neutrino}.

%%%%%%%%%%%%%%%%%%%
%%%%%%%%%%%%%%%%%%%%

\subsubsection{Direct Detection}
\label{sec:FermionicDM-DDM}
\begin{figure}[h]
    \centering
    \includegraphics[width=0.75\linewidth]{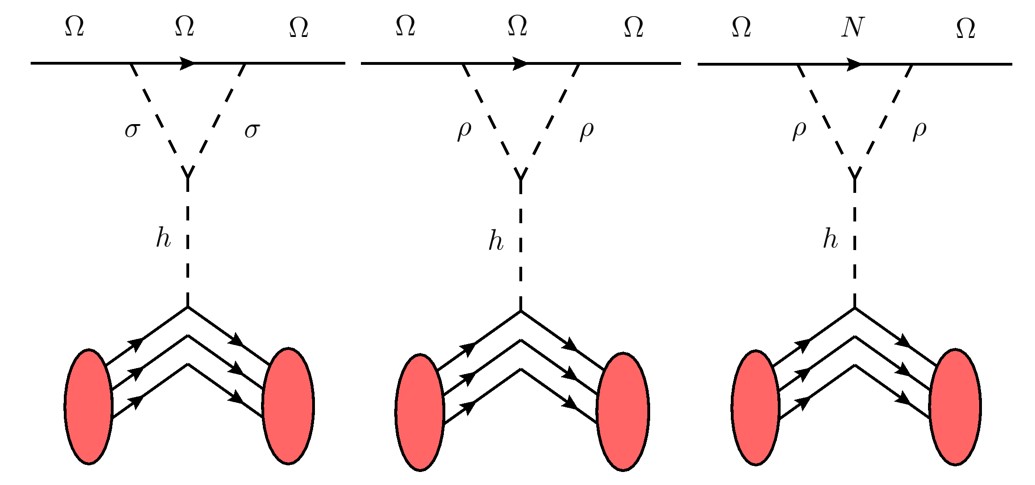}
    \caption{Penguin diagrams for the dark matter contribution to spin-independent direct detection via one-loop process.}
    \label{fig:pinguins}
\end{figure}

\begin{figure}[h!]
\centering
\subfloat[]{\includegraphics[scale=0.45]{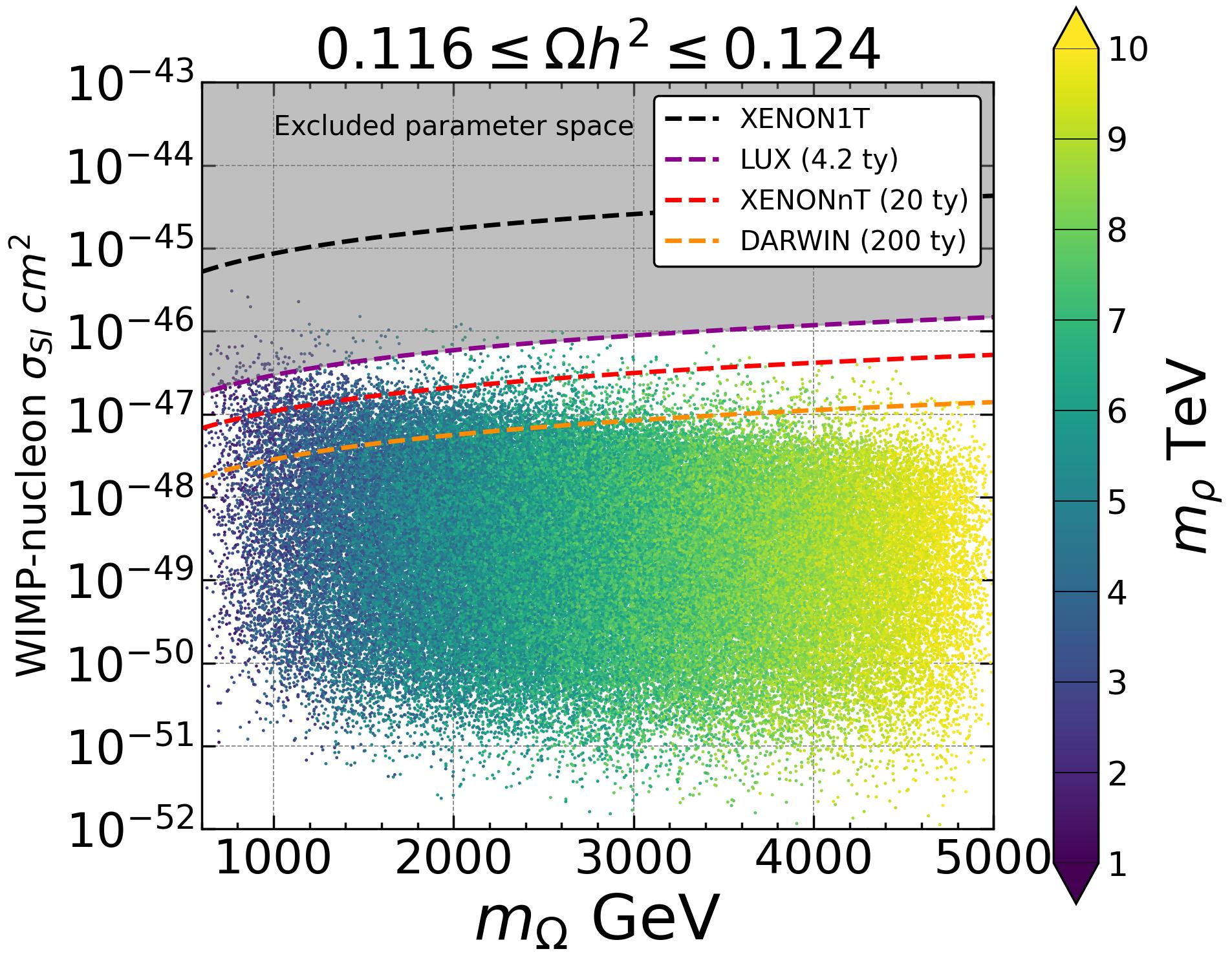}\label{fig:dd_fermion_rho}}\quad
\subfloat[]{\includegraphics[scale=0.45]{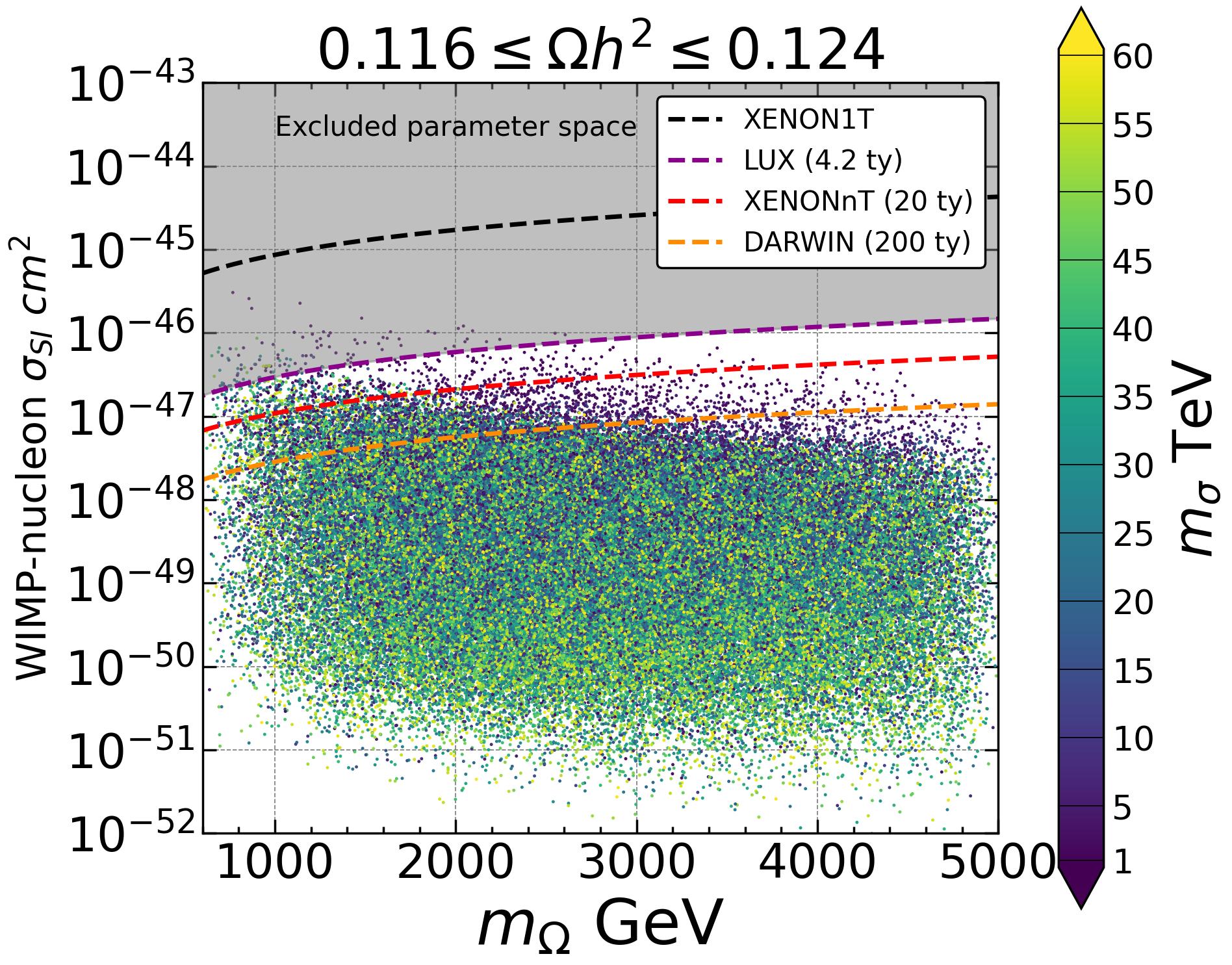}\label{fig:dd_fermion_sigma}}
\caption{Spin-independent elastic DM-nucleon scattering cross section as a function of the DM mass. All the points satisfied the saturation of relic density at 3$\sigma$. The color map indicates the mass of the $\rho$ singlet. The gray area represents the excluded parameter space given by the XENON1T and LUX-ZEPLIN at 4.2 ton-year represented by the black and purple lines respectively. While the red line shows the future projections from XENONnT at 20 ton-year and the orange line the projections for DARWIN at 200 ton-year.}
\label{dd_fermionicDM}
\end{figure}

Unlike the previous case of the scalar DM candidate $\rho$, the fermionic DM candidate $\Omega$ does not receive any tree-level contribution to its scattering off nuclei, since $\rho$–$h$ mixing is forbidden in our model. The process responsible for the DM direct detection is therefore dominated by the one-loop penguin diagrams shown in Fig.~\ref{fig:pinguins}. As a result, it is loop-suppressed with respect to $\Omega\Omega$ annihilation, which is determined by the tree-level diagram in Fig.~\ref{fermionic_DM_image}. This allows for the reconciliation of the stringent direct-detection constraints and the observable relic density. The corresponding contribution to the spin-independent scattering cross section is given by \cite{Pathak:2024sei,Amiri:2025ras}
\begin{equation}
\sigma_{SI} = \left(\frac{m_n}{v}\right)^2 \frac{(\mu_{DM-n})^2}{\pi m^4_h} f^2_n\left(\sum_{i=1}^3 g_{\Omega\Omega h,i}\right)^2,
\end{equation} where $\mu_{DM-n} = \frac{M_{DM} \, m_n}{M_{DM} + m_n}$ is the reduced mass of DM-nucleon system with $m_n$ being the mass of the nucleon in GeV and $f_{n}=0.3$ \cite{fnGiedt:2009mr}.  The effective coupling $g_{\Omega\Omega h}$ between DM and Higgs is given by
\begin{align*}
g_{\Omega \Omega h,1} &= \dfrac{\lambda_
{\rho\phi}\,v\,y_N^2m_N\left((m_\rho^2-m_N^2)+m_\rho^2\ln{\left(\dfrac{m_N^2}{m_\rho^2}\right)}\right)}{8\pi^2(m_\rho^2-m_N^2)^2},\\
g_{\Omega \Omega h,2} &= \dfrac{\lambda_
{\rho\phi}\,v\,z_\Omega^2m_\Omega\left((m_\rho^2-m_\Omega^2)+m_\rho^2\ln{\left(\dfrac{m_\Omega^2}{m_\rho^2}\right)}\right)}{8\pi^2(m_\rho^2-m_\Omega^2)^2},\\
g_{\Omega \Omega h,3} &= \dfrac{\lambda_
{\sigma\phi}\,v\,y_\Omega^2m_\Omega\left((m_\sigma^2-m_\Omega^2)+m_\sigma^2\ln{\left(\dfrac{m_\Omega^2}{m_\sigma^2}\right)}\right)}{8\pi^2(m_\sigma^2-m_\Omega^2)^2}.
\end{align*}
The spin-independent DM-nucleon scattering cross section is shown in Fig.\ref{dd_fermionicDM} as a function of the dark matter candidate mass. The relic density saturation is achieved. Plot \ref{fig:dd_fermion_rho} shows that almost all points are below the parameter space excluded by current experiments, in the mass range $1.0$ TeV $\lesssim m_\Omega \lesssim 5.0$ TeV and also it is shown a correlation between the mass of the DM candidate and the mass of the scalar singlet $\rho$ indicating that the larger the masses, the smaller the spin-independent cross section. In plot \ref{fig:dd_fermion_sigma} we obtain a similar behavior between DM mass and the spin-independent cross section in the same mass range but with no apparent correlation with the other scalar singlet $\sigma$.

\begin{figure}[h!]
\centering
\subfloat[]{\includegraphics[scale=0.4]{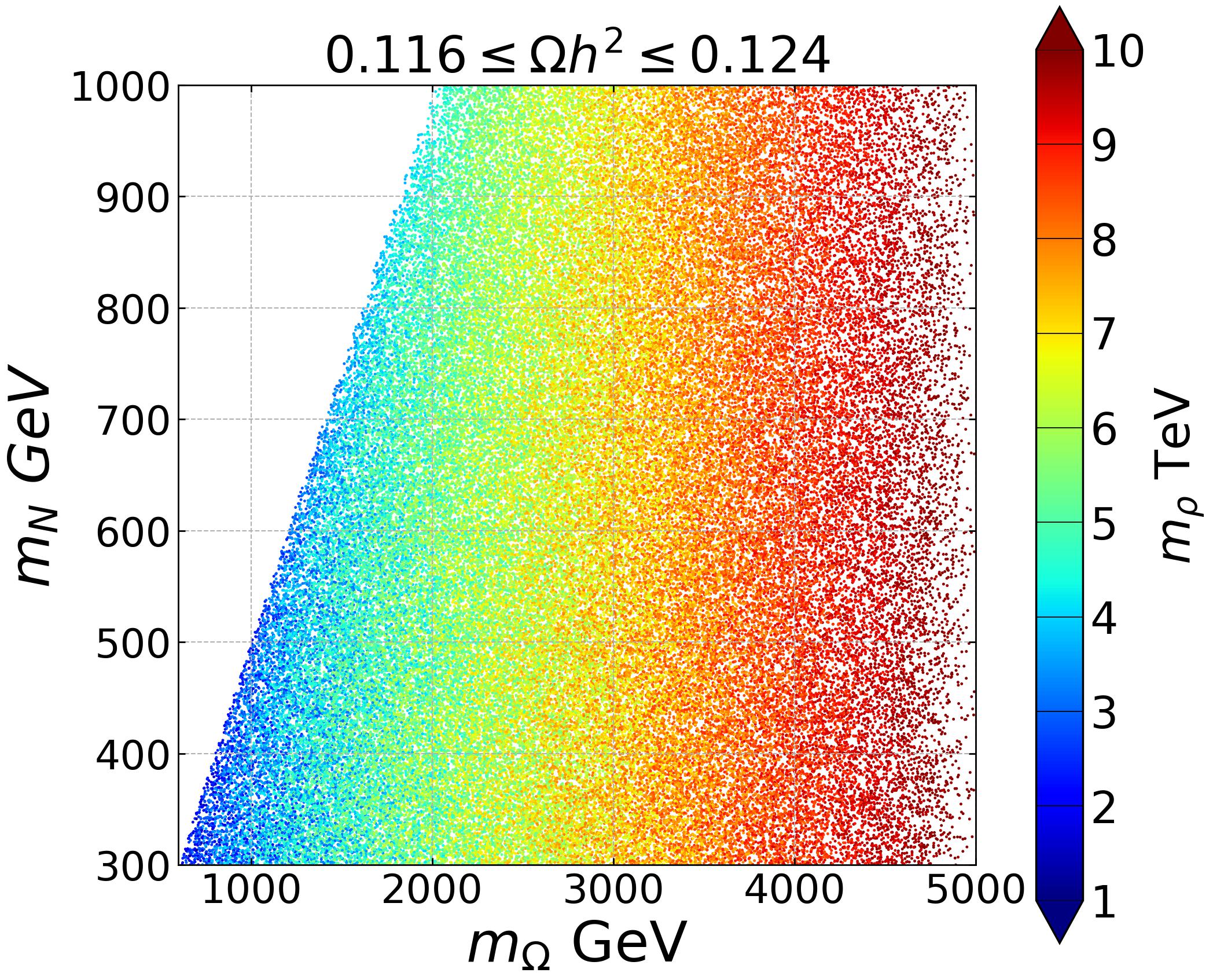}\label{fig:mN_fermion}}\quad
\subfloat[]{\includegraphics[scale=0.4]{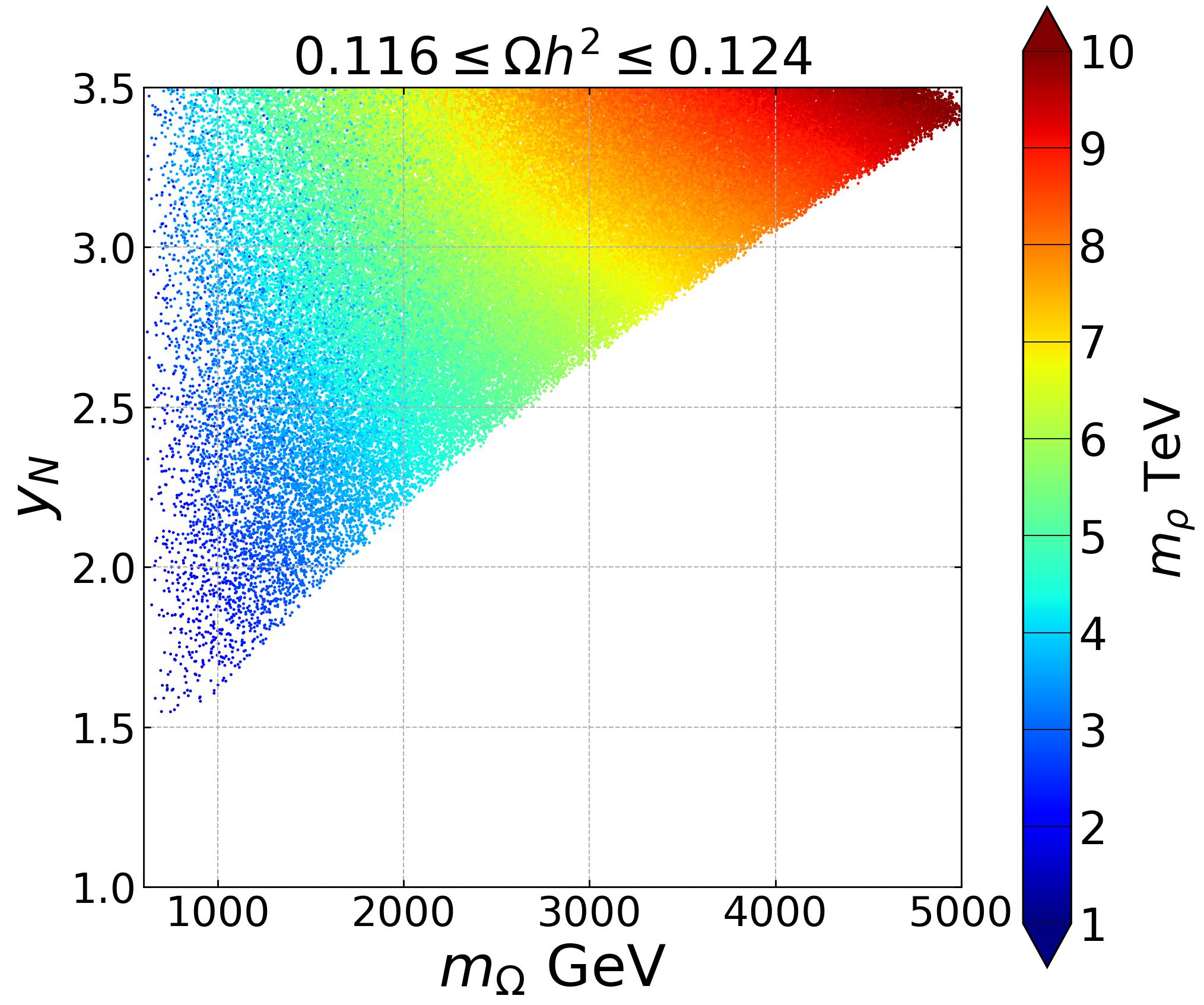}\label{fig:yuk_fermion}}
\caption{(a) Mass of the sterile heavy neutrino as a function of the DM mass and (b) Neutrino Yukawa coupling $y_N$ as a function of the DM mass. All the points satisfied the saturation of relic density at 3$\sigma$. The color map indicates the mass of the $\rho$ singlet.
}
\label{mN_yuk_fermionicDM}
\end{figure}

In Fig.\ref{fig:mN_fermion} we can observe the mass of the heavy sterile neutrinos versus the mass of the fermionic DM. In the range of $1-2$ TeV for the $m_{\Omega}$, it is possible to observe that as the mass of the heavy neutrinos increases, their quantity decreases, making it so that for $m_{\Omega}=2$ TeV there are practically no mass values of heavy neutrinos close to 1 TeV. However, it is also possible to observe that from $m_{\Omega}=2$ TeV onward, there is a direct correlation with the mass of the scalar mediator $\rho$, that is, as the $m_{\Omega}$ increases, so does the $m_{\rho}$.

In Fig.\ref{fig:yuk_fermion}, we plot the neutrino Yukawa coupling $Y_{N}$ versus the $m_{\Omega}$. We take a range of values for $Y_{N}$ between $1.5-3.5$ where we can see that this coupling is strongly restricted by the $m_{\Omega}$. For example, for values of $m_{\Omega}=2$ TeV, $Y_{N}$ can take values between 2.0 and 3.5, but if we increase the omega mass to $m_{\Omega}=4$ TeV, $Y_{N}$ can only take values between 3.0 and 3.5. However, as in the plot \ref{fig:yuk_fermion}, $m_{\Omega}$ is directly correlated with $m_{\rho}$. This analysis establishes a lower bound for the omega mass since below 1 TeV it is not possible to obtain a Yukawa coupling.

\section{Charged lepton flavor violation}
\label{clfv}

Now, we will study the charged lepton flavor violation processes due to the
mixing between the light active and heavy sterile neutrino of the model. The
branching ratio for the one-loop decay $l_{i}\rightarrow l_{j}\gamma$ is
given by~\cite{Langacker:1988up,Lavoura:2003xp}: 
\begin{eqnarray}
\text{BR}\left( l_{i}\rightarrow l_{j}\gamma \right) &=&\frac{\alpha
_{W}^{3}s_{W}^{2}m_{l_{i}}^{5}}{256\pi ^{2}m_{W}^{4}\Gamma _{i}}\left\vert
G_{ij}\right\vert ^{2},  \label{Brmutoegamma1}
\end{eqnarray}

where $\Gamma _{\mu }=3\times 10^{-19}$ GeV is the total muon decay width
and $G_{ij}$ has the following structure, 
\begin{eqnarray}
G_{ij} &\simeq &\sum_{k=1}^{3}\left( \left[ \left( 1-RR^{\dagger }\right)
U_{\nu }\right] ^{\ast }\right) _{ik}\left( \left( 1-RR^{\dagger }\right)
U_{\nu }\right) _{jk}G_{\gamma }\left( \frac{m_{\nu _{k}}^{2}}{m_{W}^{2}}%
\right) +2\sum_{l=1}^{2}\left(R^{\ast }\right) _{il}\left( R\right)
_{jl}G_{\gamma }\left( \frac{m_{N_{R_l}}^{2}}{m_{W}^{2}}\right),
\label{Brmutoegamma2} \\
G_{\gamma } (x) &=&\frac{10-43x+78x^{2}-49x^{3}+18x^{3}\ln x+4x^{4}}{%
12\left( 1-x\right) ^{4}}.  \notag
\end{eqnarray}

Here, $U_{\nu}$ is the matrix that diagonalizes the light active neutrino,
and since we are considering the diagonal charged lepton mass matrix, we
have that $U_{\text{PMNS}}= U_l^{\dagger}U_{\nu}= U_{\nu}$ (since $U_l=%
\mathbb{I}$), while the matrix $R$ is a rotation matrix given by, 
\begin{equation}
R=\frac{1}{\sqrt{2}}m_D^{*}M^{-1},  \label{eq:Rneutrino}
\end{equation}
where $M$ and $m_D$ are the  matrices from  Eq.~\eqref{eq:mD_M}. 
\begin{figure}[tbp]
\centering
\subfloat[]{\includegraphics[scale=0.4]{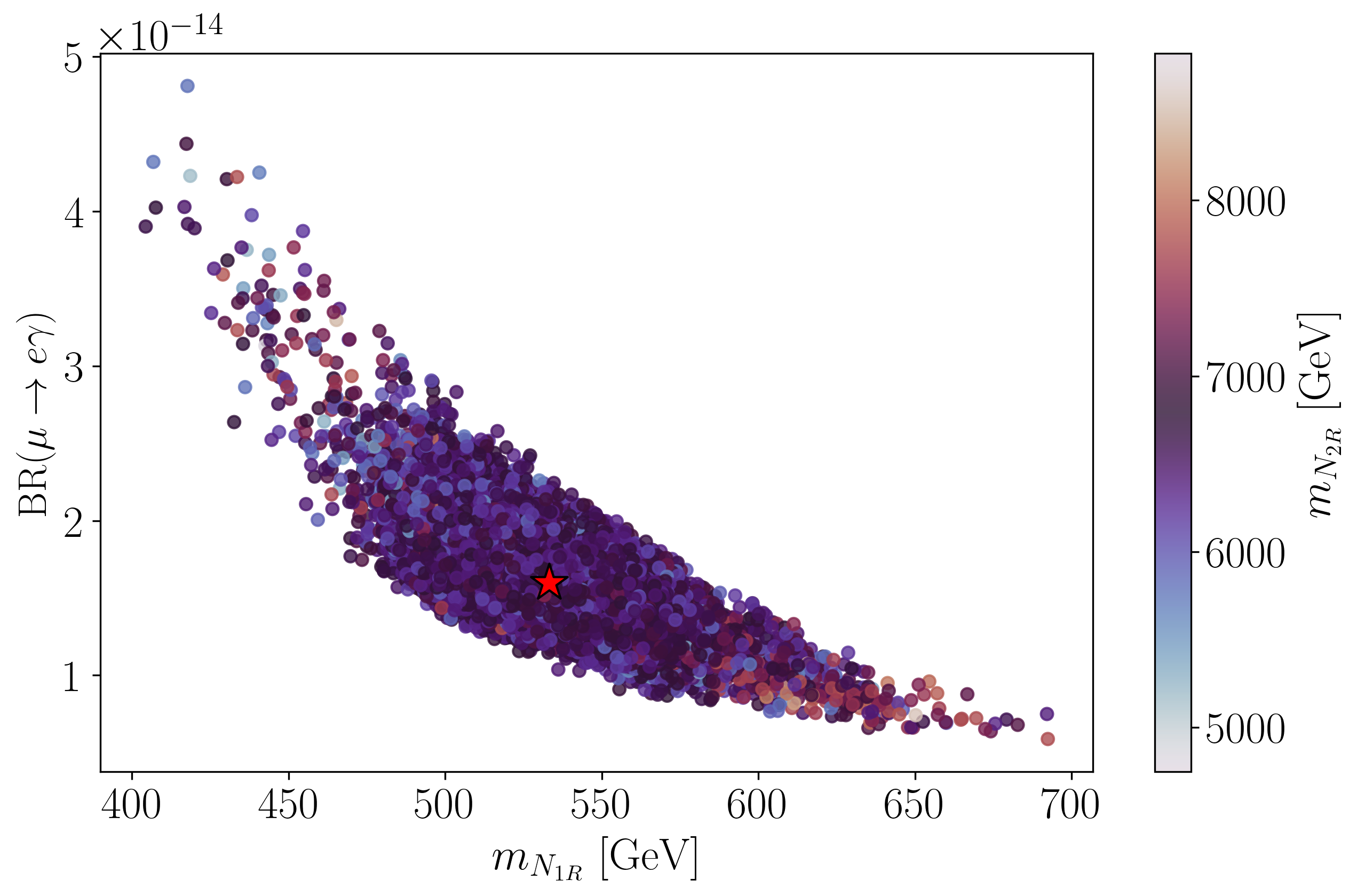}\label{fig:BRmN1-mN2}}\quad
\subfloat[]{\includegraphics[scale=0.4]{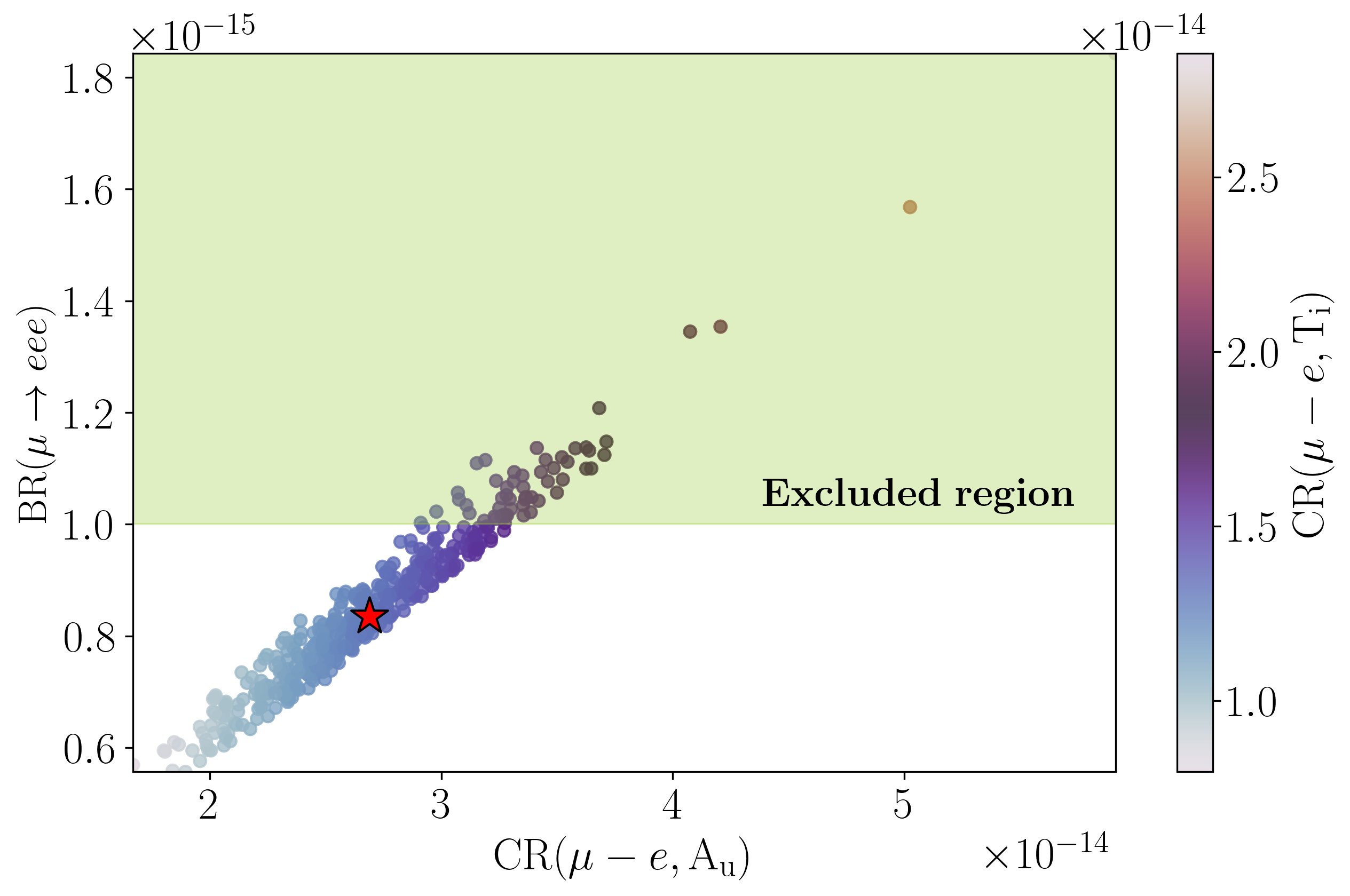}\label{fig:BReee-CRAu}}
\caption{a) Scatter plot between the branching ratio of the process $\protect\mu%
\rightarrow e\protect\gamma$ and the mass of the lightest Majorana neutrino,
considering different values of the mass of the heaviest Majorana neutrino. b) Scatter plot between the branching ratio of the process $\mu\rightarrow eee$ and $\text{CR}(\mu- e, \text{A}_{\text{u}})$, considering different values of the $\text{CR}(\mu- e, \text{T}_{\text{i}})$}.
\label{fig:scatterplot}
\end{figure}

For the processes $\mu\rightarrow e\gamma$, $\tau\rightarrow \mu\gamma$ and $\tau\rightarrow e\gamma$ our model predicts, according to Eq.~\eqref{Brmutoegamma1}, the following value for the corresponding branching ratios at the selected benchmark point:
\begin{eqnarray}
\text{BR}(\mu\rightarrow e\gamma) &\simeq & 1.6\times 10^{-14,}\\
\text{BR}(\tau\rightarrow e\gamma) &\simeq & 5.9\times 10^{-15},\\
\text{BR}(\tau\rightarrow \mu\gamma) &\simeq & 3.1\times 10^{-14},
\end{eqnarray}
that are significantly below the current experimental upper bounds \cite{MEGII:2025gzr,BaBar:2009hkt,Belle:2021ysv}
\begin{eqnarray}
\text{BR}(\mu\rightarrow e\gamma)_{\text{exp}} &\leq & 1.5\times 10^{-13}\\
\text{BR}(\tau\rightarrow e\gamma)_{\text{exp}} &\leq & 3.3\times 10^{-8} \\
\text{BR}(\tau\rightarrow \mu\gamma)_{\text{exp}} &\leq & 4.2\times 10^{-8}. 
\end{eqnarray}
In Fig.~\ref{fig:BRmN1-mN2}, a scatter plot is shown between the Branching ratio for the $\mu\rightarrow e\gamma$ process versus the mass of the lightest sterile right neutrino ($m_{N_1}$), in relation to the different values of the mass of the heaviest sterile right neutrino ($m_{N_2}$). This plot was generated by perturbing the benchmark point around 20\% and placing a restriction that each generated point is in agreement with the experimental values of the neutrino oscillation (see Table \ref{table:neutrinos_value}), representing the benchmark point of the model with the star point (red), where the value for the masses of sterile right neutrino are $m_{N_1}\lesssim 533.1$ GeV and $m_{N_2}\lesssim 6537$ GeV. In addition, we can see that each point in the plot is below the experimental limit according to the MEG II collaboration \cite{MEGII:2025gzr}, for the mass range for the lightest right handed neutrino between approximately $400\ \text{GeV}\lesssim m_{N_1}\lesssim 700\ \text{GeV}$. In turn, we can also see that the model predicts masses for the heaviest right neutrino in the $4750\ \text{GeV} \lesssim m_{N_2} \lesssim 8800\ \text{GeV}$ range.

\subsection{$\text{CR}\left(\mu-e,N\right)$}
\begin{figure}
\centering
\subfloat[]{\includegraphics[scale=0.1]{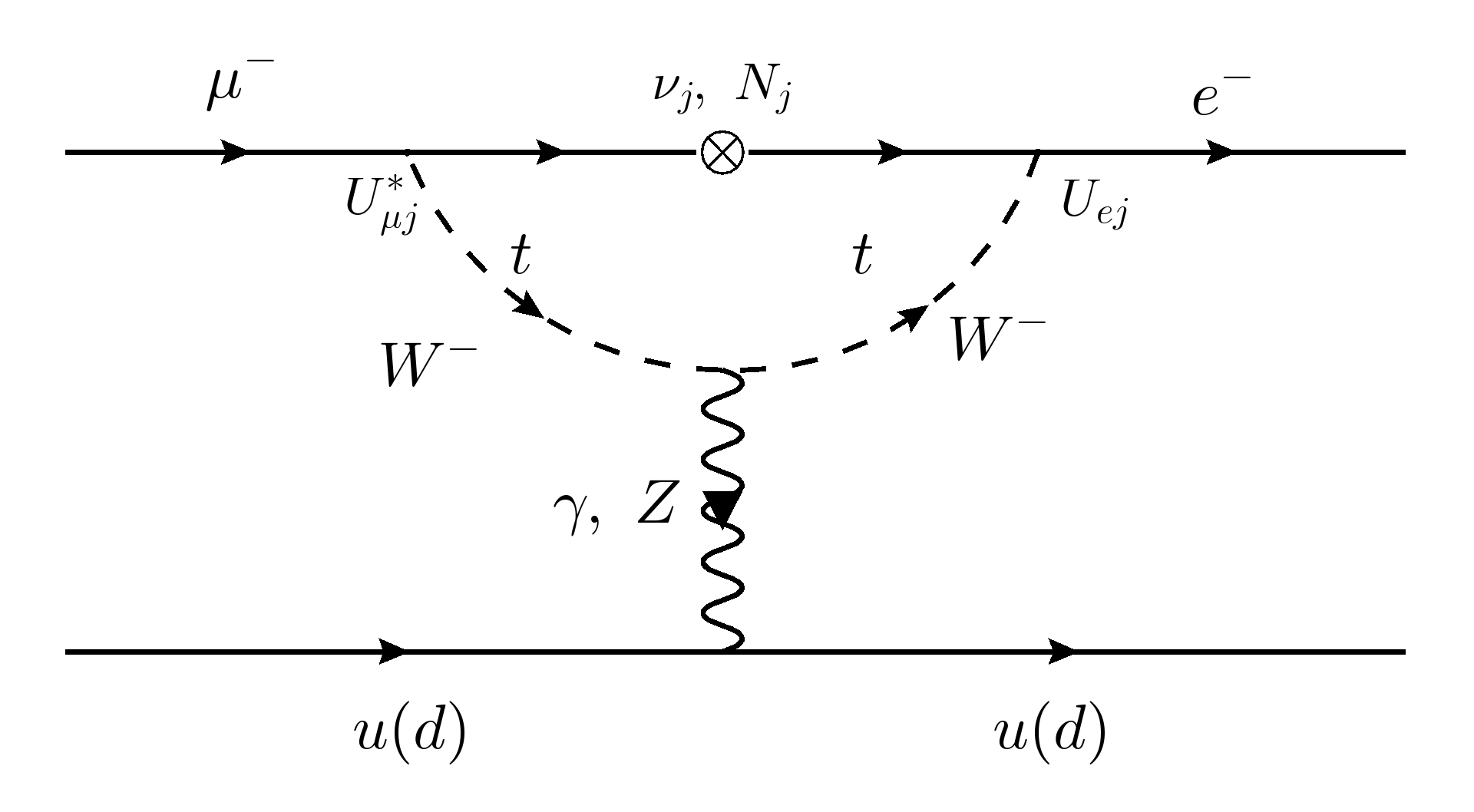}}\quad
\subfloat[]{\includegraphics[scale=0.1]{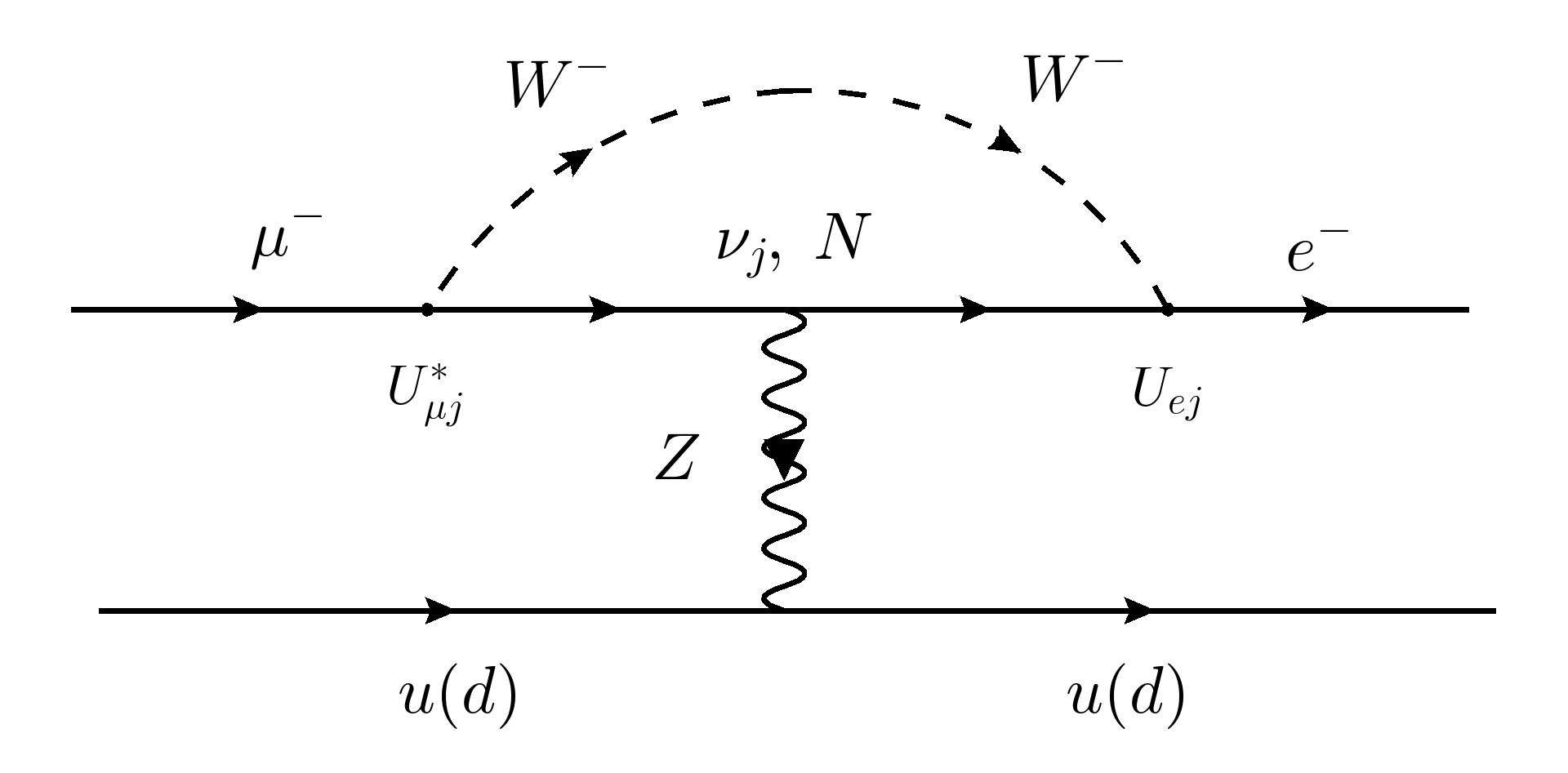}}\quad
\subfloat[]{\includegraphics[scale=0.1]{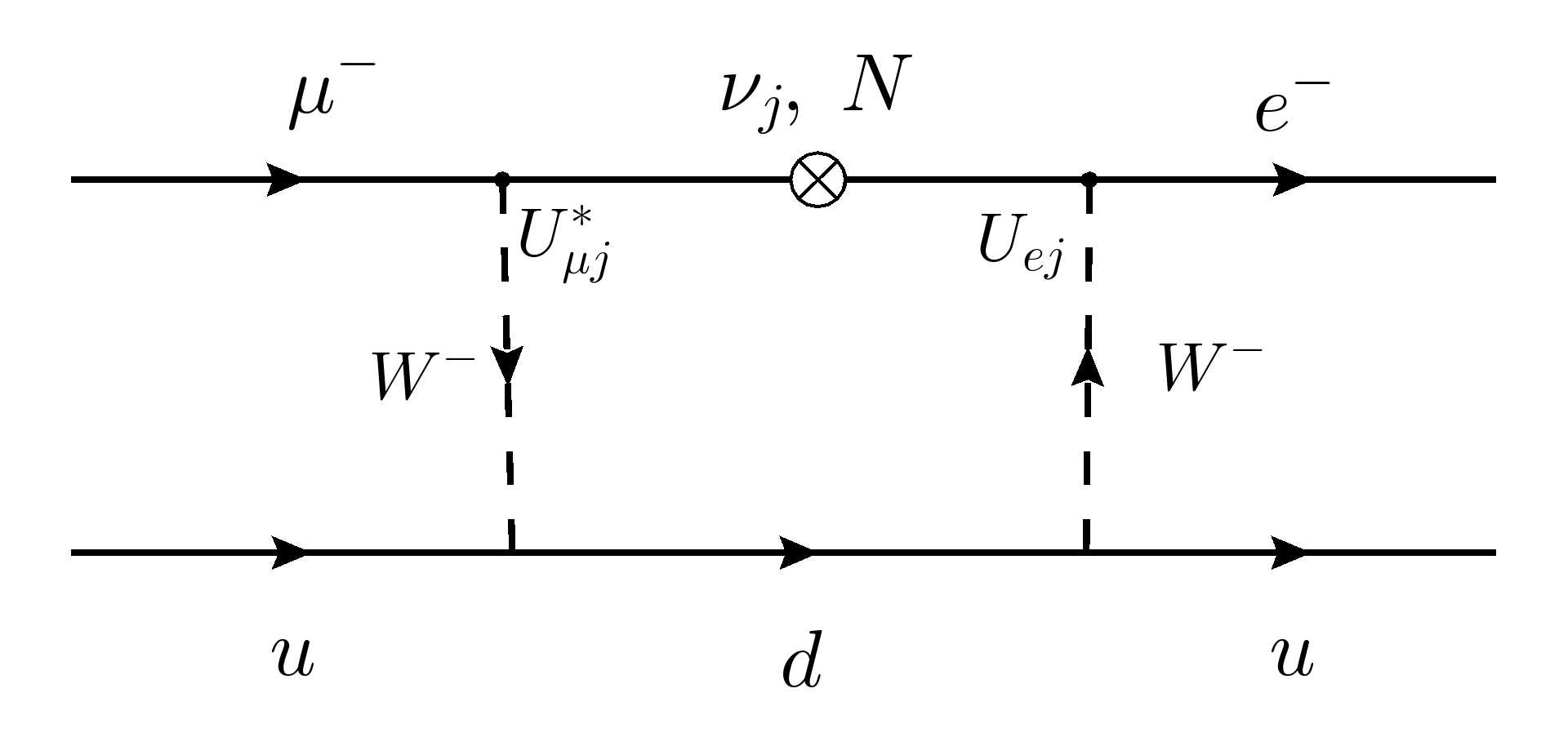}}\quad
\subfloat[]{\includegraphics[scale=0.1]{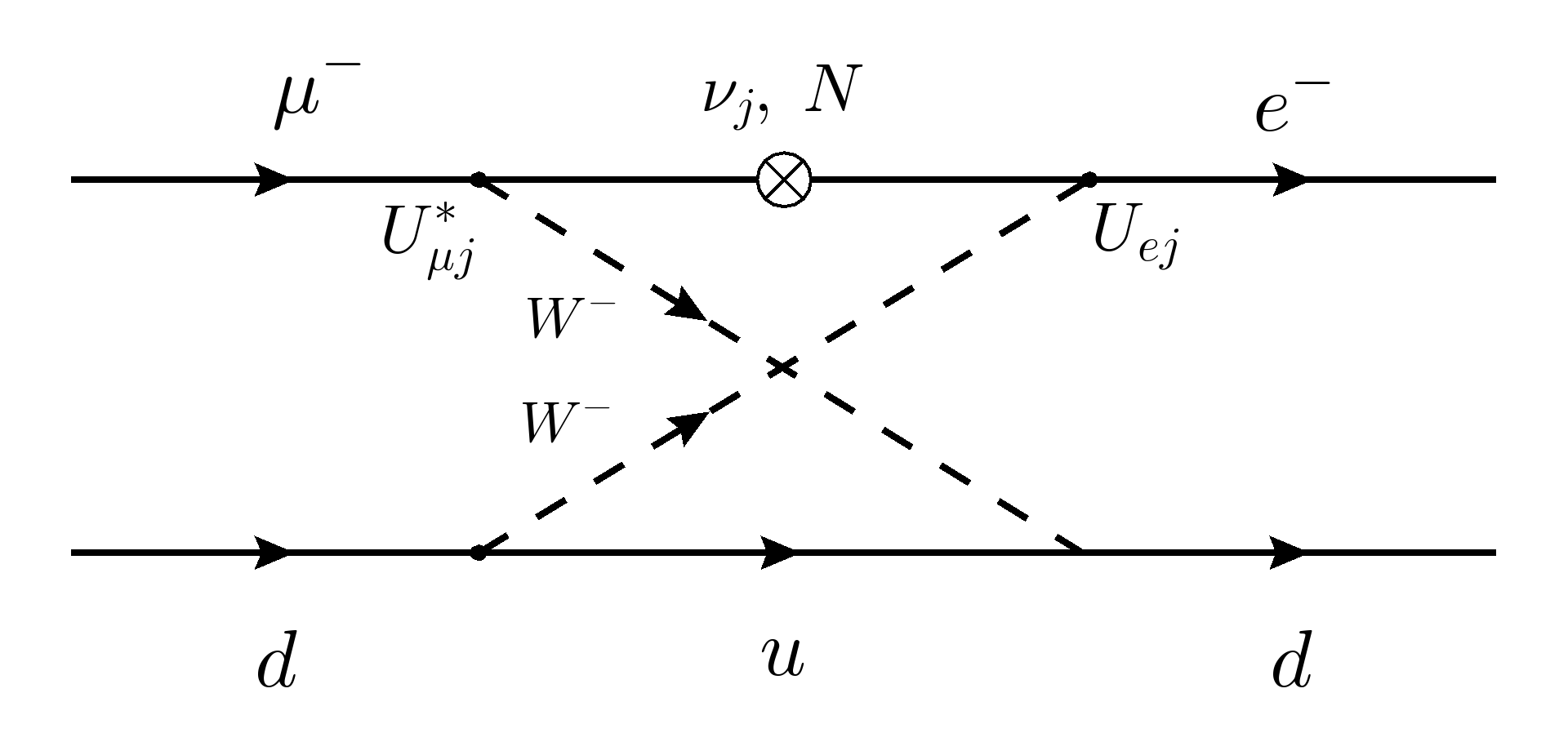}}
\caption{Penguin and box diagrams for $\mu-e$ conversion.}
\label{fig:diag-mue}
\end{figure}

The $\mu-e$ conversion refers to the decay of a muon bound to an atom, that is, when an atom captures a muon (muon atom), falling to the first state of a target nucleus N without neutrino emission. The conversion rate is defined as

\begin{equation}
\text{CR}\left(\mu-e,N\right)\equiv \frac{\Gamma\left(\mu^- +N\rightarrow e^- +N\right)}{\Gamma\left(\mu^- +N\rightarrow \text{all}\right)}.
\end{equation}
Box and penguin diagrams of Fig.~\ref{fig:diag-mue} contribute as \cite{Abada:2023zbb},
\begin{equation}
\text{CR}\left(\mu-e,N\right)= \frac{2G_F^2 \alpha_w^2m_{\mu}^5}{(4\pi)^2\Gamma_{\text{capt}}(Z)}\abs{4V^{(p)}\left(2\tilde{F}_u^{\mu e} +\tilde{F}_d^{\mu e} \right)+ 4V^{(n)}\left(\tilde{F}_u^{\mu e} +2\tilde{F}_d^{\mu e} \right)+ DG_{\gamma}^{\mu e}\frac{s_w^2}{2\sqrt{4\pi\alpha}}}^2
\end{equation}

where $\Gamma_{\text{capt(Z)}}$ denotes the capture rate of a nucleus with atomic number $Z$~\cite{Kitano:2002mt}, $G_F$ is the Fermi constant, $m_{\mu}$ the muon mass, $\alpha= e^2/(4\pi)$, with $s_w$ corresponding to the sine of the weak mixing angle. The form factors $\tilde{F}_q^{\mu e}$ $(q = u, d)$ are given by    
\begin{equation}
\tilde{F}_d^{\mu e}= Q_qs_w^2F_{\gamma}^{\mu e}+ F_Z^{\mu e}\left(\frac{I_q^3}{2}- Q_qs_w^2\right)+ \frac{1}{4}F_{\text{box}}^{\mu eqq},
\end{equation}

where $Q_q$ denotes the quark electric charge $(Q_u = 2/3,\ Q_d = -1/3)$ and $I_q^3$ is the weak isospin $(I_u^3 = 1/2, I_d^3 = -1/2)$. The quantities $F_{\gamma}^{\mu e}$ , $F_Z^{\mu e}$ and $F_{\text{box}}^{\mu eqq}$ correspond to the different form factors of the diagrams, and $G_{\gamma}^{\mu e}$ corresponds to the dipole term; all expressions are in ref.~\cite{Abada:2023zbb}. The relevant nuclear information (nuclear form factors and averages over the atomic electric field) is encoded in the form factors D, V (p), and V (n). In our analysis, we use the numerical values presented in ref.~\cite{Kitano:2002mt}.

\subsection{\boldmath $\ell_\alpha\to \ell_\beta\ell_\beta\ell_\rho$}

Next, we will see the analysis for the branching ratio of the muon decaying to three electrons, where according to~\cite{Ilakovac:1994kj,Alonso:2012ji}, it is given by: %has the following dependence,
\begin{align} \label{eq:mueee}
    \text{BR}(\mu \to eee)=& \frac{\alpha^4_w}{24576\pi^3}\frac{m^4_\mu}{M^4_W}\frac{m_\mu}{\Gamma_\mu} \times \Bigg\{ 2 \left|\frac12F^{\mu eee}_{\rm Box}+F^{\mu e}_Z-2s^2_w(F^{\mu e}_Z-F^{\mu e}_\gamma)\right|^2+4 s^4_w \left|F^{\mu e}_Z-F^{\mu e}_\gamma\right|^2 \nonumber \\
    &+ 16 s^2_w \text{Re}\left[	(F^{\mu e}_Z +\frac12F^{\mu eee}_{\rm Box})	G^{\mu e*}_\gamma \right] - 48 s^4_w \text{Re}\left[(F^{\mu e}_Z-F^{\mu e}_\gamma)	G^{\mu
    e*}_\gamma \right]	\nonumber \\ 
    &+32 s^4_w |G^{\mu e}_\gamma|^2\left[\ln \frac{m^2_\mu}{m^2_{e}} -\frac{11}{4}	\right] \Bigg\}, 
\end{align}
which contains the same form factors as those entering in CR($\mu-e$, N), although in different combinations, see Ref.~\cite{Abada:2023zbb, Lindner:2016bgg}. In Fig. \ref{fig:BReee-CRAu}, we can see the correlation between the process $\mu\rightarrow eee$ and $\text{CR}(\mu-e, A_u)$, considering different values of the process $\text{CR}(\mu-e, T_i)$. As in the previous case, the star point represents the values for the benchmark point, where, according to the plot, these values are,
\begin{align}
\text{BR}(\mu\rightarrow eee)&\simeq 8.35\times 10^{-16}, \\
\text{CR}(\mu-e, A_u)&\simeq 2.69\times 10^{-14}, \\
\text{CR}(\mu-e, T_i)&\simeq 1.3\times 10^{-14}. 
\end{align}

In turn, the light green band represents the experimental exclusion zone, where we can see that most of the values predicted by the model are below the experimental limit, including the benchmark point. From figure \ref{fig:BReee-CRAu} it can be seen that the model predicts values for the observables in the ranges: $5.57\times 10^{-16}\lesssim \text{BR}(\mu\rightarrow eee)\lesssim 1.84\times 10^{-15}$, $1.67\times 10^{-14}\lesssim \text{CR}(\mu-e, A_u)\lesssim 5.91\times 10^{-14}$ and $7.95\times 10^{-15}\lesssim \text{CR}(\mu-e, T_i)\lesssim 2.85\times 10^{-14}$.

\section{Conclusions}
\label{sec:conclusions}

We have proposed a minimal extension of the SM in which the lepton-number--violating submatrix $\mu$ of the inverse seesaw arises at \emph{two loops}. The model adds two singlet scalars $\rho,\sigma$ and two generations of neutral fermions $\nu_R, N_R, \Omega_{L,R}$, organized under a spontaneously broken global $U(1)_X$ and an exact $Z_3$. With the charge assignments in Table~\ref{tab:charge}, all required Yukawa interactions \eqref{eq:lag-neutrino} are allowed, while a tree-level $N$--$\Omega$ mixing and any tree-level (or one-loop) source of $\mu$ are forbidden. The preserved $Z_3$ also stabilizes the lightest $Z_3$-nontrivial state, providing either a scalar ($\rho$) or fermionic ($\Omega$) dark-matter candidate.

We derived the two-loop expression for $\mu$ and provided an analytic master representation in Appendix~A. The two-loop realization
naturally explains the smallness of $\mu$ without tiny Yukawas or superheavy mediators. The active-neutrino mass matrix then follows the standard inverse-seesaw form \eqref{eq:InvSeesaw-1}, 
yielding three light states and two pseudo-Dirac heavy pairs with a small intra-pair splitting $\propto \mu$. We showed that the observed neutrino oscillation data are successfully reproduced with a representative benchmark, predicting an effective Majorana neutrino mass parameter for neutrinoless double beta decay ($0\nu\beta\beta$) to be in the range $m_{ee}\in (2.1,\,4.4)\,\text{meV}$ for normal ordering.

On the dark-sector side, we analyzed both scalar and fermionic DM options. For scalar DM ($m_\rho<m_\Omega$), annihilation proceeds dominantly via the Higgs portal into $W^+W^-,\,ZZ,\,b\bar b$, with direct detection controlled by the same coupling. Current XENON1T/LZ limits already push $\lambda_{\phi\rho}\!\lesssim\!0.1$ across most of the $m_\rho\sim (1\!-\!3.5)\,\text{TeV}$ range, while future multi-ton detectors (XENONnT, DARWIN) will probe most of the remaining parameter space. Fermionic DM ($m_\Omega<m_\rho$) mainly annihilates through $t$-channel $\rho$ exchange into sterile neutrinos and is free of a tree-level Higgs portal; one-loop matching generates an effective Higgs coupling well below present bounds for $m_\Omega\sim(1\!-\!5)\,\text{TeV}$. On the other hand, this results in a significant suppression of the scattering cross section of our fermionic DM candidate off nuclei allowing reconciliation of the stringent direct DM detection constraints with the observable relic density.  

Mixings between active and exotic neutrinos leads to charged-lepton-flavor violation at testable levels. For a representative benchmark we find $\mathrm{BR}(\mu\!\to\!e\gamma)\simeq 1.6\times10^{-14}$ and correlated predictions for $\mu\!\to\! eee$ and $\mu$--$e$ conversion, within reach of next-generation experiments, while staying consistent with current bounds.

In summary, the same discrete symmetry that enforces the radiative origin of $\mu$ also stabilizes dark matter, linking neutrino masses, charged lepton flavor violation, and DM phenomenology. Three natural next steps are: (i) a 1-loop RGE study to assess vacuum stability and possible Landau poles up to a chosen UV scale; (ii) a systematic exploration of CP phases (including EDM constraints); and (iii) collider recasts targeting mono-$X$ and displaced signatures of the heavy-neutral sector. Taken together with imminent direct-detection improvements, these will decisively test the favored parameter space of this two-loop inverse-seesaw framework.

\section{Acknowledgments}
R.B.B. was supported by the USM Postgraduate Department, scholarship No. 037/2025 and ANID-Chile FONDECYT 1210378, 1261103. R.B.B acknowledges PUCV. J.M.G supported by the Universidad de Playa Ancha de Ciencias de la Educaci\'on under the Equipment Acquisition Program for R\&D\&I Units (Agreement UPA 24991). GB received support from ANID-Chile FONDECYT 1210378. The work of C.B. is supported by ANID-Chile under the grant FONDECYT Regular No. 1241855. AECH is supported by ANID-Chile FONDECYT 1241855, 1261103, ANID – Millennium Science Initiative Program $ICN2019\_044$, ANID CCTVal CIA250027 and ICTP through the Associates Programme (2026-2031). SK is supported by ANID-Chile FONDECYT 1230160, and ANID – Millennium Science Initiative Program $ICN2019\_044$. C. B. thanks UTFSM for its hospitality during a visit in which part of this work was completed. 
\appendix

\section{Two–loop integral for Inverse seesaw $\mu$-entry}
\label{App:2loop-1}

Here, we evaluate the two-loop integral 
\begin{eqnarray}
\label{eq:2loopInt-322}
    I(m_1, m_2) &=& \int \frac{d^4 p}{(2\pi)^4} \int \frac{d^4 q}{(2\pi)^4} \frac{1}{(p^2 - m_2^2)(p^2 - m_1^2)(q^2 - m_2^2)} \frac{1}{q^2 - m_1^2} \frac{1}{(p-q)^2 - m_2^2}
\end{eqnarray}
introduced in Eq.~\eqref{eq:mu-2loop-23}.

After the Wick rotation we get its Euclidean representation
\begin{eqnarray}
\label{eq:Idef-Euclead-1}
    I(m_1, m_2) &=& \int \frac{d^4 p}{(2\pi)^4} \int \frac{d^4 q}{(2\pi)^4} \frac{1}{(p^2 + m_2^2)(p^2 + m_1^2)(q^2 + m_2^2)} \frac{1}{q^2 + m_1^2} \frac{1}{(p+q)^2 + m_2^2}.
\end{eqnarray}
Using the identity
\begin{equation}
\Delta \equiv m_2^2-m_1^2,
\qquad
\frac{1}{(p^2+m_1^2)(p^2+m_2^2)}
=\frac{1}{\Delta}\!\left(\frac{1}{p^2+m_1^2}-\frac{1}{p^2+m_2^2}\right),
\end{equation}
and the analogous relation for the $q$–factor, we can split the double pole factors and reduce the two-loop integral \eqref{eq:Idef-Euclead-1} to a linear combination of single–pole ``sunset" integrals \cite{Martin2003BasisIntegrals,LaportaRemiddi2005EqualMass}:
\begin{equation}
\label{eq:I-reduction}
I(m_1,m_2)=\frac{1}{\Delta^{2}}\Big[
S(m_1,m_1,m_2)-2\,S(m_1,m_2,m_2)+S(m_2,m_2,m_2)
\Big],
\end{equation}
with the master formula
\begin{equation}
\label{eq:Sdef}
S(a,b,c)\equiv
\int\!\frac{d^4p}{(2\pi)^4}\!\int\!\frac{d^4q}{(2\pi)^4}\,
\frac{1}{(p^{2}+a^{2})(q^{2}+b^{2})\big((p+q)^{2}+c^{2}\big)}\, .
\end{equation}

A convenient Euclidean (Schwinger–parameter) representation is (see, for instance, a book \cite{Smirnov2006Book})
\begin{equation}
\label{eq:Schwinger}
S(a,b,c)=\frac{1}{(16\pi^{2})^{2}}
\int_{0}^{\infty}\!ds\,dt\,du\;
\frac{\exp\!\left[-a^{2}s-b^{2}t-c^{2}u\right]}{st+tu+us}\, ,
\end{equation}
which is manifestly finite for $a,b,c>0$  and obeys the scaling
\begin{eqnarray}
S(\lambda a,\lambda b,\lambda c)=\lambda^{2}S(a,b,c).
\end{eqnarray}
Then, we can rewrite the integral \eqref{eq:I-reduction} as
\begin{equation}
\label{eq:I-reduction-1}
I(m_1,m_2)=
\frac{1}{(16\pi^{2})^{2}}\frac{1}{m^{2}_1}
\mathcal{J}(x)
\end{equation}
where $x= m_2/m_1$ and  
\begin{eqnarray}
\label{eq:limit-1}
   \mathcal{J}(x)&=& \frac{(16\pi^{2})^{2}}{(1 - x^2)^{2}} \left[
S(1,1,x)-2\,S(1,x,x)+S(x,x,x)
\right] \leq \mbox{max}\, \mathcal{J}(x) = \mathcal{J}(1)\approx 1.56.
\end{eqnarray}
This function has no singularities and decays rapidly with increasing $x$ roughly as $x^{-4}$. For estimation, we  can use an approximation
\begin{eqnarray}
\label{eq:limit-124}
   \mathcal{J}(x)&\approx &  \frac{z}{x^4}.
\end{eqnarray}
The constant $z = \mathcal{J}(1)\approx 1.56$ was determined via numerical triple integration in  \eqref{eq:Schwinger}.

\bibliographystyle{utphys}
\bibliography{biblio2loop}

\providecommand{\href}[2]{#2}\begingroup\raggedright\begin{thebibliography}{100}

\bibitem{Minkowski:1977sc}
P.~Minkowski, ``{$\mu \to e\gamma$ at a Rate of One Out of $10^{9}$ Muon
  Decays?},'' \href{http://dx.doi.org/10.1016/0370-2693(77)90435-X}{{\em Phys.
  Lett. B} {\bfseries 67} (1977) 421--428}.

\bibitem{Yanagida:1979as}
T.~Yanagida, ``{Horizontal gauge symmetry and masses of neutrinos},'' {\em
  Conf. Proc. C} {\bfseries 7902131} (1979) 95--99.

\bibitem{Glashow:1979nm}
S.~L. Glashow, ``{The Future of Elementary Particle Physics},''
  \href{http://dx.doi.org/10.1007/978-1-4684-7197-7_15}{{\em NATO Sci. Ser. B}
  {\bfseries 61} (1980) 687}.

\bibitem{Mohapatra:1979ia}
R.~N. Mohapatra and G.~Senjanovic, ``{Neutrino Mass and Spontaneous Parity
  Nonconservation},'' \href{http://dx.doi.org/10.1103/PhysRevLett.44.912}{{\em
  Phys. Rev. Lett.} {\bfseries 44} (1980) 912}.

\bibitem{Gell-Mann:1979vob}
M.~Gell-Mann, P.~Ramond, and R.~Slansky, ``{Complex Spinors and Unified
  Theories},'' {\em Conf. Proc. C} {\bfseries 790927} (1979) 315--321,
  \href{http://arxiv.org/abs/1306.4669}{{\ttfamily arXiv:1306.4669 [hep-th]}}.

\bibitem{Schechter:1980gr}
J.~Schechter and J.~W.~F. Valle, ``{Neutrino Masses in SU(2) x U(1)
  Theories},'' \href{http://dx.doi.org/10.1103/PhysRevD.22.2227}{{\em Phys.
  Rev. D} {\bfseries 22} (1980) 2227}.

\bibitem{Schechter:1981cv}
J.~Schechter and J.~W.~F. Valle, ``{Neutrino Decay and Spontaneous Violation of
  Lepton Number},'' \href{http://dx.doi.org/10.1103/PhysRevD.25.774}{{\em Phys.
  Rev. D} {\bfseries 25} (1982) 774}.

\bibitem{Cai:2017jrq}
Y.~Cai, J.~Herrero-Garc{\'\i}a, M.~A. Schmidt, A.~Vicente, and R.~R. Volkas,
  ``{From the trees to the forest: a review of radiative neutrino mass
  models},'' \href{http://dx.doi.org/10.3389/fphy.2017.00063}{{\em Front. in
  Phys.} {\bfseries 5} (2017) 63},
  \href{http://arxiv.org/abs/1706.08524}{{\ttfamily arXiv:1706.08524
  [hep-ph]}}.

\bibitem{Jana:2019mgj}
S.~Jana, P.~K. Vishnu, and S.~Saad, ``{Minimal realizations of Dirac neutrino
  mass from generic one-loop and two-loop topologies at $d = 5$},''
  \href{http://dx.doi.org/10.1088/1475-7516/2020/04/018}{{\em JCAP} {\bfseries
  04} (2020) 018}, \href{http://arxiv.org/abs/1910.09537}{{\ttfamily
  arXiv:1910.09537 [hep-ph]}}.

\bibitem{Arbelaez:2022ejo}
C.~Arbel{\'a}ez, R.~Cepedello, J.~C. Helo, M.~Hirsch, and S.~Kovalenko, ``{How
  many 1-loop neutrino mass models are there?},''
  \href{http://dx.doi.org/10.1007/JHEP08(2022)023}{{\em JHEP} {\bfseries 08}
  (2022) 023}, \href{http://arxiv.org/abs/2205.13063}{{\ttfamily
  arXiv:2205.13063 [hep-ph]}}.

\bibitem{Tao:1996vb}
Z.-j. Tao, ``{Radiative seesaw mechanism at weak scale},''
  \href{http://dx.doi.org/10.1103/PhysRevD.54.5693}{{\em Phys. Rev. D}
  {\bfseries 54} (1996) 5693--5697},
  \href{http://arxiv.org/abs/hep-ph/9603309}{{\ttfamily arXiv:hep-ph/9603309}}.

\bibitem{Wyler:1982dd}
D.~Wyler and L.~Wolfenstein, ``{Massless Neutrinos in Left-Right Symmetric
  Models},'' \href{http://dx.doi.org/10.1016/0550-3213(83)90482-0}{{\em Nucl.
  Phys. B} {\bfseries 218} (1983) 205--214}.

\bibitem{Mohapatra:1986bd}
R.~N. Mohapatra and J.~W.~F. Valle, ``{Neutrino Mass and Baryon Number
  Nonconservation in Superstring Models},''
  \href{http://dx.doi.org/10.1103/PhysRevD.34.1642}{{\em Phys. Rev. D}
  {\bfseries 34} (1986) 1642}.

\bibitem{GonzalezGarcia:1988rw}
M.~C. Gonzalez-Garcia and J.~W.~F. Valle, ``{Fast Decaying Neutrinos and
  Observable Flavor Violation in a New Class of Majoron Models},''
  \href{http://dx.doi.org/10.1016/0370-2693(89)91131-3}{{\em Phys. Lett. B}
  {\bfseries 216} (1989) 360--366}.

\bibitem{Akhmedov:1995ip}
E.~K. Akhmedov, M.~Lindner, E.~Schnapka, and J.~W.~F. Valle, ``{Left-right
  symmetry breaking in NJL approach},''
  \href{http://dx.doi.org/10.1016/0370-2693(95)01504-3}{{\em Phys. Lett. B}
  {\bfseries 368} (1996) 270--280},
  \href{http://arxiv.org/abs/hep-ph/9507275}{{\ttfamily arXiv:hep-ph/9507275}}.

\bibitem{Akhmedov:1995vm}
E.~K. Akhmedov, M.~Lindner, E.~Schnapka, and J.~W.~F. Valle, ``{Dynamical
  left-right symmetry breaking},''
  \href{http://dx.doi.org/10.1103/PhysRevD.53.2752}{{\em Phys. Rev. D}
  {\bfseries 53} (1996) 2752--2780},
  \href{http://arxiv.org/abs/hep-ph/9509255}{{\ttfamily arXiv:hep-ph/9509255}}.

\bibitem{Malinsky:2005bi}
M.~Malinsky, J.~C. Romao, and J.~W.~F. Valle, ``{Novel supersymmetric SO(10)
  seesaw mechanism},''
  \href{http://dx.doi.org/10.1103/PhysRevLett.95.161801}{{\em Phys. Rev. Lett.}
  {\bfseries 95} (2005) 161801},
  \href{http://arxiv.org/abs/hep-ph/0506296}{{\ttfamily arXiv:hep-ph/0506296}}.

\bibitem{Ma:2009gu}
E.~Ma, ``{Radiative inverse seesaw mechanism for nonzero neutrino mass},''
  \href{http://dx.doi.org/10.1103/PhysRevD.80.013013}{{\em Phys. Rev. D}
  {\bfseries 80} (2009) 013013},
  \href{http://arxiv.org/abs/0904.4450}{{\ttfamily arXiv:0904.4450 [hep-ph]}}.

\bibitem{Malinsky:2009df}
M.~Malinsky, T.~Ohlsson, Z.-z. Xing, and H.~Zhang, ``{Non-unitary neutrino
  mixing and CP violation in the minimal inverse seesaw model},''
  \href{http://dx.doi.org/10.1016/j.physletb.2009.07.038}{{\em Phys. Lett. B}
  {\bfseries 679} (2009) 242--248},
  \href{http://arxiv.org/abs/0905.2889}{{\ttfamily arXiv:0905.2889 [hep-ph]}}.

\bibitem{Bazzocchi:2010dt}
F.~Bazzocchi, ``{Minimal Dynamical Inverse See Saw},''
  \href{http://dx.doi.org/10.1103/PhysRevD.83.093009}{{\em Phys. Rev. D}
  {\bfseries 83} (2011) 093009},
  \href{http://arxiv.org/abs/1011.6299}{{\ttfamily arXiv:1011.6299 [hep-ph]}}.

\bibitem{Law:2012mj}
S.~S.~C. Law and K.~L. McDonald, ``{Inverse seesaw and dark matter in models
  with exotic lepton triplets},''
  \href{http://dx.doi.org/10.1016/j.physletb.2012.06.044}{{\em Phys. Lett. B}
  {\bfseries 713} (2012) 490--494},
  \href{http://arxiv.org/abs/1204.2529}{{\ttfamily arXiv:1204.2529 [hep-ph]}}.

\bibitem{Das:2012ze}
A.~Das and N.~Okada, ``{Inverse seesaw neutrino signatures at the LHC and
  ILC},'' \href{http://dx.doi.org/10.1103/PhysRevD.88.113001}{{\em Phys. Rev.
  D} {\bfseries 88} (2013) 113001},
  \href{http://arxiv.org/abs/1207.3734}{{\ttfamily arXiv:1207.3734 [hep-ph]}}.

\bibitem{Abada:2014vea}
A.~Abada and M.~Lucente, ``{Looking for the minimal inverse seesaw
  realisation},'' \href{http://dx.doi.org/10.1016/j.nuclphysb.2014.06.003}{{\em
  Nucl. Phys. B} {\bfseries 885} (2014) 651--678},
  \href{http://arxiv.org/abs/1401.1507}{{\ttfamily arXiv:1401.1507 [hep-ph]}}.

\bibitem{Fraser:2014yha}
S.~Fraser, E.~Ma, and O.~Popov, ``{Scotogenic Inverse Seesaw Model of Neutrino
  Mass},'' \href{http://dx.doi.org/10.1016/j.physletb.2014.08.069}{{\em Phys.
  Lett. B} {\bfseries 737} (2014) 280--282},
  \href{http://arxiv.org/abs/1408.4785}{{\ttfamily arXiv:1408.4785 [hep-ph]}}.

\bibitem{Ahriche:2016acx}
A.~Ahriche, S.~M. Boucenna, and S.~Nasri, ``{Dark Radiative Inverse Seesaw
  Mechanism},'' \href{http://dx.doi.org/10.1103/PhysRevD.93.075036}{{\em Phys.
  Rev. D} {\bfseries 93} no.~7, (2016) 075036},
  \href{http://arxiv.org/abs/1601.04336}{{\ttfamily arXiv:1601.04336
  [hep-ph]}}.

\bibitem{CarcamoHernandez:2013krw}
A.~E. C{\'a}rcamo~Hern{\'a}ndez, R.~Martinez, and F.~Ochoa, ``{Fermion masses
  and mixings in the 3-3-1 model with right-handed neutrinos based on the $S_3$
  flavor symmetry},''
  \href{http://dx.doi.org/10.1140/epjc/s10052-016-4480-3}{{\em Eur. Phys. J. C}
  {\bfseries 76} no.~11, (2016) 634},
  \href{http://arxiv.org/abs/1309.6567}{{\ttfamily arXiv:1309.6567 [hep-ph]}}.

\bibitem{Das:2017ski}
A.~Das, T.~Nomura, H.~Okada, and S.~Roy, ``{Generation of a radiative neutrino
  mass in the linear seesaw framework, charged lepton flavor violation, and
  dark matter},'' \href{http://dx.doi.org/10.1103/PhysRevD.96.075001}{{\em
  Phys. Rev. D} {\bfseries 96} no.~7, (2017) 075001},
  \href{http://arxiv.org/abs/1704.02078}{{\ttfamily arXiv:1704.02078
  [hep-ph]}}.

\bibitem{CarcamoHernandez:2017owh}
A.~E. C{\'a}rcamo~Hern{\'a}ndez, S.~Kovalenko, J.~W.~F. Valle, and C.~A.
  Vaquera-Araujo, ``{Predictive Pati-Salam theory of fermion masses and
  mixing},'' \href{http://dx.doi.org/10.1007/JHEP07(2017)118}{{\em JHEP}
  {\bfseries 07} (2017) 118}, \href{http://arxiv.org/abs/1705.06320}{{\ttfamily
  arXiv:1705.06320 [hep-ph]}}.

\bibitem{CarcamoHernandez:2018hst}
A.~E. C{\'a}rcamo~Hern{\'a}ndez, S.~Kovalenko, J.~W.~F. Valle, and C.~A.
  Vaquera-Araujo, ``{Neutrino predictions from a left-right symmetric flavored
  extension of the standard model},''
  \href{http://dx.doi.org/10.1007/JHEP02(2019)065}{{\em JHEP} {\bfseries 02}
  (2019) 065}, \href{http://arxiv.org/abs/1811.03018}{{\ttfamily
  arXiv:1811.03018 [hep-ph]}}.

\bibitem{CarcamoHernandez:2018iel}
A.~E. C{\'a}rcamo~Hern{\'a}ndez, H.~N. Long, and V.~V. Vien, ``{The first
  $\Delta(27)$ flavor 3-3-1 model with low scale seesaw mechanism},''
  \href{http://dx.doi.org/10.1140/epjc/s10052-018-6284-0}{{\em Eur. Phys. J. C}
  {\bfseries 78} no.~10, (2018) 804},
  \href{http://arxiv.org/abs/1803.01636}{{\ttfamily arXiv:1803.01636
  [hep-ph]}}.

\bibitem{Bertuzzo:2018ftf}
E.~Bertuzzo, S.~Jana, P.~A.~N. Machado, and R.~Zukanovich~Funchal, ``{Neutrino
  Masses and Mixings Dynamically Generated by a Light Dark Sector},''
  \href{http://dx.doi.org/10.1016/j.physletb.2019.02.023}{{\em Phys. Lett. B}
  {\bfseries 791} (2019) 210--214},
  \href{http://arxiv.org/abs/1808.02500}{{\ttfamily arXiv:1808.02500
  [hep-ph]}}.

\bibitem{Mandal:2019oth}
S.~Mandal, N.~Rojas, R.~Srivastava, and J.~W.~F. Valle, ``{Dark matter as the
  origin of neutrino mass in the inverse seesaw mechanism},''
  \href{http://dx.doi.org/10.1016/j.physletb.2021.136609}{{\em Phys. Lett. B}
  {\bfseries 821} (2021) 136609},
  \href{http://arxiv.org/abs/1907.07728}{{\ttfamily arXiv:1907.07728
  [hep-ph]}}.

\bibitem{Das:2019pua}
A.~Das, S.~Goswami, K.~N. Vishnudath, and T.~Nomura, ``{Constraining a general
  U(1)$^\prime$ inverse seesaw model from vacuum stability, dark matter and
  collider},'' \href{http://dx.doi.org/10.1103/PhysRevD.101.055026}{{\em Phys.
  Rev. D} {\bfseries 101} no.~5, (2020) 055026},
  \href{http://arxiv.org/abs/1905.00201}{{\ttfamily arXiv:1905.00201
  [hep-ph]}}.

\bibitem{CarcamoHernandez:2019eme}
A.~E. C{\'a}rcamo~Hern{\'a}ndez and S.~F. King, ``{Littlest Inverse Seesaw
  Model},'' \href{http://dx.doi.org/10.1016/j.nuclphysb.2020.114950}{{\em Nucl.
  Phys. B} {\bfseries 953} (2020) 114950},
  \href{http://arxiv.org/abs/1903.02565}{{\ttfamily arXiv:1903.02565
  [hep-ph]}}.

\bibitem{CarcamoHernandez:2019pmy}
A.~E. C{\'a}rcamo~Hern{\'a}ndez, J.~Marchant~Gonz{\'a}lez, and U.~J.
  Salda{\~n}a-Salazar, ``{Viable low-scale model with universal and inverse
  seesaw mechanisms},''
  \href{http://dx.doi.org/10.1103/PhysRevD.100.035024}{{\em Phys. Rev. D}
  {\bfseries 100} no.~3, (2019) 035024},
  \href{http://arxiv.org/abs/1904.09993}{{\ttfamily arXiv:1904.09993
  [hep-ph]}}.

\bibitem{CarcamoHernandez:2019vih}
A.~E. C{\'a}rcamo~Hern{\'a}ndez, Y.~Hidalgo~Vel{\'a}squez, and N.~A.
  P{\'e}rez-Julve, ``{A 3-3-1 model with low scale seesaw mechanisms},''
  \href{http://dx.doi.org/10.1140/epjc/s10052-019-7325-z}{{\em Eur. Phys. J. C}
  {\bfseries 79} no.~10, (2019) 828},
  \href{http://arxiv.org/abs/1905.02323}{{\ttfamily arXiv:1905.02323
  [hep-ph]}}.

\bibitem{CarcamoHernandez:2019lhv}
A.~E. C{\'a}rcamo~Hern{\'a}ndez, D.~T. Huong, and H.~N. Long, ``{Minimal model
  for the fermion flavor structure, mass hierarchy, dark matter, leptogenesis,
  and the electron and muon anomalous magnetic moments},''
  \href{http://dx.doi.org/10.1103/PhysRevD.102.055002}{{\em Phys. Rev. D}
  {\bfseries 102} no.~5, (2020) 055002},
  \href{http://arxiv.org/abs/1910.12877}{{\ttfamily arXiv:1910.12877
  [hep-ph]}}.

\bibitem{Hernandez:2021uxx}
A.~E.~C. Hern{\'a}ndez and I.~Schmidt, ``{A renormalizable left-right symmetric
  model with low scale seesaw mechanisms},''
  \href{http://dx.doi.org/10.1016/j.nuclphysb.2022.115696}{{\em Nucl. Phys. B}
  {\bfseries 976} (2022) 115696},
  \href{http://arxiv.org/abs/2101.02718}{{\ttfamily arXiv:2101.02718
  [hep-ph]}}.

\bibitem{Hernandez:2021xet}
A.~E.~C. Hern{\'a}ndez, D.~T. Huong, and I.~Schmidt, ``{Universal inverse
  seesaw mechanism as a source of the SM fermion mass hierarchy},''
  \href{http://dx.doi.org/10.1140/epjc/s10052-022-10011-x}{{\em Eur. Phys. J.
  C} {\bfseries 82} no.~1, (2022) 63},
  \href{http://arxiv.org/abs/2109.12118}{{\ttfamily arXiv:2109.12118
  [hep-ph]}}.

\bibitem{Hernandez:2021kju}
A.~E.~C. Hern{\'a}ndez, C.~Espinoza, J.~C. G{\'o}mez-Izquierdo, and
  M.~Mondrag{\'o}n, ``{Fermion masses and mixings, dark matter, leptogenesis
  and $g-2$ muon anomaly in an extended 2HDM with inverse seesaw},''
  \href{http://dx.doi.org/10.1140/epjp/s13360-022-03432-w}{{\em Eur. Phys. J.
  Plus} {\bfseries 137} no.~11, (2022) 1224},
  \href{http://arxiv.org/abs/2104.02730}{{\ttfamily arXiv:2104.02730
  [hep-ph]}}.

\bibitem{Nomura:2021adf}
T.~Nomura, H.~Okada, and P.~Sanyal, ``{A radiatively induced inverse seesaw
  model with hidden U(1) gauge symmetry},''
  \href{http://dx.doi.org/10.1140/epjc/s10052-022-10662-w}{{\em Eur. Phys. J.
  C} {\bfseries 82} no.~8, (2022) 697},
  \href{http://arxiv.org/abs/2103.09494}{{\ttfamily arXiv:2103.09494
  [hep-ph]}}.

\bibitem{Hernandez:2021mxo}
A.~E.~C. Hern{\'a}ndez, H.~N. Long, M.~L. Mora-Urrutia, N.~H. Thao, and V.~V.
  Vien, ``{Fermion masses and mixings and $g-2$ muon anomaly in a 3-3-1 model
  with $D_4$ family symmetry},''
  \href{http://dx.doi.org/10.1140/epjc/s10052-022-10639-9}{{\em Eur. Phys. J.
  C} {\bfseries 82} no.~8, (2022) 769},
  \href{http://arxiv.org/abs/2104.04559}{{\ttfamily arXiv:2104.04559
  [hep-ph]}}.

\bibitem{Abada:2021yot}
A.~Abada, N.~Bernal, A.~E.~C. Hern{\'a}ndez, X.~Marcano, and G.~Piazza,
  ``{Gauged inverse seesaw from dark matter},''
  \href{http://dx.doi.org/10.1140/epjc/s10052-021-09535-5}{{\em Eur. Phys. J.
  C} {\bfseries 81} no.~8, (2021) 758},
  \href{http://arxiv.org/abs/2107.02803}{{\ttfamily arXiv:2107.02803
  [hep-ph]}}.

\bibitem{Abada:2023zbb}
A.~Abada, N.~Bernal, A.~E. C{\'a}rcamo~Hern{\'a}ndez, S.~Kovalenko, and T.~B.
  de~Melo, ``{Three-loop inverse scotogenic seesaw models},''
  \href{http://dx.doi.org/10.1007/JHEP05(2024)035}{{\em JHEP} {\bfseries 05}
  (2024) 035}, \href{http://arxiv.org/abs/2312.14105}{{\ttfamily
  arXiv:2312.14105 [hep-ph]}}.

\bibitem{Bonilla:2023egs}
C.~Bonilla, A.~E. Carcamo~Hernandez, B.~Saez~D{\i}az, S.~Kovalenko, and
  J.~Marchant~Gonzalez, ``{Dark matter from a radiative inverse seesaw majoron
  model},'' \href{http://dx.doi.org/10.1016/j.physletb.2023.138282}{{\em Phys.
  Lett. B} {\bfseries 847} (2023) 138282},
  \href{http://arxiv.org/abs/2306.08453}{{\ttfamily arXiv:2306.08453
  [hep-ph]}}.

\bibitem{Bonilla:2023wok}
C.~Bonilla, A.~E. Carcamo~Hernandez, S.~Kovalenko, H.~Lee, R.~Pasechnik, and
  I.~Schmidt, ``{Fermion mass hierarchy in an extended left-right symmetric
  model},'' \href{http://dx.doi.org/10.1007/JHEP12(2023)075}{{\em JHEP}
  {\bfseries 12} (2023) 075}, \href{http://arxiv.org/abs/2305.11967}{{\ttfamily
  arXiv:2305.11967 [hep-ph]}}.

\bibitem{Binh:2024lez}
V.~H. Binh, C.~Bonilla, A.~E. C{\'a}rcamo~Hern{\'a}ndez, D.~T. Huong, V.~K.~N.,
  H.~N. Long, P.~N. Thu, and I.~Schmidt, ``{Phenomenology of 3-3-1 models with
  a radiative inverse seesaw mechanism},''
  \href{http://dx.doi.org/10.1103/PhysRevD.110.075022}{{\em Phys. Rev. D}
  {\bfseries 110} no.~7, (2024) 075022},
  \href{http://arxiv.org/abs/2404.13373}{{\ttfamily arXiv:2404.13373
  [hep-ph]}}.

\bibitem{CarcamoHernandez:2024hll}
A.~E. C{\'a}rcamo~Hern{\'a}ndez, D.~T. Huong, H.~N. Long, and
  D.~Salinas-Arizmendi, ``{Fermion masses and mixings and charged lepton flavor
  violation in a 3-3-1 model with inverse seesaw},''
  \href{http://dx.doi.org/10.1093/ptep/ptaf067}{{\em PTEP} {\bfseries 6} (2025)
  063}, \href{http://arxiv.org/abs/2412.18550}{{\ttfamily arXiv:2412.18550
  [hep-ph]}}.

\bibitem{Huong:2025uwx}
D.~T. Huong, A.~E. C{\'a}rcamo~Hern{\'a}ndez, H.~T. Hung, T.~T. Hieu, and N.~A.
  P{\'e}rez-Julve, ``{Extended IDM theory with low scale seesaw mechanisms},''
  \href{http://arxiv.org/abs/2502.19488}{{\ttfamily arXiv:2502.19488
  [hep-ph]}}.

\bibitem{Pilaftsis:1997jf}
A.~Pilaftsis, ``{CP violation and baryogenesis due to heavy Majorana
  neutrinos},'' \href{http://dx.doi.org/10.1103/PhysRevD.56.5431}{{\em Phys.
  Rev. D} {\bfseries 56} (1997) 5431--5451},
  \href{http://arxiv.org/abs/hep-ph/9707235}{{\ttfamily arXiv:hep-ph/9707235}}.

\bibitem{Gu:2010xc}
P.-H. Gu and U.~Sarkar, ``{Leptogenesis with Linear, Inverse or Double
  Seesaw},'' \href{http://dx.doi.org/10.1016/j.physletb.2010.09.062}{{\em Phys.
  Lett. B} {\bfseries 694} (2011) 226--232},
  \href{http://arxiv.org/abs/1007.2323}{{\ttfamily arXiv:1007.2323 [hep-ph]}}.

\bibitem{Dolan:2018qpy}
M.~J. Dolan, T.~P. Dutka, and R.~R. Volkas, ``{Dirac-Phase Thermal Leptogenesis
  in the extended Type-I Seesaw Model},''
  \href{http://dx.doi.org/10.1088/1475-7516/2018/06/012}{{\em JCAP} {\bfseries
  06} (2018) 012}, \href{http://arxiv.org/abs/1802.08373}{{\ttfamily
  arXiv:1802.08373 [hep-ph]}}.

\bibitem{Dib:2019jod}
C.~Dib, S.~Kovalenko, I.~Schmidt, and A.~Smetana, ``{Low-scale seesaw from
  neutrino condensation},''
  \href{http://dx.doi.org/10.1016/j.nuclphysb.2019.114910}{{\em Nucl. Phys. B}
  {\bfseries 952} (2020) 114910},
  \href{http://arxiv.org/abs/1904.06280}{{\ttfamily arXiv:1904.06280
  [hep-ph]}}.

\bibitem{Blanchet:2009kk}
S.~Blanchet, T.~Hambye, and F.-X. Josse-Michaux, ``{Reconciling leptogenesis
  with observable mu ---{\ensuremath{>}} e gamma rates},''
  \href{http://dx.doi.org/10.1007/JHEP04(2010)023}{{\em JHEP} {\bfseries 04}
  (2010) 023}, \href{http://arxiv.org/abs/0912.3153}{{\ttfamily arXiv:0912.3153
  [hep-ph]}}.

\bibitem{Blanchet:2010kw}
S.~Blanchet, P.~S.~B. Dev, and R.~N. Mohapatra, ``{Leptogenesis with TeV Scale
  Inverse Seesaw in SO(10)},''
  \href{http://dx.doi.org/10.1103/PhysRevD.82.115025}{{\em Phys. Rev. D}
  {\bfseries 82} (2010) 115025},
  \href{http://arxiv.org/abs/1010.1471}{{\ttfamily arXiv:1010.1471 [hep-ph]}}.

\bibitem{Hirsch:2004higgsMajoron}
M.~Hirsch, J.~C. Romao, J.~W.~F. Valle, and A.~Villanova~del Moral, ``{Higgs
  boson decays into majorons},''
  \href{http://arxiv.org/abs/hep-ph/0407269}{{\ttfamily arXiv:hep-ph/0407269
  [hep-ph]}}.

\bibitem{Bonilla:2015uwa}
C.~Bonilla, J.~W.~F. Valle, and J.~C. Rom{\~a}o, ``{Neutrino mass and invisible
  Higgs decays at the LHC},''
  \href{http://dx.doi.org/10.1103/PhysRevD.91.113015}{{\em Phys. Rev. D}
  {\bfseries 91} no.~11, (2015) 113015},
  \href{http://arxiv.org/abs/1502.01649}{{\ttfamily arXiv:1502.01649
  [hep-ph]}}.

\bibitem{PDG2024}
{Particle Data Group}, ``{Review of Particle Physics},'' {\em Phys. Rev. D}
  {\bfseries 110} (2024) 030001.

\bibitem{ATLAS:2023invH}
{ATLAS Collaboration}, ``{Combination of searches for invisible decays of the
  Higgs boson using 139 fb$^{-1}$ of proton-proton collision data at
  $\sqrt{s}=13$ TeV collected with the ATLAS experiment},''
  \href{http://arxiv.org/abs/2301.10731}{{\ttfamily arXiv:2301.10731
  [hep-ex]}}.

\bibitem{PDGBBN2025}
{Particle Data Group}, ``{Big Bang Nucleosynthesis (PDG mini-review, 2025
  update)}.''
  \url{https://ccwww.kek.jp/pdg/2025/reviews/rpp2025-rev-bbang-nucleosynthesis.pdf},
  2025.
\newblock Entropy-conservation temperature-ratio arguments for decoupled
  species in the BSM discussion.

\bibitem{deSalas:2016Neff}
P.~F. de~Salas and S.~Pastor, ``{Relic neutrino decoupling with flavour
  oscillations revisited},'' \href{http://arxiv.org/abs/1606.06986}{{\ttfamily
  arXiv:1606.06986 [hep-ph]}}.

\bibitem{PDGNeutrinosCosmology2025}
{Particle Data Group}, ``{Neutrinos in Cosmology (PDG mini-review, 2025
  update)}.''
  \url{https://ccwww.kek.jp/pdg/2025/reviews/rpp2025-rev-neutrinos-in-cosmology.pdf},
  2025.
\newblock See Eq. (26.1) for the definition of $N_{\rm eff}$ and discussion of
  the SM value.

\bibitem{Husdal:2016gstar}
L.~Husdal, ``{On Effective Degrees of Freedom in the Early Universe},''
  \href{http://arxiv.org/abs/1609.04979}{{\ttfamily arXiv:1609.04979
  [hep-ph]}}.

\bibitem{Drees:2015qcdEoS}
M.~Drees, F.~Hajkarim, and E.~R. Schmitz, ``{The Effects of QCD Equation of
  State on the Relic Density of WIMP Dark Matter},''
  \href{http://arxiv.org/abs/1503.03513}{{\ttfamily arXiv:1503.03513
  [hep-ph]}}.

\bibitem{Planck:2018params}
{Planck Collaboration}, ``{Planck 2018 results. VI. Cosmological parameters},''
  \href{http://arxiv.org/abs/1807.06209}{{\ttfamily arXiv:1807.06209
  [astro-ph.CO]}}.

\bibitem{Hall:2009freezein}
L.~J. Hall, K.~Jedamzik, J.~March-Russell, and S.~M. West, ``{Freeze-In
  Production of FIMP Dark Matter},''
  \href{http://arxiv.org/abs/0911.1120}{{\ttfamily arXiv:0911.1120 [hep-ph]}}.
  Published as JHEP 03 (2010) 080.

\bibitem{BHATTACHARYYA_2016}
G.~BHATTACHARYYA and D.~DAS, ``Scalar sector of two-higgs-doublet models: A
  minireview,'' \href{http://dx.doi.org/10.1007/s12043-016-1252-4}{{\em
  Pramana} {\bfseries 87} no.~3, (Aug., 2016) }.
  \url{http://dx.doi.org/10.1007/s12043-016-1252-4}.

\bibitem{Maniatis_2006}
M.~Maniatis, A.~von Manteuffel, O.~Nachtmann, and F.~Nagel, ``Stability and
  symmetry breaking in the general two-higgs-doublet model,''
  \href{http://dx.doi.org/10.1140/epjc/s10052-006-0016-6}{{\em The European
  Physical Journal C} {\bfseries 48} no.~3, (Oct., 2006) 805–823}.
  \url{http://dx.doi.org/10.1140/epjc/s10052-006-0016-6}.

\bibitem{Catano:2012kw}
M.~E. Catano, R.~Martinez, and F.~Ochoa, ``{Neutrino masses in a 331 model with
  right-handed neutrinos without doubly charged Higgs bosons via inverse and
  double seesaw mechanisms},''
  \href{http://dx.doi.org/10.1103/PhysRevD.86.073015}{{\em Phys. Rev. D}
  {\bfseries 86} (2012) 073015},
  \href{http://arxiv.org/abs/1206.1966}{{\ttfamily arXiv:1206.1966 [hep-ph]}}.

\bibitem{deSalas:2020pgw}
P.~F. de~Salas, D.~V. Forero, S.~Gariazzo, P.~Mart{\'\i}nez-Mirav{\'e},
  O.~Mena, C.~A. Ternes, M.~T{\'o}rtola, and J.~W.~F. Valle, ``{2020 global
  reassessment of the neutrino oscillation picture},''
  \href{http://dx.doi.org/10.1007/JHEP02(2021)071}{{\em JHEP} {\bfseries 02}
  (2021) 071}, \href{http://arxiv.org/abs/2006.11237}{{\ttfamily
  arXiv:2006.11237 [hep-ph]}}.

\bibitem{Esteban:2024eli}
I.~Esteban, M.~C. Gonzalez-Garcia, M.~Maltoni, I.~Martinez-Soler, J.~P.
  Pinheiro, and T.~Schwetz, ``{NuFit-6.0: updated global analysis of
  three-flavor neutrino oscillations},''
  \href{http://dx.doi.org/10.1007/JHEP12(2024)216}{{\em JHEP} {\bfseries 12}
  (2024) 216}, \href{http://arxiv.org/abs/2410.05380}{{\ttfamily
  arXiv:2410.05380 [hep-ph]}}.

\bibitem{KamLAND-Zen:2024eml}
{\bfseries KamLAND-Zen} Collaboration, S.~Abe {\em et~al.}, ``{Search for
  Majorana Neutrinos with the Complete KamLAND-Zen Dataset},''
  \href{http://arxiv.org/abs/2406.11438}{{\ttfamily arXiv:2406.11438
  [hep-ex]}}.

\bibitem{Planck:2018vyg}
{\bfseries Planck} Collaboration, N.~Aghanim {\em et~al.}, ``{Planck 2018
  results. VI. Cosmological parameters},''
  \href{http://dx.doi.org/10.1051/0004-6361/201833910}{{\em Astron. Astrophys.}
  {\bfseries 641} (2020) A6}, \href{http://arxiv.org/abs/1807.06209}{{\ttfamily
  arXiv:1807.06209 [astro-ph.CO]}}. [Erratum: Astron.Astrophys. 652, C4
  (2021)].

\bibitem{XENON1T:2018voc}
{\bfseries XENON} Collaboration, E.~Aprile {\em et~al.}, ``{Dark Matter Search
  Results from a One Ton-Year Exposure of XENON1T},''
  \href{http://dx.doi.org/10.1103/PhysRevLett.121.111302}{{\em Phys. Rev.
  Lett.} {\bfseries 121} no.~11, (2018) 111302},
  \href{http://arxiv.org/abs/1805.12562}{{\ttfamily arXiv:1805.12562
  [astro-ph.CO]}}.

\bibitem{XENONnT:2020kmp}
{\bfseries XENON} Collaboration, E.~Aprile {\em et~al.}, ``{Projected WIMP
  sensitivity of the XENONnT dark matter experiment},''
  \href{http://dx.doi.org/10.1088/1475-7516/2020/11/031}{{\em JCAP} {\bfseries
  11} (2020) 031}, \href{http://arxiv.org/abs/2007.08796}{{\ttfamily
  arXiv:2007.08796 [physics.ins-det]}}.

\bibitem{LZ:2024zvo}
{\bfseries LZ} Collaboration, J.~Aalbers {\em et~al.}, ``{Dark Matter Search
  Results from 4.2 Tonne-Years of Exposure of the LUX-ZEPLIN (LZ)
  Experiment},'' \href{http://arxiv.org/abs/2410.17036}{{\ttfamily
  arXiv:2410.17036 [hep-ex]}}.

\bibitem{DARWIN:2016hyl}
{\bfseries DARWIN} Collaboration, J.~Aalbers {\em et~al.}, ``{DARWIN: towards
  the ultimate dark matter detector},''
  \href{http://dx.doi.org/10.1088/1475-7516/2016/11/017}{{\em JCAP} {\bfseries
  11} (2016) 017}, \href{http://arxiv.org/abs/1606.07001}{{\ttfamily
  arXiv:1606.07001 [astro-ph.IM]}}.

\bibitem{Bernal:2017xat}
N.~Bernal, A.~E. C{\'a}rcamo~Hern{\'a}ndez, I.~de~Medeiros~Varzielas, and
  S.~Kovalenko, ``{Fermion masses and mixings and dark matter constraints in a
  model with radiative seesaw mechanism},''
  \href{http://dx.doi.org/10.1007/JHEP05(2018)053}{{\em JHEP} {\bfseries 05}
  (2018) 053}, \href{http://arxiv.org/abs/1712.02792}{{\ttfamily
  arXiv:1712.02792 [hep-ph]}}.

\bibitem{Farina_2010}
M.~Farina, D.~Pappadopulo, and A.~Strumia, ``Cdms stands for constrained dark
  matter singlet,''
  \href{http://dx.doi.org/10.1016/j.physletb.2010.04.025}{{\em Physics Letters
  B} {\bfseries 688} no.~4–5, (May, 2010) 329–331}.
  \url{http://dx.doi.org/10.1016/j.physletb.2010.04.025}.

\bibitem{Giedt_2009}
J.~Giedt, A.~W. Thomas, and R.~D. Young, ``Dark matter, constrained minimal
  supersymmetric standard model, and lattice qcd,''
  \href{http://dx.doi.org/10.1103/physrevlett.103.201802}{{\em Physical Review
  Letters} {\bfseries 103} no.~20, (Nov., 2009) }.
  \url{http://dx.doi.org/10.1103/PhysRevLett.103.201802}.

\bibitem{XENON:2018voc}
{\bfseries XENON} Collaboration, E.~Aprile {\em et~al.}, ``{Dark Matter Search
  Results from a One Tonne$\times$Year Exposure of XENON1T},''
  \href{http://dx.doi.org/10.1103/PhysRevLett.121.111302}{{\em Phys. Rev.
  Lett.} {\bfseries 121} no.~11, (2018) 111302},
  \href{http://arxiv.org/abs/1805.12562}{{\ttfamily arXiv:1805.12562
  [hep-ex]}}.

\bibitem{XENON:2023cxc}
{\bfseries XENON} Collaboration, E.~Aprile {\em et~al.}, ``{First Dark Matter
  Search with Nuclear Recoils from the XENONnT Experiment},''
  \href{http://dx.doi.org/10.1103/PhysRevLett.131.041003}{{\em Phys. Rev.
  Lett.} {\bfseries 131} no.~4, (2023) 041003},
  \href{http://arxiv.org/abs/2303.14729}{{\ttfamily arXiv:2303.14729
  [hep-ex]}}.

\bibitem{LZ:2020sensitivity}
{\bfseries LZ} Collaboration, D.~S. Akerib {\em et~al.}, ``Projected wimp
  sensitivity of the lux-zeplin (lz) dark matter experiment,''
  \href{http://dx.doi.org/10.1103/PhysRevD.101.052002}{{\em Phys. Rev. D}
  {\bfseries 101} (2020) 052002},
  \href{http://arxiv.org/abs/1802.06039}{{\ttfamily arXiv:1802.06039}}.

\bibitem{DARWIN:2016whitepaper}
{\bfseries DARWIN} Collaboration, J.~Aalbers {\em et~al.}, ``Darwin: towards
  the ultimate dark matter detector,''
  \href{http://dx.doi.org/10.1088/1475-7516/2016/11/017}{{\em JCAP} {\bfseries
  2016} no.~11, (2016) 017}, \href{http://arxiv.org/abs/1606.07001}{{\ttfamily
  arXiv:1606.07001}}.

\bibitem{Cirelli:2024ssz}
M.~Cirelli, A.~Strumia, and J.~Zupan, ``{Dark Matter},''
  \href{http://arxiv.org/abs/2406.01705}{{\ttfamily arXiv:2406.01705
  [hep-ph]}}.

\bibitem{Goyal:2016zeh}
A.~Goyal and M.~Kumar, ``{Fermionic Dark Matter in a simple $t$-channel
  model},'' \href{http://dx.doi.org/10.1088/1475-7516/2016/11/001}{{\em JCAP}
  {\bfseries 11} (2016) 001}, \href{http://arxiv.org/abs/1609.03364}{{\ttfamily
  arXiv:1609.03364 [hep-ph]}}.

\bibitem{Pathak:2024sei}
G.~Pathak, P.~Das, and M.~K. Das, ``{Neutrino mass genesis in scoto-inverse
  seesaw with modular $A_4$},''
  \href{http://dx.doi.org/10.1140/epjc/s10052-025-14263-1}{{\em Eur. Phys. J.
  C} {\bfseries 85} no.~5, (2025) 569},
  \href{http://arxiv.org/abs/2411.13895}{{\ttfamily arXiv:2411.13895
  [hep-ph]}}.

\bibitem{Amiri:2025ras}
A.~Amiri, B.~Diaz~Saez, and K.~M{\"o}hling, ``{Freeze-in at Low Reheating and
  Direct Detection of Fermion Dark Matter},''
  \href{http://arxiv.org/abs/2511.21520}{{\ttfamily arXiv:2511.21520
  [hep-ph]}}.

\bibitem{fnGiedt:2009mr}
J.~Giedt, A.~W. Thomas, and R.~D. Young, ``{Dark matter, the CMSSM and lattice
  QCD},'' \href{http://dx.doi.org/10.1103/PhysRevLett.103.201802}{{\em Phys.
  Rev. Lett.} {\bfseries 103} (2009) 201802},
  \href{http://arxiv.org/abs/0907.4177}{{\ttfamily arXiv:0907.4177 [hep-ph]}}.

\bibitem{Langacker:1988up}
P.~Langacker and D.~London, ``{Lepton Number Violation and Massless
  Nonorthogonal Neutrinos},''
  \href{http://dx.doi.org/10.1103/PhysRevD.38.907}{{\em Phys. Rev. D}
  {\bfseries 38} (1988) 907}.

\bibitem{Lavoura:2003xp}
L.~Lavoura, ``{General formulae for f(1) ---\ensuremath{>} f(2) gamma},''
  \href{http://dx.doi.org/10.1140/epjc/s2003-01212-7}{{\em Eur. Phys. J. C}
  {\bfseries 29} (2003) 191--195},
  \href{http://arxiv.org/abs/hep-ph/0302221}{{\ttfamily arXiv:hep-ph/0302221}}.

\bibitem{MEGII:2025gzr}
{\bfseries MEG II} Collaboration, K.~Afanaciev {\em et~al.}, ``{New limit on
  the {\ensuremath{\mu}}+-{\ensuremath{>}}e+{\ensuremath{\gamma}}decay with the
  MEG II experiment},'' \href{http://arxiv.org/abs/2504.15711}{{\ttfamily
  arXiv:2504.15711 [hep-ex]}}.

\bibitem{BaBar:2009hkt}
{\bfseries BaBar} Collaboration, B.~Aubert {\em et~al.}, ``{Searches for Lepton
  Flavor Violation in the Decays $\tau^\pm \to e^\pm \gamma$ and $\tau^\pm \to
  \mu^\pm \gamma$},''
  \href{http://dx.doi.org/10.1103/PhysRevLett.104.021802}{{\em Phys. Rev.
  Lett.} {\bfseries 104} (2010) 021802},
  \href{http://arxiv.org/abs/0908.2381}{{\ttfamily arXiv:0908.2381 [hep-ex]}}.

\bibitem{Belle:2021ysv}
{\bfseries Belle} Collaboration, A.~Abdesselam {\em et~al.}, ``{Search for
  lepton-flavor-violating tau-lepton decays to $\ell\gamma$ at Belle},''
  \href{http://dx.doi.org/10.1007/JHEP10(2021)019}{{\em JHEP} {\bfseries 10}
  (2021) 19}, \href{http://arxiv.org/abs/2103.12994}{{\ttfamily
  arXiv:2103.12994 [hep-ex]}}.

\bibitem{Kitano:2002mt}
R.~Kitano, M.~Koike, and Y.~Okada, ``{Detailed calculation of lepton flavor
  violating muon electron conversion rate for various nuclei},''
  \href{http://dx.doi.org/10.1103/PhysRevD.76.059902}{{\em Phys. Rev. D}
  {\bfseries 66} (2002) 096002},
  \href{http://arxiv.org/abs/hep-ph/0203110}{{\ttfamily arXiv:hep-ph/0203110}}.
  [Erratum: Phys.Rev.D 76, 059902 (2007)].

\bibitem{Ilakovac:1994kj}
A.~Ilakovac and A.~Pilaftsis, ``{Flavor violating charged lepton decays in
  seesaw-type models},''
  \href{http://dx.doi.org/10.1016/0550-3213(94)00567-X}{{\em Nucl. Phys. B}
  {\bfseries 437} (1995) 491},
  \href{http://arxiv.org/abs/hep-ph/9403398}{{\ttfamily arXiv:hep-ph/9403398}}.

\bibitem{Alonso:2012ji}
R.~Alonso, M.~Dhen, M.~B. Gavela, and T.~Hambye, ``{Muon conversion to electron
  in nuclei in type-I seesaw models},''
  \href{http://dx.doi.org/10.1007/JHEP01(2013)118}{{\em JHEP} {\bfseries 01}
  (2013) 118}, \href{http://arxiv.org/abs/1209.2679}{{\ttfamily arXiv:1209.2679
  [hep-ph]}}.

\bibitem{Lindner:2016bgg}
M.~Lindner, M.~Platscher, and F.~S. Queiroz, ``{A Call for New Physics : The
  Muon Anomalous Magnetic Moment and Lepton Flavor Violation},''
  \href{http://dx.doi.org/10.1016/j.physrep.2017.12.001}{{\em Phys. Rept.}
  {\bfseries 731} (2018) 1--82},
  \href{http://arxiv.org/abs/1610.06587}{{\ttfamily arXiv:1610.06587
  [hep-ph]}}.

\bibitem{Martin2003BasisIntegrals}
S.~P. Martin, ``Evaluation of two-loop self-energy basis integrals using
  differential equations,'' {\em Physical Review D} {\bfseries 68} (2003)
  075002, \href{http://arxiv.org/abs/hep-ph/0307101}{{\ttfamily
  arXiv:hep-ph/0307101}}.

\bibitem{LaportaRemiddi2005EqualMass}
S.~Laporta and E.~Remiddi, ``Analytic treatment of the two-loop equal-mass
  sunrise graph,'' {\em Nuclear Physics B} {\bfseries 704} (2005) 349--386,
  \href{http://arxiv.org/abs/hep-ph/0406160}{{\ttfamily arXiv:hep-ph/0406160}}.

\bibitem{Smirnov2006Book}
V.~A. Smirnov, {\em Evaluating Feynman Integrals}.
\newblock Springer, 2006.

\end{thebibliography}\endgroup

\end{document}